\documentclass[a4paper,11pt]{article}
\pdfoutput=1 
\usepackage{jheppub} 
\usepackage{dcolumn}
\usepackage{bm}

\usepackage{amsmath}
\usepackage{graphicx}
\usepackage{siunitx}
\usepackage{slashed}
\usepackage{booktabs}
\usepackage{braket}
\usepackage{tikz-feynman} 
\usetikzlibrary{positioning}

\newcommand{\Lagr}{\mathcal{L}}

\preprint{MIT-CTP/5908}

\begin{document}

\title{Testing Viability of Benchmark Dark Matter Models for the Galactic Center Excess}
\author{Yongao Hu$^1$,}
\author{Cari Cesarotti$^{1, 2}$,}
\author{and Tracy R.~Slatyer$^1$}

\affiliation{$^1$Center for Theoretical
 Physics -- a Leinweber Institute, Massachusetts Institute of Technology, 77 Massachusetts Avenue, Cambridge, Massachusetts, USA}
 \affiliation{$^2$Theoretical Physics Department, CERN, 1211 Geneva 23, Switzerland}
 
 \emailAdd{yongao@mit.edu}
 \emailAdd{carissa.cesarotti@cern.ch}
 \emailAdd{tslatyer@mit.edu}

\date{\today}

\abstract{We examine the status of benchmark simplified dark matter models that have been proposed to explain the GeV gamma-ray Galactic Center excess. We constrain the available parameter space using updated observations from indirect detection, cosmology, direct detection, and accelerators. We show that there is still unconstrained parameter space in both classes of models we consider (a secluded dark sector with a vector portal coupling, and a two-Higgs doublet with a pseudoscalar mediator), and discuss the prospects for potential improvement of these constraints in future experiments.}

\maketitle

\section{Introduction}
The origin of the Galactic Center excess (GCE) is a long-standing puzzle. First identified in 2009 by Goodenough and Hooper \cite{Goodenough:2009gk}, the excess is spectrally broad with a peak around $1-3$ GeV in energy, and spatially extended out to at least $10-20^\circ$ from the Galactic Center \cite{Hooper:2013rwa, Daylan:2014rsa, Buschmann:2020adf}. The signal is an ``excess'' in the sense that it cannot be readily absorbed into standard models of the diffuse Galactic gamma-ray emission arising from cosmic rays interacting with the gas and starlight (e.g.~\cite{Calore:2014xka}). However, the inferred spatial and spectral properties of the GCE do inherit substantial uncertainties from the modeling of this background, and---in part for this reason---its physical origin remains unclear. 

It is plausible that the GCE originates from familiar but uncatalogued sources, such as an unknown population of pulsars in the stellar bulge of the Milky Way \cite{Goodenough:2009gk,Abazajian:2010zy} (or in principle, other gamma-ray-emitting stars (e.g.~\cite{Martin:2017sno}), although we are not aware of another source category that has been confirmed to produce an appropriate spectrum). Some studies have suggested the detection of point-like structure in the GCE that would support this hypothesis (e.g.~\cite{Lee:2015fea,Bartels:2015aea}), or that detected gamma-ray sources might contribute significantly to the GCE (e.g.~\cite{Malyshev:2024obk}). However, the initial searches for sources below Fermi's detection threshold have been shown to be susceptible to false positives in the presence of background mismodeling \cite{Leane:2020nmi,Leane:2020pfc}, and more recent studies with more sophisticated modeling and analysis techniques have generally found reduced or absent evidence for GCE-associated point sources compared to the initial claims \cite{Buschmann:2020adf, List:2020mzd,List:2021aer,Mishra-Sharma:2021oxe,Caron:2022akb, List:2025qbx}.\footnote{This does not exclude the pulsar hypothesis, as a sufficiently large population of dim pulsars (see e.g.~\cite{Gautam:2021wqn} for a model) would be indistinguishable from diffuse emission. However, on their face these results do not appear compatible with inferred luminosity functions for pulsar populations in globular clusters, which tend to predict more bright pulsars (e.g.~\cite{Hooper:2016rap, Dinsmore:2021nip, Amerio:2024qor}).} Perhaps the strongest argument for a pulsar (or other stellar) origin for the GCE is the claim that its morphology traces that of the Galactic stellar bulge (e.g.~\cite{Macias:2016nev,Bartels:2017vsx, Macias:2019omb, Song:2024iup}). However, the robustness of this conclusion is disputed in the literature, with the preferred morphology depending to some degree on the modeling of the Galactic diffuse emission (e.g.~\cite{DiMauro:2021raz,Cholis:2021rpp,McDermott:2022zmq,Zhong:2024vyi, Ramirez:2024oiw}). A recent study has also argued that given the merger history of the Milky Way, a bulge-like morphology may be natural for a signal sourced from the dark matter (DM) halo, making it difficult to use morphological information to confirm or refute the pulsar hypothesis \cite{Muru:2025vpz}. 

The GCE has garnered great interest from the particle physics community as a possible signal of new physics, notably DM annihilation. The GCE peaks at the Galactic Center and has an inferred spatial profile broadly consistent with expectations for the Milky Way's DM halo from theory and simulations \cite{Hussein:2025xwm, McKeown:2025zzf} (subject to the uncertainties on both the predicted and inferred morphology described above). The annihilation cross section needed to fit the data is similar to that required to obtain the correct DM relic abundance through thermal freeze-out, albeit with substantial uncertainties \cite{Daylan:2014rsa,Calore:2014nla}, and (within those uncertainties) remains broadly consistent with null results from gamma-ray observations of dwarf galaxies \cite{McDaniel:2023bju}. The spectrum is consistent with a range of Standard Model (SM) final states for the annihilation process (e.g.~\cite{Daylan:2014rsa,Calore:2014xka, Agrawal:2014oha, elor_multi-step_2015,hooper_systematic_2020}), provided the DM mass is a few 10s of GeV or heavier. 
However, no counterpart signal has been observed in direct detection experiments or at particle accelerators; this significantly constrains the range of viable DM models for the excess. 

This situation has prompted the community to broadly explore classes of DM models that could fit the GCE while evading other bounds (e.g.~\cite{carena_return_2019,hooper_systematic_2020,Cline_2014,Berlin:2014pya,Berlin:2014tja,ipek_renormalizable_2014,Martin:2014sxa,Izaguirre:2014vva,liu_signals_2015,Pospelov:2007mp}). However, many of these analyses were done roughly a decade ago, based on the best limits available at that time; since then, direct detection limits have advanced by around 2 orders of magnitude \cite{LUX:2013afz,LZ:2024zvo}, accelerators (including but not limited to the LHC) have collected much more data, and new analyses have been performed using updated cosmological and astrophysical datasets. Some of these analyses have claimed to set stringent limits on classes of models explaining the GCE, e.g.~the gamma-ray line analysis of Ref.~\cite{foster_search_2023} constrains Higgs portal models. It is thus timely to reconsider the status of these scenarios, to better understand the plausibility (or otherwise) of the DM hypothesis for the GCE.

The outline of this work is as follows. We first introduce and summarize two major classes of benchmark simplified models proposed as viable DM scenarios for the GCE, and then discuss theoretical and observational tests of these scenarios.\footnote{We use $(+---)$ as our metric choice. Unless otherwise specified, $\hbar=c=1$.} In Section \ref{sec:matching_gce}, we will discuss the DM models that can create the GCE signature. Section \ref{sec:constraints} will describe the various constraints on the DM models. Section \ref{sec:results} will report the updated results and the available parameter space in the DM models. Lastly, we will discuss the potential improvements on the constraints in future experiments in Section \ref{sec:forecast}. 

\section{Dark matter models for the GCE}
\label{sec:matching_gce}

\subsection{General considerations from the GCE rate and spectrum}

There are significant uncertainties in the cross section required to fit the GCE \cite{Calore:2014nla}, but smaller uncertainties in the cross section required to obtain the correct cosmological density of DM \cite{Steigman:2012nb}. Given the similarity of the two cross sections (in the $s$-wave-dominated case) within the uncertainties \cite{Daylan:2014rsa,Calore:2014nla}, we will focus in this work on models for the GCE that simultaneously obtain the correct relic density through the thermal freeze-out mechanism. Note that we do not claim this is the only viable mechanism or that our models represent an exhaustive set of possibilities. We focus on this scenario as a minimal solution to two outstanding questions in the SM phenomenology.

In this light, when we impose a cross section constraint, we require the DM annihilation cross section to be at the value required to match the thermal relic cross section (\qty{\sim 4.4e-26}{\cm^3\s^{-1}} for Dirac DM, corresponding to \qty{\sim 2.2e-26}{\cm^3\s^{-1}} for Majorana DM) \cite{murgia_fermilat_2020,Steigman:2012nb}. Further, the annihilation must be predominantly through an $s$-wave process at low velocity \cite{hooper_systematic_2020}.

The spectrum of the GCE can be well described by scenarios where the DM annihilates to quarks that subsequently hadronize, and/or where cascade decays occur in the dark sector
\cite{elor_multi-step_2015}. Hadronization (of quarks or gluons) generically produces neutral pions which decay to gamma-rays with a near-unity branching ratio and which can produce approximately the correct spectrum. The $b\bar{b}$ final state has often been used as a benchmark, but other final states with hadronic decays can also provide a good fit to the data \cite{elor_multi-step_2015,Calore:2014xka}. For scenarios of this type, the GCE is generally best fitted by DM particle candidates of 30-\qty{70}{\GeV} in mass \cite{liu_signals_2015,hooper_systematic_2020,ipek_renormalizable_2014}. This will inform the mass range we consider. However, given the spectral uncertainties, we will not place a stringent prior on the branching ratios to different SM final states, provided there is a substantial branching ratio to hadronically decaying states \cite{Abdughani:2021oit,hooper_systematic_2020,Cuoco:2016jqt}. We show some example spectra in Appendix \ref{sec:spectra}. 

We note that some papers have found good fits to the spectrum with an even wider range of final states. For example, Ref.~\cite{Agrawal:2014oha} finds that DM masses up to 300 GeV can give an acceptable fit. Ref.~\cite{DiMauro:2021qcf} finds that a good fit is achieved for 60 GeV thermal relic DM annihilating to muons with a thermal relic cross section, once inverse Compton scattering of charged particles on the Galactic radiation field is included, and that scenarios of this type ameliorate tension with upper limits on the antiproton flux. Leptonic final states with a significant inverse Compton contribution are discussed more broadly in e.g.~Ref.~\cite{Kaplinghat:2015gha}. We will focus in this work on models where the DM mass is below 100 GeV and the final state is hadronic, but will comment briefly on these broader possibilities as appropriate.

In addition to the model-dependent constraints that will be the main focus of this work, there are tests of the DM hypothesis for the GCE that are nearly model-independent: in particular, searches for an identical counterpart signal in gamma-rays from dwarf galaxies. If we specify that the final state is comprised of quarks, this furthermore automatically implies a minimum flux of antiprotons from dark matter annihilation. While neither of these limits is truly model-independent (see e.g. \cite{Berlin:2025fwx} for an example of a class of models that evades bounds from dwarf galaxies) they are more generic than bounds from terrestrial experiments. Thus one might ask about the consistency of our broad requirements (DM mass below 100 GeV and hadronic final states) with these limits.

Current limits on hadronic final states from dwarf galaxy observations constrain cross sections comparable to those needed to fit the GCE \cite{McDaniel:2023bju}; for example, for 50-100 GeV DM annihilating to $b$ quarks, the limit is in the range $1-2\times10^{-26}$ cm$^3$/s for Majorana DM (which is slightly weaker than expected, but not to a degree that reflects a significant excess). There are prospects for improving the sensitivity of dwarf galaxy searches with either a successor gamma-ray instrument, or with the discovery of new dwarf galaxies whose locations can then be studied in archival data from the Fermi Gamma-Ray Space Telescope. We will keep the standard thermal relic cross section as our main benchmark but also discuss the impact on the allowed parameter space of a slightly smaller cross section, which would reduce tension with the dwarf limits.

There is a long-standing claim of a possible GCE counterpart signal in antiproton observations from AMS-02 (e.g.~\cite{Cuoco:2016eej, Cui:2016ppb, Cholis:2019ejx}), although the significance of that signal was found to be marginal by some later studies (e.g.~\cite{Boudaud:2019efq, Heisig:2020nse}). Ref.~\cite{DiMauro:2021qcf} found there is little evidence for an excess in more recent AMS-02 antiproton data \cite{AGUILAR20211}, and that hadronic final states are in some tension with these newer data (preferring a cosmic-ray halo of  smaller extent than favored by other cosmic-ray data). In the bulk of this work we will not focus on alleviating this tension, but we will briefly discuss variations on the secluded hypercharge model that could potentially achieve this goal. 

Similarly, studies of the M31 (Andromeda) galaxy in the radio band have been used to set constraints on DM annihilation \cite{Egorov:2013exa, Chan:2016wpu, Egorov:2022fkc, Weikert:2023cyv} (with one study claiming a possible GCE counterpart \cite{Chan:2021dyy}), via searching for synchrotron emission from charged annihilation products diffusing in magnetic fields. Given the uncertainties in the diffusion modeling, these constraints do not yet seem to be in robust tension with the DM interpretation of the GCE \cite{Weikert:2023cyv}.

\subsection{Constraints from the cosmological history}

In this paper, we consider the \emph{freeze-out} model of DM production. The freeze-out mechanism assumes that in the radiation-dominated era, DM particles were in thermal equilibrium with SM particles until the DM annihilation  rate dropped below the Hubble expansion rate; at that point the particles thermally decouple and the DM annihilation `freezes out' \cite{early_universe_kolb}. 

The DM particle comoving density then roughly remains constant afterwards, resulting in what we term \emph{thermal relics}, whose annihilation cross-section explains both the DM abundance and the GCE. Successful freeze-out (obtaining the observed relic density) constrains the interaction between DM and SM particles. 

 In this paper, we will denote this scenario as the \emph{standard cosmological history} (also referred to as the ``WIMP Next Door'' in Ref.~\cite{Evans:2017kti}). In particular, this assumption requires the coupling between the DM and SM sector to be large enough to fully equilibrate the two sectors prior to freeze-out. More complicated cosmological histories are also able to produce the thermal relic density \cite{Alenezi:2025kwl,Chu:2011be}, but they might lead to a greater discrepancy between the cross sections required to match the observed density and to produce the GCE, and would require a more in-depth calculation; we do not consider such scenarios in this work. 

\subsection{Models}
There are multiple classes of theories that can produce an excess of gamma-rays at GeV energies with features similar to the GCE. Based on the considerations above, we restrict our attention to models that (i) feature an $s$-wave annihilation channel which reproduces the thermal relics cross section $\langle\sigma v\rangle_{\text{th}}\simeq4.4\times10^{-26}\text{ cm}^3\text{s}^{-1}$ (for Dirac DM), (ii) yield SM final states that hadronize or decay producing abundant photons, in order to reproduce the characteristic 1–3 GeV bump in the GCE for dark-matter masses near 30–70 GeV, and (iii) keep the dark and SM sectors in full thermal equilibrium until freeze-out, thereby fixing a minimal portal coupling. The final condition for viability is that the model features (iv) sufficient suppression of all other signals to evade the current bounds from indirect detection experiments, direct detection experiments, and from LHC searches; determining the remaining parameter space given these bounds will be the main focus of this work. 

Over the past decade, the constraints in (iv) have become significantly more stringent. For example, one early work mapping out simplified models for the GCE \cite{Berlin:2014tja} considered scenarios where the DM annihilates directly to SM particles through a mediator, and identified 16 distinct scenarios for the DM and mediator spins that could explain the GCE without violating constraints from colliders or direct detection. However, seven of those scenarios have now been excluded by updated direct-detection limits \cite{LZ:2024zvo}, at least in the regime assumed in Fig.~9 of that work;\footnote{This corresponds to taking the mediator mass to be significantly heavier than the DM mass, and the couplings of the mediator to the SM to be flavor-universal (+ proportional to the SM particle mass in the case of a scalar mediator) \cite{privcommDan}.} eight of the remaining nine scenarios correspond to scalar mediators, with the exception involving a vector mediator with axial-vector couplings to SM fermions. Thus recent improvements in terrestrial searches can certainly constrain the space of possibilities for a DM origin of the GCE. At the same time, recent studies have identified models for the GCE that remain compatible with all constraints (e.g.~for two very recent examples, see \cite{Koechler:2025ryv, Hooper:2025fda}).

In this work, we update the status of two classes of models that are relatively simple, UV-complete, and were found to meet all requirements when they were first proposed as explanations for the GCE. One major class is the secluded dark sector, where the dark sector is coupled to the SM sector through a metastable mediator with a lower mass than the DM candidate \cite{Cline_2014, Martin:2014sxa,hooper_systematic_2020}. As a result, secluded models can naturally evade both direct detection and collider production with an extremely small cross section \cite{Pospelov:2007mp}, by reducing the coupling of the mediator to the SM without reducing the annihilation rate of the DM into mediators.  One such minimal example on which we focus in this paper is \emph{the secluded hypercharge model}, where a dark gauge boson of a U(1)$_\text{D}$ symmetry kinetically mixes with the gauge boson of the SM U(1)$_\text{Y}$ symmetry \cite{hooper_systematic_2020}. This model is the simplest UV-complete vector portal and has been well studied in the context of the GCE (e.g.~\cite{Cline_2014,Martin:2014sxa,Escudero:2017yia,hooper_systematic_2020}). 

The other model we consider is a two-Higgs doublet model (2HDM) with an additional pseudoscalar mediator that couples to a fermionic DM candidate (\emph{the 2HDM+$a$ model}). The 2HDM+$a$ model is the minimal pseudoscalar portal that survives direct detection at one loop, making it of much interest in explaining the GCE \cite{ipek_renormalizable_2014,hooper_systematic_2020,bauer2018introduction}. Further, the 2HDM+$a$ model is UV-complete, with an extended Higgs sector which is a common feature in beyond the Standard Model (BSM) models \cite{Branco:2011iw,Cabrera:2019gaq} and well motivated by supersymmetry \cite{gunion2000higgs,weinberg2000quantum3}. It has also been discussed as a possible explanation for tentative hints in favor of a 95 GeV resonance at the LHC (e.g.~\cite{Arcadi:2023smv}). A closely related model, where the DM is a pseudo-Nambu-Goldstone boson, has been discussed in the context of the GCE and LHC excesses by Ref.~\cite{Biekotter:2021ovi}.

In the following sections, we will explain the calculation details of both the secluded hypercharge model and the 2HDM+$a$ model. 

\subsubsection{Secluded hypercharge model}
Here, we offer a review of the secluded hypercharge model. We consider an additional $U(1)_D$ symmetry. The particle content of this secluded sector is a heavy dark Dirac fermion $\chi$ (charged under $U(1)_D$) and a lighter vector mediator $\hat{D}$, which is the gauge boson of $U(1)_D$. The $\hat{D}$ kinetically mixes with the gauge boson of $U(1)_Y$  with a mixing parameter $\varepsilon$ \cite{holdom_1986,bauer2018introduction}:
\begin{equation}
    \label{eq:kinetic_mixing_hyper}
    \Lagr \supset-\frac{1}{4}\hat{B}^{\mu \nu}\hat{B}_{\mu \nu} -\frac{1}{4}\hat{D}^{\mu \nu}\hat{D}_{\mu \nu}-\frac{\varepsilon}{2}\hat{D}^{\mu \nu}\hat{B}_{\mu \nu},
\end{equation}
where the $U(1)_Y$ field strength is $\hat{B}_{\mu \nu}$ and the $U(1)_D$ field strength is $\hat{D}_{\mu \nu}$.

We add a mass term for the dark gauge boson of mass $\hat{m}_D$: 
\begin{equation}
    \Lagr\supset \frac{1}{2}\hat{m}_D^2\hat{D}^\mu\hat{D}_\mu.
\end{equation}

We do not include a dynamical dark Higgs that gives mass to the $\hat{D}$ in the model. If the $\hat{D}$ gains its mass due to the Higgs mechanism, we assume that the dark Higgs has a higher mass than the dark gauge boson and the dark fermion, making it less relevant to the low-energy physics that is relevant to the GCE. It is also possible to give the $\hat{D}$ a mass through the  St\"uckelberg mechanism, without invoking a dynamical dark Higgs \cite{Ruegg:2003ps}. 
Our results are largely insensitive to the mass-generating mechanism for the dark gauge boson (although a sufficiently light dark Higgs field would impact the phenomenology).

After the electroweak symmetry breaking, the neutral gauge sector
contains $\hat Z_\mu = c_w \hat W_{3\mu} - s_w \hat B_\mu$ and $\hat D_\mu$. We use the shorthand $c_a = \cos a$ and $s_a = \sin a$, and $w$ the Weinberg angle. We can define the small mixing angle $\xi$ and the auxiliary angle $x$ with
\begin{equation}
  \label{angle_mixing}
  \tan 2\xi =
    \frac{2\varepsilon s_w}
         {1 - \hat m_D^{\,2}/\hat m_Z^{\,2}},
  \qquad
  \sin x = \varepsilon. 
\end{equation}

Diagonalizing the mass matrix gives the physical masses of the gauge bosons $Z$ and $Z'$ \cite{holdom_1986,bauer2018introduction,Alenezi:2025kwl}:
\begin{equation}
  \label{interaction_diagonalization}
  \begin{pmatrix}
     \hat A \\ \hat Z \\ \hat D
  \end{pmatrix}
  =
  \begin{pmatrix}
       1 &
      -\dfrac{c_w s_x s_\xi}{c_x} &
      -\dfrac{c_w s_x c_\xi}{c_x} \\[6pt]
       0 &
       c_\xi+\dfrac{s_w s_x s_\xi}{c_x} &
       \dfrac{s_w s_x c_\xi}{c_x}-s_\xi \\[6pt]
       0 &
       \dfrac{s_\xi}{c_x} &
       \dfrac{c_\xi}{c_x}
  \end{pmatrix}
  \begin{pmatrix}
     A \\ Z \\ Z'
  \end{pmatrix}.
\end{equation}

We can read off the physical masses as \cite{bauer2018introduction}:
\begin{equation}
    m_A^2=0,\quad m_Z^2=\hat m_Z^2\left[1+s_x^2s_w^2\left(1+\frac{\hat m_D^2}{\hat m_Z^2}\right)\right],\quad m_{Z'}^2=\hat m_D^2\left(1+s_x^2c_w^2\right).
\end{equation}

We can then obtain the covariant derivative \cite{holdom_1986,Alenezi:2025kwl}: 
\begin{equation}
\label{eq:covariant_full}
\begin{aligned}
D_\mu &\equiv \partial_\mu - i\hat g_D Q_D \hat D_\mu - i g_Y Q_Y \hat B_\mu - i g T^{a} W^{a}_\mu\\
&=
\partial_\mu
- i e QA_\mu- i g T^{a} W^{a}_\mu - i\Bigl[
      e\varepsilon s_\xi Q
    + \tfrac{e}{s_w c_w}\bigl(c_\xi+\varepsilon s_ws_\xi\bigr)(T_3-s_w^{2}Q)
    + g_D Q_D s_\xi
  \Bigr]Z_\mu\\
&
- i\Bigl[
      e\varepsilon c_\xi Q
    + \tfrac{e}{s_w c_w}\bigl(s_\xi-\varepsilon s_w c_\xi\bigr)(T_3-s_w^{2}Q)
    + g_D Q_D c_\xi
  \Bigr]Z'_\mu.
\end{aligned}
\end{equation}
where $g_D$ is the coupling of $Z'$, $Q_D$ is the dark charge, and $\hat{g}_D=g_D/\sqrt{1-\varepsilon^2}$. For the SM parameters, $g_Y$ is the hypercharge coupling constant, $Q_Y$ is the hypercharge, $e$ is the electric charge, $g$ is the weak isospin coupling constant, and $T^a$ labels the $SU(2)$ generators. 

Absorbing $Q_D$ into $g_D$, we can then obtain the following interaction term for SM fermions $\psi$ and the dark fermion $\chi$
\cite{holdom_1986,bauer2018introduction,Alenezi:2025kwl}:
\begin{equation}
\label{eq:hypercharge_lagr}
\begin{aligned}
& \Lagr \supset Z_\mu\Bigl[\bar{\psi}\gamma^\mu\Bigl(e\varepsilon s_\xi Q+\tfrac{e}{s_w c_w}(c_\xi+\varepsilon s_w s_\xi)(T_3-s_w^{2}Q)\Bigr)\psi+\bar{\psi}\gamma^\mu\gamma^{5}\Bigl(\tfrac{e}{s_w c_w}(c_\xi+\varepsilon s_w s_\xi)T_3\Bigr)\psi\Bigr]\\
&+Z'_\mu\Bigl[\bar{\psi}\gamma^\mu\Bigl(e\varepsilon c_\xi Q+\tfrac{e}{s_w c_w}(s_\xi-\varepsilon s_w c_\xi)(T_3-s_w^{2}Q)\Bigr)\psi+\bar{\psi}\gamma^\mu\gamma^{5}\Bigl(\tfrac{e}{s_w c_w}(s_\xi-\varepsilon s_w c_\xi)T_3\Bigr)\psi\Bigl]\\
&+A_\mu\Bigl(eQ\bar{\psi}\gamma^\mu\psi\Bigr)+g_D\Bigl(Z_\mu s_\xi+ Z_\mu ' c_\xi\Bigr) \bar{\chi}\gamma^\mu\chi.
\end{aligned}
\end{equation}

\begin{figure}
    \centering
    \begin{tikzpicture}
  \begin{feynman}
    \vertex (chi1) at (0, 1)   {$\chi$};
    \vertex (chi2) at (0,-1)   {$\chi$};
    \vertex (v1)   at (1, 0);
    \vertex (v2)   at (3, 0) ;
    \vertex (f1)   at (4, 1)   {$f$};
    \vertex (f2)   at (4,-1)   {$\bar f$};
    
    \diagram*{
      (chi1) -- [fermion] (v1) -- [fermion] (chi2),
      (v1)   -- [photon, edge label=$Z'/Z$] (v2) ,
      (v2)   -- [fermion]       (f1),
      (v2)   -- [anti fermion]  (f2),
    };
  \end{feynman}
  1) Annihilation (s‑channel) across the middle
  \draw[->, thick]
    ($ (chi1) + (-1,1) $) 
  -- 
    ($ (chi2)  + (-1,-1) $)
  node[midway, above=2pt, rotate = 90] {Direct detection};

  \draw[->, thick]
    ($ (f1)   + ( 0.5,1) $)
  --
   ($ (chi1) + (-0.5,1) $) 
  node[midway, above=2pt] {Collider production};

  \draw[->, thick]
    ($ (chi2) + (-0.5,-1) $) 
  -- 
    ($ (f2)   + ( 0.5,-1) $)
  node[midway, below=2pt] {Annihilation, Indirect detection};
\end{tikzpicture}
    \caption{Feynman diagram illustrating $\chi$ interactions within the secluded hypercharge model. The process shows an $s$-channel annihilation into SM fermions, and the arrows indicate the relationship to collider production and direct detection.}
    \label{fig:hypercharge_arrows}
\end{figure}

\subsubsection*{Summary of parameters in the secluded hypercharge model}
There are four parameters in the secluded hypercharge model: the kinetic mixing parameter $\varepsilon$, the dark sector coupling $g_D$, the dark vector mediator mass $m_{Z'}$, and the dark fermion mass $m_\chi$. If we assume a standard cosmological history, $g_D$ is determined by requiring that the annihilation produces the correct DM thermal relic density, and not independent \cite{Alenezi:2025kwl} (a detailed discussion is offered in Section \ref{sec:secluded_relic}).  

The parameter space we consider is:
\begin{itemize}
    \item mass of the DM fermion $10\text{ GeV}\leq m_\chi\leq 100\text{ GeV}$ to produce the signature of the GCE,
    \item mass of the dark mediator $m_{Z'}<m_\chi$ at each $m_\chi$ to remain in the secluded regime, 
    \item and kinetic mixing $\qty{e-9}{}\leq\varepsilon\leq\qty{e-4}{}$ so that the cosmological history can remain standard (i.e. allowing the dark sector to reach equilibrium with the SM), while evading direct detection constraints. 
\end{itemize}

\subsubsection{Two-Higgs doublet model}
Next, we offer a review on the two-Higgs doublet model. We consider an extended Higgs sector with two complex doublets (2HDM). The most general CP-conserving 2HDM potential is \cite{ipek_renormalizable_2014,bauer2018introduction}: 
\begin{equation}
    \label{2hdm}
\begin{aligned}
& V(H_1, H_2)_{\textsc{2HDM}}= \lambda_1\left(H_1^\dag H_1-\frac{v_1^2}{2}\right)^2+\lambda_2\left(H_2^\dag H_2-\frac{v_2^2}{2}\right)^2+\lambda_3\left[\left(H_1^\dag H_1-\frac{v_1^2}{2}\right)+\left(H_2^\dag H_2-\frac{v_2^2}{2}\right)\right]^2\\
&+\lambda_4\left[\left(H_1^\dag H_1\right)\left(H_2^\dag H_2\right)-\left(H_1^\dag H_2\right)\left(H_2^\dag H_1\right)\right]+\lambda_5\left[\operatorname{Re}\left(H_1^\dag H_2\right)-\frac{v_1v_2}{2}\right]^2+\lambda_6\left[\operatorname{Im}\left(H_1^\dag H_2\right)\right]^2.
\end{aligned}
\end{equation}
where $\lambda_a$ are the constants of the 2HDM, and $v_i$ is the corresponding vacuum expectation value (vev) of each Higgs doublet $H_i$. 

In unitary gauge we can write \cite{ipek_renormalizable_2014}
\begin{equation}
    \label{2hdm,coupling}
    H_i=\frac{1}{\sqrt{2}}\begin{pmatrix}
        \sqrt{2}\phi_i^+ \\
        v_i+\rho_i+i\eta_i
    \end{pmatrix},
\end{equation}
\begin{equation}
    \label{2hdm,auxillary}
    v_1^2+v_2^2=v_{\textsc{SM}}^2,
\end{equation}
\begin{equation}
    \label{2hdm,auxillary2}
    \tan{\beta}=\frac{v_2}{v_1}.
\end{equation}

After symmetry breaking, we will have 5 physical states: CP-even $h$ and $H$, CP-odd $A$, and charged $H^\pm$. We have a pair of neutral Higgs bosons with mixing parameter $\alpha$,
\begin{equation}
    \label{2hdm_mix_2}
    \begin{pmatrix}
h \\ H
\end{pmatrix}
=
\begin{pmatrix}
    -\sin\alpha & \cos\alpha \\
    \cos\alpha & \sin\alpha
\end{pmatrix}^{-1}
\begin{pmatrix}
    \rho_1 \\ \rho_2
\end{pmatrix},
\end{equation}
a pseudoscalar boson,
\begin{equation}
    \label{2hdm mass}
    A_0=\sin{\beta}\eta_1-\cos{\beta}\eta_2,
\end{equation}
and a pair of charged Higgs bosons,
\begin{equation}
    \label{2hdm,charged}
    H^\pm=\sin{\beta}\phi_1^\pm-\cos{\beta}\phi_2^\pm.
\end{equation}

\subsubsection*{2HDM with a pseudoscalar mediator (2HDM+$a$)}
To add a pseudoscalar portal for the GCE, we introduce a DM candidate Dirac fermion $\chi$ with mass $m_\chi$ coupled to a real singlet pseudoscalar mediator $a_0$, which then couples to the SM by mixing with the pseudoscalar $A_0$ with mixing parameter $B$ \cite{ipek2015galactic,Bauer:2017ota}: 
\begin{equation}
    \label{2hdm+pseudo,dark}
    \Lagr_D\supset g_\chi a_0\bar{\chi}i\gamma^5\chi,
\end{equation}
\begin{equation}
    \label{2hdm+pseudo,couple}
    V=V_{\textsc{2HDM}}+\frac{1}{2}m_{a_0}^2a_0^2+\frac{\lambda_a}{4}a_0^4+i Ba_0H_1^\dag H_2+\textrm{h.c.}\,.
\end{equation}

We assume $\Lagr_D$ and $V$ are CP-conserving---thus $a_0$ does not have a vev, and we do not need to worry about large mass corrections or mixing of pseudoscalar and scalar mediators. We can define the mixing angle $\theta$ needed to rotate to the mass eigenstates $A$ and $a$:
\begin{equation}
    \label{2hdm_mix}
    \begin{pmatrix}
A_0 \\ a_0
\end{pmatrix}
=
\begin{pmatrix}
    \cos\theta & \sin\theta \\
    -\sin\theta & \cos\theta
\end{pmatrix}
\begin{pmatrix}
    A \\ a
\end{pmatrix},
\end{equation}
\begin{equation}
    \label{2hdm,mix,theta}
    \tan{2\theta}=\frac{2Bv_{\textsc{SM}}}{{m_{A_0}^2}/{m_{a_0}^2}},
\end{equation}
\begin{equation}
    \label{2hdm,mix,mass}
    m_{a,A}^2=\frac{1}{2}\left[m_{A_0}^2+m_{a_0}^2\pm\sqrt{(m_{A_0}^2-m_{a_0}^2)^2+4B^2v_{\textsc{SM}}^2}\right].
\end{equation}

The Lagrangian can then be simplified as:
\begin{equation}
    \label{2hdm_portal}
    V_\text{port} = \frac{1}{2v_{\textsc{SM}}} (m_{A}^{2} - m_{a}^{2}) \left[ \sin{4\theta}\, a A + \sin^{2}{2\theta}\, (A^{2} - a^{2}) \right] \left[ \sin(\beta - \alpha) h + \cos(\beta - \alpha) H \right],
\end{equation}
\begin{equation}
    \label{2hdm+pseudo}
\Lagr_D\supset g_\chi(\cos{\theta}a+\sin{\theta}A)\bar{\chi}i\gamma^5\chi    .
\end{equation}

The interaction vertices involving heavier SM fermions ($t$, $b$, and $\tau$) are \cite{Bauer:2017ota}: 
\begin{equation}
    \Lagr \supset \sum_{f=t, b,\tau}\frac{m_f}{v_\text{SM}}\bar f\bigl[ i\xi_f(\cos\theta A - \sin\theta a)\gamma_5\bigr]f.
\end{equation}

There are multiple types of 2HDM models where down-type quarks, up-type quarks, and leptons couple to different doublets. In this paper, we consider the type-II 2HDM model, which yields similar Higgs sector phenomenology as the Minimally Supersymmetric Standard Model (MSSM) \cite{weinberg2000quantum3}. In the type-II 2HDM model, $H_1$ couples to down-type quarks and leptons while $H_2$ couples to up-type quarks \cite{Bauer:2017ota,bauer2018introduction}:
\begin{equation}
    \xi_t=\cot\beta,\quad \xi_b=\xi_\tau=\tan\beta.
\end{equation}

In the type-II 2HDM model, $\tan\beta$, or the ratio between two vevs, controls the strength of coupling to down-type quarks compared to up-type quarks: the higher $\tan\beta$ is, the less the mediator couples to top quarks compared to bottom quarks and leptons.

To ensure that the light $h$ boson acts as the SM Higgs boson to give the familiar behavior of the SM, and that the electroweak precision limit is met \cite{atlascollaboration20232hdm+1}, we set $m_h\ll m_A=m_H=m_{H^\pm}$, and work in the decoupling limit $\cos(\beta-\alpha)=0$ \cite{abe2018lhc,su2019exploring}. Moreover, we set  $m_a\ll m_A$, so that interactions involving the $A$ mediator are suppressed at the lower energy scale relevant to the GCE. To keep the couplings perturbative, $m_A=m_H=m_{H^\pm}\leq\mathcal{O}(\qty{1}{\TeV})$ \cite{Bauer:2017ota}.

\subsubsection*{Summary of parameters in the 2HDM+$a$ model}
Following the recommendation from the LHC Working Group \cite{LHCHiggsCrossSectionWorkingGroup:2016ypw}, the list of parameters that we set are: $\lambda_1=\lambda_2=\lambda_3=3$, $m_h=125\text{ GeV}$, $v_{\textsc{SM}}=246\text{ GeV}$, $m_H=m_A=m_{H^\pm}=800\text{ GeV}$. 

The parameters we vary are: 
\begin{itemize}
    \item the dark Dirac fermion mass $\qty{10}{\GeV}\leq m_\chi\leq\qty{70}{\GeV}$ to facilitate matching the gamma-ray spectrum for the GCE,
    \item the light pseudoscalar mediator mass $m_a\leq\qty{125}{\GeV}$, 
    \item the ratio of the two vevs $5\leq\tan\beta\leq 40$; we will see that at lower $\tan \beta$ values there are stringent indirect-detection limits, while at higher $\tan \beta$ the bottom and tau Yukawa interactions would start to become non-perturbative, 
    \item the dark sector pseudoscalar coupling $0.5\leq g_\chi\leq 1$ so that the DM sector remains perturbative while being able to reproduce the thermal relic density, 
    \item and the mixing angle $\theta$ between pseudoscalar mediators $A$ and $a$, with $\qty{e-3}\leq\theta\leq 1$, so that the relevant interactions remain perturbative and we are still able to reproduce the thermal relic density. 
\end{itemize}

\section{Constraints}
\label{sec:constraints}
Given the advancement in direct detection constraints, indirect detection constraints, and collider constraints over the last decade, let us now examine whether these models retain unconstrained phase space in the regime consistent with the GCE. 

\subsection{The secluded hypercharge model}

\subsubsection{Relic density calculation}
\label{sec:secluded_relic}
First, we calculate the appropriate annihilation cross section to reproduce the thermal relic density. In the secluded regime, $\chi\bar{\chi} \rightarrow Z'Z'$ is the dominant annihilation channel (Fig. \ref{fig:hypercharge_feynman}(a)). 

\begin{figure}[ht]
\centering

\begin{minipage}{0.3\textwidth}
\centering
    \begin{tikzpicture}
\begin{feynman}
  \vertex (chi1) at (-1.5,  1) {$\chi$};
  \vertex (chi2) at (-1.5, -1) {$\bar\chi$};
  \vertex (v1)    at (0,1);
  \vertex (v2)   at (0,-1);
  \vertex (z1)   at ( 1.5,  1) {$Z'$};
  \vertex (z2)   at ( 1.5, -1) {$Z'$};

  \diagram*{
    (chi1) -- [fermion]      (v1) -- [boson] (z1),
    (chi2) -- [anti fermion] (v2) -- [boson] (z2),
    (v1) -- [fermion] (v2)
  };
\end{feynman}
\node at (0, -1.5) {(\textbf{a})};
\end{tikzpicture}
\end{minipage}
\begin{minipage}{0.3\textwidth}
\centering
\begin{tikzpicture}
    \begin{feynman}
        \vertex (z) at (-1,0) {$Z'$};
        \vertex (v) at (0.5,0);
        \vertex (f1) at (1.5,1) {$f$};
        \vertex (f2) at (1.5,-1) {$f'$};

        \diagram*{(z) -- [boson] (v) -- [fermion] (f1), (v) -- [anti fermion] (f2)};
    \end{feynman}
    \node at (0, -1.5) {(\textbf{b})};
\end{tikzpicture}
\end{minipage}

\caption{The processes to keep the equilibrium in early universe and achieving the relic abundance. (a): Dominant annihilation in dark sector in secluded regime $\chi\bar{\chi}\to Z' Z'$. (b): Decaying of dark boson back into the SM sector $Z' \to f\bar{f}$.}
\label{fig:hypercharge_feynman}
\end{figure}

The $s$-wave thermally-averaged annihilation cross section is \cite{Alenezi:2025kwl}
\begin{equation}
    \label{eq:secluded_z}
    \langle\sigma v\rangle=\frac{g_D^4}{16\pi m_\chi^2}\frac{ \left(1-\frac{m_{Z'}^2}{m_\chi^2} \right)^{\frac{3}{2}}}{ \left(1-\frac{m_{Z'}^2}{2m_\chi^2} \right)^2}.
\end{equation}
This expression is valid for $m_{Z'}/m_\chi \lesssim 0.95$ which covers most of our parameter space for the secluded regime. At $m_{Z'}\approx m_\chi$, velocity dependence becomes relevant and modifies the annihilation cross section (Eq.~\eqref{eq:secluded_z} is evaluated in the zero-velocity limit).  Near threshold $m_{Z'}\to m_\chi$, the center-of-mass energy is approximately $s = 4m_\chi^{2}(1+v^{2}/4)$, so the final-state velocity becomes
\begin{equation}
    \beta_f = \sqrt{1-m_{Z’}^{2}/s}\simeq\sqrt{\Delta+v^{2}/4},
\end{equation}
with $\Delta\equiv1-m_{Z’}^{2}/m_\chi^{2}$.  When $\Delta\to0$, the phase-space factor no longer vanishes at finite velocity. Instead, $\beta_f\propto v$ and the $s$-wave cross section scales as $\langle\sigma v\rangle\propto v$. 

Under a standard cosmological history, the WIMP thermally-averaged annihilation cross section for thermal relics if the DM candidate is a Majorana fermion for $m_\chi\geq\qty{10}{\GeV}$ has to be \cite{Steigman:2012nb}: 
\begin{equation}
    \label{dm_density_ma}
    \langle\sigma v\rangle_\text{Majorana}\sim\qty{2.2e-26}{\cm^{3}\s^{-1}}.
\end{equation}

For the dark Dirac fermion in our model, we need to double the annihilation cross section to reach the same thermal relic density \cite{Steigman:2012nb}:
\begin{equation}
    \label{dm_density_z}
    \langle\sigma v\rangle\sim\qty{4.4e-26}{\cm^{3}\s^{-1}}\approx \qty{3.8e-9}{\GeV^{-2}}.
\end{equation}

We can then calculate the dark sector coupling $g_D$ as a function of $m_\chi$ and $m_{Z'}$. As we are assuming the $Z'$ to stay in thermal equilibrium with the SM throughout freeze-out, we can eliminate the $Z^\prime$ from the Boltzmann evolution equation, and effectively recover the standard calculation for the thermal relic density of the DM (the $Z^\prime$ is effectively just another field in the SM thermal bath). This argument does not rely on the $Z^\prime$ remaining relativistic, and so remains valid even for $m_{Z'}$ close to $m_\chi$.

\subsubsection{Thermalization}
Next, we consider the constraint of the standard cosmological history. As the GCE requires models where $m_\chi > m_{Z'}$, the main decay process that we consider is $Z' \rightarrow f^+ f^-$ (Fig. \ref{fig:hypercharge_feynman}(b)) to SM fermions $f$. 

For the secluded hypercharge model, the tree level decay width of $Z'\to f\bar{f}$ is calculated to be \cite{bauer2018introduction,hooper_systematic_2020}: 
\begin{equation}
    \label{eq:decay_z}
    \Gamma_\text{HC, FO} = \sum_{f} \frac{N_C^fm_D}{12\pi}\sqrt{1-\frac{4m_f^2}{m_D^2}}\left[ (g_V^{f})^{2} \left(1+\frac{2m_f^2}{m_D^2} \right)
    +  (g_A^{f})^{2} \left(1-\frac{4m_f^2}{m_D^2} \right)\right].
 \end{equation}
where $N_C^f$ is the color factor of the SM fermion $f$, $g_V^f$ is the vector coupling of fermion $f$ with the $Z'$, and $g_A^f$ is the axial vector coupling of fermion $f$ with the $Z'$ from \eqref{eq:hypercharge_lagr}.  

In the standard cosmological history, the dark sector remains in equilibrium with the SM until the freeze-out temperature, which yields a lower bound for the kinetic mixing between the SM and DM sector. We estimate this by approximating the freeze-out temperature $T\sim {m_\chi}/{20}$ \cite{bauer2018introduction} to calculate the Hubble constant $H$ at the freeze-out temperature in the radiation-dominated era \cite{early_universe_kolb}. The relativistic degrees of freedom $g_\ast$ are taken as a function of temperature from \cite{Steigman:2012nb}, and we approximate the Hubble rate during radiation domination as:
\begin{equation}
    \label{eq:hubbles_rd}
    H_\text{FO} = 1.66 g_\ast^{1/2} \frac{T_\text{FO}^2}{m_\text{Pl}}.
\end{equation}

To ensure that the DM and the SM sector remain at equilibrium at freeze-out, we require that $\Gamma_\text{HC, FO}\geq H_\text{FO}$ and calculate the freeze-out temperature as a function of the kinetic mixing parameter $\varepsilon$. We then obtain a lower limit of $\varepsilon$ for dark sector to remain in equilibrium in early universe. (This approach is similar to the estimate given in Ref.~\cite{Pospelov:2007mp}; we obtain a somewhat weaker limit on $\varepsilon$ than that work, possibly because we include more SM final states when computing the decay width.)

Refs.~\cite{Evans:2017kti, Alenezi:2025kwl} provide a more in-depth analysis of the cosmological history in the case of a light mediator, $m_{Z'} \lesssim 0.1 m_\chi$. Ref.~\cite{Evans:2017kti} argued that for $T \gg m_{Z'}$, it is not the decay lifetime of the $Z'$ that dominates the coupling to the SM, but instead $2\rightarrow 2$ processes of the form SM SM $\rightarrow Z^\prime$ SM. Ref.~\cite{Alenezi:2025kwl} pointed out more recently that in-medium effects suppress the $Z'$ coupling to the SM for $T \gg m_{Z'}$, so in fact the dominant processes coupling the two sectors  require the involvement of one or more DM particles; neglecting the direct interactions of $Z'$ bosons with the SM, they find that full thermalization only occurs for significantly higher values of $\varepsilon$ (compared to the result neglecting in-medium effects, where dark photon emission from the SM is important). Thus our $\Gamma_\text{HC, FO}$-based estimate would underestimate  the true thermalization bound on $\varepsilon$ for $m_{Z'} \lesssim 0.1 m_\chi$, but for higher $Z'$ mass we do expect these decays (and their inverse decays) to dominate the energy transfer between the sectors, and our estimate should thus be valid.  Furthermore, we find that the light-mediator region $m_{Z'} \lesssim 0.1 m_\chi$ is already quite constrained by direct detection, in agreement with Ref.~\cite{Alenezi:2025kwl}.

A more detailed calculation of the thermalization bound for $m_{Z'} \gtrsim 0.1 m_\chi$ can be performed by solving the Boltzmann equations governing the number density and energy density of the two sectors. We neglect plasma effects because this regime corresponds to the $Z'$ mass being larger than or equal to the temperature of the thermal bath around freezeout; a careful treatment of plasma effects would be needed to capture the transition to the light-$Z'$ regime studied in Ref.~\cite{Alenezi:2025kwl}, and might be an interesting direction for future work. We adapt the Boltzmann equations given in Ref.~\cite{Fitzpatrick:2020vba} to the case where $m_{Z'} < m_\chi$:
\begin{alignat}{2}
\frac{d n_\chi}{dt} + 3 H n_\chi  &=&& \,\, - \frac{1}{2} \langle \sigma v \rangle_{\chi \bar{ \chi}   \rightarrow Z^{\prime} Z^{\prime}} \left[n_\chi^2    - \frac{ n_{\chi,0}(T')^2 }{n_{Z',0}(T')^2}  n_{Z^{\prime}}^{2}   \right]   - \frac{1}{2} \langle \sigma v \rangle_{\chi \bar{\chi} \rightarrow f \bar{f} } \left[ n_\chi^2 - n_{\chi,0}^2(T) \right] \,, 
 \label{eq:Boltz_chi}
\end{alignat}
\begin{alignat}{2}
    \frac{d n_{Z^{\prime}}}{dt} + 3 H n_{Z^{\prime}} &=&& \,\,  \frac{1}{2} \langle \sigma v \rangle_{\chi \bar{ \chi}  \rightarrow Z^{\prime} Z^{\prime} } \left[n_\chi^2    - \frac{ n_{\chi,0}(T')^2 }{n_{Z',0}(T')^2}  n_{Z^{\prime}}^{2}   \right] -  \left[\Gamma(T') n_{Z
    ^{\prime}} -\Gamma(T) n_{Z',0}(T) \right] \, ,
	\label{eq:Boltz_Ap}
\end{alignat}
\begin{alignat}{1}
    &\frac{d(\rho_\chi + \rho_{Z'})}{dt} + 3 H(\rho_\chi + \rho_{Z'} + P_\chi + P_{Z'}) \nonumber \\
    & \qquad = \, - \langle \sigma v \delta E \rangle_{\chi f \to \chi f} n_\chi n_f  - m_{Z'} \Gamma(0) \left[n_{Z'} - n_{Z',0}(T)\right] - \frac{1}{2} m_{\chi} \langle \sigma v \rangle_{\chi \overline{\chi} \to f \overline{f}} \left[n_\chi^2 - n_{\chi,0}^2(T) \right] \,,
    \label{eq:Boltz_rho}
\end{alignat}

As in that work, we take $n_\chi$ ($\rho_\chi$, $P_\chi$) to denote the total density (total energy density, total pressure) in $\chi+\bar{\chi}$; furthermore, we assume that the $Z'$ and DM share a common temperature $T'$ (which may differ from the SM temperature $T$), due to the lack of $\varepsilon$ suppression in the relevant elastic scattering rates. $0$ subscripts denote equilibrium densities (at zero chemical potential). $\Gamma$ is the $Z'$ decay rate to SM particles; at zero temperature, the rate $\Gamma(0)$ is given by Eq.~\ref{eq:decay_z}. At higher temperatures, for $T' \lesssim m_{Z'}$,  $\Gamma(T') = \Gamma(0) K_1(m_{Z'}/T') / K_2(m_{Z'}/T')$ \cite{Evans:2017kti}(here $K_1$ and $K_2$ denote modified Bessel functions of the 2nd kind). 

Note that Ref.~\cite{Fitzpatrick:2020vba} uses the zero-temperature width in their Boltzmann equations. This is exact for the energy-density transfer equation (Eq.~\ref{eq:Boltz_rho}), because the suppression to the width for finite $T'$ arises from time dilation, but the $Z'$ total energy (and hence the energy transferred via decay) is enhanced by the same factor as the width is suppressed (i.e.~the Lorentz $\gamma$ factor). Thus the rate of energy transfer is simply $m_{Z'} \Gamma(0)$ independent of the temperature. However, in the number-density transfer equation, $\Gamma$ should be evaluated at the dark-sector temperature $T'$ (this distinction is important, as it ensures that fast decays separately drive $T'\rightarrow T$ and $n_{Z'} \rightarrow n_{Z',0}(T)$).

Relative to Ref.~\cite{Fitzpatrick:2020vba}, we ignore processes that involve more than 2 particles in the initial state, since (unlike in that work) the leading 2-body processes are not kinematically suppressed. We do not include $2\rightarrow 2$ scatterings between the $Z'$ and the SM bath, because while 
Ref.~\cite{Evans:2017kti} argued that such scatterings should dominate over $Z'$ decay for $T \gtrsim m_{Z'}$, Ref.~\cite{Alenezi:2025kwl} argued that in exactly this regime, the $Z'$ should decouple from the SM due to plasma effects. We tested the effect of including elastic scattering between the DM and the SM bath (associated with the $\langle \sigma v \delta E\rangle$ term in Eq.~\ref{eq:Boltz_rho}, and computed as in Ref.~\cite{Fitzpatrick:2020vba}), which is the dominant mechanism equalizing the temperature of the sectors in the analysis of Ref.~\cite{Alenezi:2025kwl}, but found that close to the thermalization floor, the effect of this term on the relic density is negligible (relative to the case where $Z'$ decay is included but this term is not). We also tested the impact of omitting the DM annihilation directly to SM particles (controlled by $\langle \sigma v\rangle_{\chi \bar{\chi} \rightarrow f \bar{f}}$ in the equations above, and found that close to the thermalization floor the impact was negligible (due to the presence of the much stronger annihilation to $Z'Z'$).

We solve these equations down to a SM temperature of $T=m_\chi/200$, initially at a large value of $\varepsilon=10^{-5}$, to establish the correct coupling $\alpha_D\equiv g_D^2/(4\pi)$ to obtain the desired relic density ($\Omega_c h^2=0.12$) in the fully-thermalized regime. (We compare this with the $\alpha_D$ value required to obtain a cross section of $\langle \sigma v\rangle = 4.4\times 10^{-26}$ cm$^3$/s, and find good agreement.) We then lower $\varepsilon$ while holding $m_\chi$, $m_{Z'}$ and $\alpha_D$ fixed, and find the minimum value of $\varepsilon$ such that (i) $d \ln(\Omega_c h^2)/d\ln\varepsilon < 0.2$ (i.e. the relic density is slowly varying with respect to $\varepsilon$), and (ii) changing the initial dark-sector abundances at $T=m_\chi$ by a large factor (specifically, between their equilibrium values and those values multiplied by a factor of $e^{-5}$) affects the final relic density by less than $1\%$, indicating that thermalization has erased information about the initial conditions. We then further confirmed that at these threshold values, the computed relic density matches the high-$\varepsilon$ result (with the same $\alpha_D$) to within $5\%$ in all cases. The precise numerical thresholds used to define thermalization here are somewhat arbitrary, but changing them within reasonable bounds would only change the cutoff value of $\varepsilon$ at the $\mathcal{O}(1)$ level. 

\subsubsection{CMB}

The cosmic microwave background (CMB) is a relatively model-independent constraint on GCE models \cite{Kawasaki_2021}, directly constraining the annihilation cross section $\langle\sigma v\rangle$. 

We use the updated CMB limit from the Planck Collaboration \cite{Alenezi:2025kwl,planck_collaboration_planck_2021}:
\begin{equation}
   \label{eq:cmb_planck}
   \frac{f_\text{eff}S_0\langle\sigma v\rangle}{m_\chi}<\qty{14e-28}{\cm^3\s^{-1}\GeV^{-1}},
\end{equation}
where $f_\text{eff}$ is the efficiency factor, and $S_0$ is the Sommerfeld enhancement. 

The efficiency factor $f_\text{eff}$ can be calculated from the individual efficiency factors for different decay channels of $Z'$ \cite{Alenezi:2025kwl}:
\begin{equation}
    \label{eq:feff}
    f_{\text{eff}}^{\text{net}}=\sum_{\ell}
\mathrm{Br}\bigl(Z'\to \ell\ell\bigr)
f_{\text{eff}}^{VV\to4\ell}\bigl(m_{\chi}\bigr)
+\sum_{X\neq \ell}
\mathrm{Br}\bigl(Z'\to X\,X\bigr)
f_{\text{eff}}^{XX}\Bigl(\tfrac{m_{\chi}}{2}\Bigr).
\end{equation}

We obtain the $Z'$ branching ratio from Ref.~\cite{Ilten:2018crw} and the individual $f_\text{eff}$ factors from Ref.~\cite{Slatyer:2015jla}.  

We estimate the Sommerfeld enhancement $S_0$ to the cross section, significant when $m_{Z'} \ll m_\chi$, using the analytic solution for the Hulth\'{e}n potential \cite{Alenezi:2025kwl,Cassel:2009wt, Tulin:2013teo}: 
\begin{equation}
    \label{eq:sommerfeld}
    S_0(g_D,r,v)
= \frac{g_D^2}{2v}
  \frac{
    \sinh\!\bigl(\frac{6v}{\pi\,r}\bigr)
  }{
    \cosh\!\bigl(\frac{6v}{\pi\,r}\bigr)
    - \cosh\!\Bigl(
        \sqrt{\bigl(\tfrac{6v}{\pi\,r}\bigr)^2
               - \tfrac{24\alpha_D}{r}}
      \Bigr)
  },
\end{equation}
where $r={m_{Z'}}/{m_\chi}$, and  $v$ is the velocity of the DM at the CMB epoch $v_\text{CMB}$, which can be in the range of $\sim 10^{-5}$ to $\sim10^{-11}$\cite{Alenezi:2025kwl}. Note that the precise value of $v_\text{CMB}$ does not affect the calculation because in our parameter space, $r\gg v_\text{CMB}$, and the Sommerfeld enhancement is ``saturated'' (i.e.~velocity-independent) for the vast majority of parameter space apart from the few very fine-tuned regions at the tips of the  resonance peaks.

\subsubsection{Direct detection}
Direct detection experiments constrain high kinetic mixing $\varepsilon$ as they are sensitive to the process $\chi f\to \chi f$ (Fig. \ref{fig:hypercharge-dd}). 
\begin{figure}[ht]
\centering

\begin{minipage}{0.3\textwidth}
    \centering
    \begin{tikzpicture}
\begin{feynman}
  \vertex (f1)  at (-1,  -1) {$f$};
  \vertex (f2)  at (2, -1) {$f$};
  \vertex (a)   at ( 0.5,  -0.7);
  \vertex (b)   at ( 0.5,  0.7);
  \vertex (chi1) at ( -1,  1) {$\chi$};
  \vertex (chi2)at ( 2, 1) {$\chi$};

  \diagram*{
    (f1)  -- [fermion] (a) -- [fermion] (f2),
    (a)   -- [boson, edge label = $Z'/Z$](b),
    (chi1)  -- [fermion] (b) -- [fermion] (chi2)
  };
\end{feynman}

\end{tikzpicture}
\end{minipage}
\caption{The direct detection process of $\chi f\to \chi f$ in the secluded hypercharge model. }
\label{fig:hypercharge-dd}
\end{figure}

At $m_{Z'}/m_\chi\gtrsim0.01$, the effect of recoil energy on the DM-nucleon scattering is negligible \cite{Alenezi:2025kwl}. We can then calculate the direct detection cross section of DM-nucleon scattering for the secluded hypercharge model \cite{Cline_2014,Alenezi:2025kwl}:
\begin{equation}
    \begin{aligned}
    \sigma_N & = \frac{\mu_{\chi N}^2}{\pi} \left[\sum_{i = Z, Z'} V_{\chi,i}\frac{Z_N (2 V_{u,i}+V_{d,i})+ \left(A_N - Z_N \right)( V_{u,i}+2V_{d,i})}{A_Nm_i^2}\right]^2,\\
    & \approx \qty{1.67e-30}{\cm^{2}}g_D^2\varepsilon^2 (m_{Z'}/\text{GeV})^{-4}
    \end{aligned}
    \label{eq:hypercharge scattering cross section}
\end{equation}
where in the 2nd line we have assumed $m_D \ll m_Z$, $m_\chi \gg m_N$, and $\epsilon \ll 1$, and we use the full expression in the 1st line when computing our constraints. Here $A_N$ is the mass number, $Z_N$ is the atomic number, $m_N\approx0.939$ GeV is the mass of nucleon, $\mu_{\chi N}=m_\chi m_N/(m_\chi+m_N)$ is the reduced mass, and the vector couplings
\begin{align}
V_{q,Z} &= e\varepsilon s_\xi Q_q+\tfrac{e}{s_w c_w}(c_\xi+\varepsilon s_w s_\xi)(T_{3,q}-s_w^{2}Q_q),\\
V_{q,Z'} &= e\varepsilon c_\xi Q_q+\tfrac{e}{s_w c_w}(s_\xi-\varepsilon s_w c_\xi)(T_{3,q}-s_w^{2}Q_q),\\
V_{\chi,Z} &= g_D s_\xi,\\
V_{\chi,Z'} &= g_D c_\xi,
\end{align}
where $q=u,d$.

We will compare this to the updated LUX-ZEPLIN result \cite{LZ:2024zvo}.

\subsubsection{Accelerator and beam dump experiments}
In high intensity experiments, $Z'$ can be produced and detected through their visible decays into the SM charged particles (we are never in the regime where the $Z'$ decays invisibly into the dark sector, because of our condition that $m_\chi > m_{Z'}$). Accelerator and beam dump experiments, including $e^+e^-$ colliders, the LHC, beam dump experiments, and meson factories, thus provide exclusion limits for the kinetic mixing parameter $\varepsilon$ against the mass of the dark mediator $m_{Z'}$. We obtain the updated limits on the $Z'$ parameter space from \cite{Batell:2022dpx}. 

\subsubsection{Gamma-ray line search}

We can also consider the gamma-ray line search using data from the Fermi Gamma-Ray Space Telescope experiment \cite{foster_search_2023}, which tightly constrains processes producing mono-energetic gamma-ray photons. In the secluded hypercharge model, an on-shell $Z'$ boson cannot decay into two on-shell photons by the Landau-Yang theorem \cite{Yang:1950rg}. In the narrow regime with $m_Z<m_{Z'}<m_\chi\leq 100\text{ GeV}$, the on-shell $Z'$ can potentially decay into $Z+\gamma$. In this case, the branching ratio for the decay $Z'$ into $Z+\gamma$ is loop-suppressed. This does not, however, necessarily render the gamma-ray line constraint irrelevant, since the line-search limits can be about $2-3$ orders of magnitude stronger than those from the continuum. That said, in this on-shell decay regime,  broadening of the signal into a box spectrum weakens the line-search sensitivity. The photon will also be lower-energy than in the $\gamma \gamma$ case, meaning the relevant backgrounds are higher and the sensitivity is lower. Moreover, following Ref.~\cite{Jackson:2009kg}, we estimate the branching ratio to be lower than $\sim10^{-4}$, which leaves the process unconstrained by the gamma-ray line search. 

Outside this regime, for $m_{Z'} < m_Z$, we only need to consider processes with the DM candidate $\chi$ in the initial state. All such processes involving direct annihilation to SM particles, including $\gamma +X$ \cite{Jackson:2009kg}, are suppressed by a factor of $\varepsilon^2$ relative to the $\bar{\chi} \chi \rightarrow Z' Z'$ annihilation that sources the continuum spectrum, and thus are expected to be negligible for the parameter space allowed by other constraints (which we will find require $\varepsilon \lesssim 10^{-4}$).

\subsubsection{Model variations}

The limits discussed for the secluded hypercharge model would broadly generalize to other secluded models with a vector mediator associated with the $U(1)$ symmetry such as the $B-L$ portal, the Baryon portal, and the $L_i-L_j$ portal, in which we gauge baryon number minus lepton number, the baryon number, and the difference of two lepton families respectively \cite{hooper_systematic_2020}. To obtain the limits in these models, we would need to adjust the couplings $g_V^f$ and $g_A^f$ between the SM fermion $f$ and the dark gauge boson $Z'$ from the secluded hypercharge model. 

If a non-hadronic final state is desired (e.g.~to avoid constraints from antiprotons as in Ref.~\cite{DiMauro:2021qcf}), this might favor scenarios where the $Z'$ kinetically mixes with the gauge boson corresponding to the difference of two lepton families, in particular $L_e - L_\mu$ or $L_\mu - L_\tau$. Ref.~\cite{DiMauro:2021qcf} identifies muon-rich final states as the preferred channel (for a DM mass around 60 GeV and a thermal relic cross section). In the $L_\mu - L_\tau$ case, if the $Z'$ is lighter than $2m_\tau \approx 3.5$ GeV, annihilations will be essentially entirely to muons; this setup would also alleviate direct-detection constraints due to the mediator's lack of a direct coupling to quarks and gluons. Ref.~\cite{Koechler:2025ryv} studies such leptophilic models for the GCE and finds that the $L_\mu - L_e$ model is particularly favorable. 

\subsection{The 2HDM+$a$ model}

\subsubsection{Relic density calculation}
First, for the 2HDM+$a$ model, we calculate the annihilation rate needed to match  the thermal relic cross section. The dominant annihilation process goes through an $s$-channel $a$ exchange (Fig. \ref{fig:2hdm_anni})\cite{ipek_renormalizable_2014}. Note that even if kinematically allowed, the secluded $t$-channel decay of $\chi\bar{\chi}\to a a$ is $p$-wave suppressed \cite{Shelton:2015aqa,Bell:2017irk,hooper_systematic_2020}. In the non-relativistic limit the $s$-channel annihilation gives the following cross section \cite{ipek_renormalizable_2014}:\footnote{Note this expression corrects a typo in the phase space factor from Ref.~\cite{ipek_renormalizable_2014}.}
\begin{equation}
    \label{2hdm:thermal}
    \langle\sigma v\rangle=\frac{g_\chi^2\sin^2{2\theta}\tan^2{\beta}}{8\pi}\frac{m_\chi^2}{m_a^4}\left[ \left(1 - \frac{4m_\chi^2}{m_a^2} \right)^2 + \frac{\Gamma_a^2}{m_a^2} \right]^{-1}\sum_fN_C^f\frac{m_f^2}{v^2_\text{SM}}\sqrt{1-\frac{m_f^2}{m_\chi^2}},
\end{equation}
\begin{figure}[ht]
\centering
\begin{minipage}{0.3\textwidth}
    \centering
    \begin{tikzpicture}
\begin{feynman}
  \vertex (f1)  at (-1,  1) {$\chi$};
  \vertex (f2)  at (-1, -1) {$\bar\chi$};
  \vertex (a)   at ( 0,  0);
  \vertex (b)   at ( 1,  0);
  \vertex (chi) at ( 2,  1) {$f$};
  \vertex (chib)at ( 2, -1) {$\bar f$};

  \diagram*{
    (f1)  -- [fermion] (a) -- [fermion] (f2),
    (a)   -- [scalar, edge label = $a$] (b),
    (b)   -- [fermion] (chi),
    (b)   -- [anti fermion] (chib),
  };
\end{feynman}

\end{tikzpicture}
\end{minipage}
\caption{The $s$-channel $a$ exchange annihilation of the DM fermion $\chi$ in 2HDM+$a$ model.}
\label{fig:2hdm_anni}
\end{figure}
where the decay width of $a$ is
\begin{equation}
    \Gamma_a = \frac{g_\chi^2\cos^2\theta m_a}{8 \pi}\sqrt{1 - \frac{4 m_\chi^2}{m_a^2}}+\sum_f\frac{N_c^fm_f^2\xi_f^2\sin^2\theta m_a}{8 \pi v^2_\text{SM}}\sqrt{1 - \frac{4 m_f^2}{m_a^2}}.
    \label{eqn:a-decay}
\end{equation}

As discussed above, the thermal relic cross section for Dirac fermion dark matter is given by $\langle\sigma v\rangle \approx 4.4\times 10^{-26}$ cm$^3$/s (Eq.~\ref{dm_density_z}) in the relevant mass range. The original paper on this model (Ref.~\cite{ipek_renormalizable_2014}) quoted a band of $\langle\sigma v\rangle = 1-5 \times 10^{-26}$ cm$^3$/s, but based on a cross-comparison we believe that this is the equivalent cross section if the DM candidate were a Majorana fermion (i.e. divided by a factor of 2). We thus show the region with the cross section for Dirac fermion DM:
\begin{equation}
    \label{thermal_relic}
    \langle\sigma v\rangle \sim(2-10)\times10^{-26}\text{ cm$^3$s$^{-1}$}.
\end{equation}

We have checked that this band is equivalent to the band shown in Ref.~\cite{ipek_renormalizable_2014}. 

\subsubsection{Direct detection}
An advantage of the pseudoscalar model is that it can evade direct detection at tree level. The tree level interaction is spin-dependent \cite{Dienes:2013xya} and suppressed by momentum transfer \cite{Abe:2018emu,Abe:2019wjw}. A spin-independent interaction is generated at one-loop level (Fig. \ref{fig:2hdm_direct}). 

\begin{figure*}[htbp]
    \centering
    \begin{minipage}{0.4\textwidth}
    \centering
    \begin{tikzpicture}
      \begin{feynman}
        \vertex (TL) at (-1,1) {$\chi$};   
        \vertex (TR) at (3,1) {$\chi$};   
        \vertex (T1) at (0,1);
        \vertex (T2) at (2,1);
        \vertex (BL) at (-1,-1) {$q$}; 
        \vertex (B1) at (0,-1);
        \vertex (B2) at (2,-1);
        \vertex (BR) at (3,-1) {$q$}; 
    
        \diagram*{
          (TL) -- [fermion] (TR), 
          (BL) -- [fermion] (BR),
          (T1) -- [scalar, dashed, edge label=$a$] (B1),
          (T2) -- [scalar, dashed, edge label'=$a$] (B2),};
      \end{feynman}
       \node at (1, -1.5) {(\textbf{a})};
    \end{tikzpicture}      
    \end{minipage}
      \begin{minipage}{0.4\textwidth}
        \centering
        \begin{tikzpicture}
          \begin{feynman}
            \vertex (h2)     at (0, -0.6);
            \vertex (v2)     at (0, 0);
            \vertex (t1)     at ( 1, 0.6);
            \vertex (t3)     at ( -1, 0.6);
            \vertex (chi)    at ( 2, 1) {$\chi$};
            \vertex (chibar) at ( -2, 1) {$\chi$};
            \vertex (q)      at (2, -1) {$q$};
            \vertex (qbar)   at (-2, -1) {$q$};

            \diagram*{
              (h2) -- [scalar, edge label' = $h$] (v2),
              (v2) -- [scalar, edge label'=$a$] (t1),
              (t1) -- [anti fermion] (t3),
              (t3) -- [scalar, edge label'=$a$] (v2),
              (t1) -- [fermion]      (chi),
              (t3) -- [anti fermion] (chibar),
              (h2) -- [fermion] (q),
              (h2) -- [anti fermion] (qbar),
            };
          \end{feynman}
          \node at (0, -1.5) {(\textbf{b})};
        \end{tikzpicture}
    \end{minipage}
    \caption{The representative one-loop diagrams for direct detection for the 2HDM+$a$ model, describing the scattering of the DM particle $\chi$ with quarks $q$. (a) describes the box diagram with exchange of the pseudoscalar $a$; (b) is the triangle diagram involving Higgs exchange.}
    \label{fig:2hdm_direct}
\end{figure*}
 We can write down the one-loop effective interaction between $\chi$ and $q$ from Fig. \ref{fig:2hdm_direct}(a) \cite{ipek_renormalizable_2014}:
 \begin{equation}
     \Lagr_\text{(a)}=\sum_{\text{down type }q} \frac{m_q^2 g_\chi^2\tan^2\beta\sin^22\theta}{128\pi^2m_a^2(m_\chi^2-m_q^2)}\left(F\left(\frac{m_\chi^2}{m_a^2}\right)-F\left(\frac{m_q^2}{m_a^2}\right)\right)\frac{m_\chi m_q}{v^2_\text{SM}}\chi\bar{\chi}q\bar{q},
 \end{equation}
 with the corresponding loop integral
 \begin{equation}
     F(x)=\frac{2}{3x}\left(4+ f_+(x)+f_-(x)\right),
 \end{equation}
and
\begin{equation}
    f_\pm(x)=\frac{1}{x}\left(1 \pm \frac{3}{\sqrt{1 - 4x}}\right)\left(\frac{1 \pm \sqrt{1 - 4x}}{2}\right)^3 \log\left(\frac{1 \pm \sqrt{1 - 4x}}{2}\right),
\end{equation}
and from Fig. \ref{fig:2hdm_direct}(b) \cite{ipek_renormalizable_2014}:
\begin{equation}
    \Lagr_\text{(b)}=\frac{(m_A^2-m_a^2)\sin^22\theta g_\chi^2}{64\pi^2m_h^2m_a^2}G\left(\frac{m_\chi^2}{m_a^2},0\right)\frac{m_\chi m_q}{v^2_\text{SM}}\chi\bar{\chi}q\bar{q},
\end{equation}
with the corresponding loop integral
\begin{equation}
G(x,y) = -4i \int_0^1 dz \frac{z}{\mathcal{F}^{1/2}(x,y,z)} 
\times \ln\left( \frac{\mathcal{F}^{\frac{1}{2}}(x,y,z) + iy(1-z)}{\mathcal{F}^{1/2}(x,y,z) - iy(1-z)} \right),
\label{eq:loop-g}
\end{equation}
and
\begin{equation}
    \mathcal{F}(x,y,z) = y \left( 4(1-z) + 4xz^2 - y(1-z)^2 \right).
\end{equation}
Note that $G(0,0)=1$. 

For $\tan\beta\lesssim100$, Fig. \ref{fig:2hdm_direct}(b) dominates over Fig. \ref{fig:2hdm_direct}(a) \cite{ipek2015galactic}. We can then simplify the expression, taking $\bra{N}\sum_q m_qq\bar{q}\ket{N}\approx \qty{330}{\MeV}$ \cite{Ellis:2000ds}:
\begin{equation}
\label{2hdm+psuedo,direct}
\begin{aligned}
    \sigma_{\text{SI}}& = \frac{\mu_{\chi N}^2}{\pi}\left[\frac{(m_A^2-m_a^2)\sin^22\theta g_\chi^2}{64\pi^2m_h^2m_a^2}G\left(\frac{m_\chi^2}{m_a^2},0\right)\frac{m_\chi}{v^2_\text{SM}}\bra{N}\sum_q m_qq\bar{q}\ket{N}\right]^2\\
    & \approx\qty{2.2e-49}{\cm^2}\left(\frac{m_A}{\qty{800}{\GeV}}\right)^4\left(\frac{m_a}{\qty{50}{\GeV}}\right)^{-4}\left(\frac{m_\chi}{\qty{30}{\GeV}}\right)^2\left(\frac{\theta}{0.1}\right)^4\left(\frac{g_\chi}{0.5}\right)^4\left(\frac{\bra{N}\sum_q m_qq\bar{q}\ket{N}}{\qty{330}{\MeV}}\right)^2.
\end{aligned}
\end{equation}

We will compare this prediction to the latest LUX-ZEPLIN result \cite{LZ:2024zvo}. 

\subsubsection{Gamma-ray line search}
The gamma-ray line search from the Fermi Gamma-Ray Space Telescope experiment \cite{foster_search_2023} offers a tight constraint on the one-loop process $\chi\bar{\chi}\to\gamma\gamma$ (Fig. \ref{fig:2hdm_gamma}) which produces mono-energetic gamma-ray photons. 
\begin{figure}[ht]
\centering
\begin{minipage}{0.5\textwidth}
        \centering

        \begin{tikzpicture}
          \begin{feynman}
=            \vertex (chi1)   at (-2, 0.8) {$\chi$};
            \vertex (chi2) at (-2, -0.8) {$\bar{\chi}$};
            \vertex (h2)     at (-1, 0);
            \vertex (v2)     at (0, 0);
            \vertex (t1)     at ( 1,  0.8);
            \vertex (t3)     at ( 1, -0.8);
            \vertex (gamma)    at ( 2,  0.8) {$\gamma$};
            \vertex (gamma2) at ( 2, -0.8) {$\gamma$};

            \diagram*{
              (chi1) -- [fermion] (h2),
              (chi2) -- [anti fermion] (h2),
              (h2) -- [scalar, edge label=$a$] (v2),
              (v2) -- [fermion, edge label=$f$] (t1),
              (t1) -- [fermion, edge label=$f$] (t3),
              (t3) -- [fermion, edge label=$f$] (v2),
              (t1) -- [boson]      (gamma),
              (t3) -- [boson] (gamma2),
            };
          \end{feynman}
        \end{tikzpicture}
    \end{minipage}
\caption{The representative Feynman diagram for one-loop $\chi\bar{\chi}\to\gamma\gamma$ process with a loop involving SM fermions $f$. This process gives a monochromatic gamma-ray line that can be constrained with data from the Fermi Gamma-Ray Space Telescope. }
\label{fig:2hdm_gamma}
\end{figure}

We consider the effective Lagrangian \cite{Hektor_2017}:
\begin{equation}
    \label{2hdm:coy}
    \Lagr\supset i{y_\chi}a\bar{\chi}\gamma^5\chi+\sum_fi{y_f}a\bar{f}\gamma^5f.
\end{equation}

This effective Lagrangian is related to the type II 2HDM+$a$ model by
\begin{equation}
    \label{2hdm:effective}
    y_f = \begin{cases}
    \frac{m_f}{v_\text{SM}}\frac{\sin{\theta}}{\tan{\beta}} & \text{up-type quarks,} \\
    \frac{m_f}{v_\text{SM}}\sin{\theta}\tan{\beta} & \text{down-type quarks and leptons,}
    \end{cases}
\end{equation}
\begin{equation}
    \label{2hdm_effective_gx}
    y_\chi=g_\chi\cos\theta.
\end{equation}

We can write down the one-loop level cross section: 
\begin{equation}
    \label{2hdm:gamma}
    \langle\sigma_{\chi\bar{\chi}\to\gamma\gamma}\rangle=\frac{y_\chi^2\alpha^2}{4\pi^3}\frac{|\sum_fN_C^fQ_f^2y_fm_f\mathcal{F}(m_\chi^2/m_a^2)|^2}{(4m_\chi^2-m_a^2)^2+m_a^2\Gamma_a^2},
\end{equation}
where $Q_f$ is the electric charge of fermion $f$, and
\begin{equation}
    \label{2hdm:gamma,f}
    \mathcal{F}=\begin{cases}
    \arcsin^2{(\sqrt{x})} & \text{if } x\le1\\
    -\frac{1}{4}\left[\log{\left(\frac{1+\sqrt{1-1/x}}{1-\sqrt{1-1/x}}\right)}-i\pi\right]^2 & \text{if } x > 1
    \end{cases}.
\end{equation}

We will compare this to the Fermi gamma-ray line search constraint from Ref.~\cite{foster_search_2023}.

\subsubsection{$B$-physics}
The branching ratio of the rare decay $B_s^0\rightarrow \mu^+ \mu^-$ constrains the parameters of the 2HDM+$a$ model \cite{Athron:2024rir,ipek_renormalizable_2014}. 

The relationship between the 2HDM branching ratio and the SM branching ratio for $B_s^0\rightarrow \mu^+ \mu^-$ is \cite{ipek_renormalizable_2014,Skiba:1992mg,Altmannshofer:2017wqy,MartinezSantos:2010xda}:\footnote{Ref.~\cite{ipek_renormalizable_2014} quotes this expression for $m_a\ll m_Z$, but we believe that the expression is valid for $m_a\ll m_A$.}
\begin{equation}
\label{eq:2hdm:bmumu}
\text{Br}\left(B_s^0 \to \mu^+ \mu^- \right) \simeq \text{Br}\left(B_s^0 \to \mu^+ \mu^- \right)_{\text{SM}} \times \left| 1 + \frac{m_b m_{B_s} \tan^2\beta \sin^2\theta}{m_{B_s}^2 - m_a^2} \frac{f(x_t, y_t, r)}{Y(x_t)} \right|^2,
\end{equation}
where $x_t = {m_t^2}/{m_W^2}$, $y_t = {m_t^2}/{m_{H^\pm}^2}$, $r = {m_{H^\pm}^2}/{m_W^2}$, 
\begin{equation}
f(x, y, r) = \frac{x}{8} \left[ - \frac{r(x - 1) - x}{(r - 1)(x - 1)} \log r + \frac{x \log x}{(x - 1)} - \frac{y \log y}{(y - 1)} + \frac{x \log y}{(r - x)(x - 1)} \right],
\end{equation}
and $Y(x)$ the Inami-Lim function:
\begin{equation}
Y(x) = \frac{x}{8} \left[ \frac{x - 4}{x - 1} \log x + \frac{3x \log x}{(x - 1)^2} \right].
\end{equation}

We will compare this to the SM prediction of $(3.66\pm 0.14)\times10^{-9}$ and the measured value of  $3.09 ^{+ 0.46 + 0.15}_{-0.43 -0.11}\times10^{-9}$ \cite{Aaij_2022}. We take the one-sided 95\% exclusion limit as the upper bound of the branching ratio. 

Note that other flavor anomalies might constrain the class of 2HDM models as well. Variants of the 2HDM model, such as a type-II 2HDM model with an extended Yukawa structure \cite{Celis:2012dk}, have been explored to explain several persistent flavor anomalies, including potential deviations from the SM predictions observed in $b\to c\tau^-\nu$ and $b\to sl^+l^-$ transitions \cite{Crivellin:2013wna,Athron:2024rir}.
Other loop-induced processes like $b\to s\gamma$ require a relatively heavy charged Higgs (typically $M_{H^\pm}\gtrsim600 $GeV in Type II scenarios) \cite{Arbey:2017gmh}. The mass of the charged Higgs is not the main focus of this paper; we require the charged Higgs to be heavy enough to evade the other constraints while preserving the perturbativity of the 2HDM+$a$ model. 

\subsubsection{Invisible Higgs decay}

In the 2HDM+$a$ model, it is possible for the SM-like Higgs $h$ to decay into dark sector particles, thereby creating an invisible Higgs decay. The invisible Higgs decay is constrained by the LHC \cite{ATLAS:2023tkt}. 
In this section, we consider the effective Lagrangian
\begin{equation}
    \Lagr\supset \lambda h a a + i y_\chi a\bar{\chi}\gamma^5 \chi,
    \label{eqn:inv-lagr}
\end{equation}
where 
\begin{equation}
    \lambda = \frac{(m_A^2 - m_a^2) \sin^2{2\theta}}{2v_\text{SM}},\quad y_\chi=g_\chi\cos\theta.
\end{equation}

The usual tree-level two-body decay $h\rightarrow aa$ (Fig. \ref{fig:haa}(a)) only meaningfully constrains the part of the parameter space where $m_a<\frac{1}{2}m_h$.  Therefore, to calculate the constraints at higher $m_a$, where kinematically it might still be possible for the SM-like Higgs $h$ to decay into the lighter $\chi$ through an off-shell $a$, we also consider the following decay channels:
\begin{itemize}
    \item the one-loop decay $h\rightarrow a a \chi\text{ (loop)}\rightarrow \chi\bar{\chi}$ for $m_\chi<\frac{1}{2}m_h$ (Fig. \ref{fig:haa}(b)), 
    \item the three-body decay $h\rightarrow a^* a\rightarrow a \chi \bar{\chi}$ for $m_a+2m_\chi<m_h$ (Fig. \ref{fig:haa}(c)), 
    \item the four-body decay $h\rightarrow a^* a^*\rightarrow  \chi \bar{\chi}\chi \bar{\chi}$ for $m_\chi<\frac{1}{4}m_h$ (Fig. \ref{fig:haa}(d)).
\end{itemize}

\begin{figure*}[htbp]
    \centering
        \begin{minipage}{0.23\textwidth}
        \centering
        
        
        \begin{tikzpicture}
          \begin{feynman}
            \vertex (h1) at (-1,  0) {$h$};
            \vertex (v1) at ( 0,  0);
            \vertex (a1) at ( 1,  0.8) {$a$};
            \vertex (a2) at ( 1, -0.8) {$a$};

            \diagram*{
              (h1) -- [scalar] (v1),
              (v1) -- [scalar] (a1),
              (v1) -- [scalar] (a2),
            };
          \end{feynman}

          \node at (0, -1.5) {(\textbf{a})};
        \end{tikzpicture}
    \end{minipage}
    \begin{minipage}{0.23\textwidth}
        \centering
        

        \begin{tikzpicture}
          \begin{feynman}
            \vertex (h2)     at (-1, 0)    {$h$};
            \vertex (v2)     at (0, 0);
            \vertex (t1)     at ( 1,  0.8);
            \vertex (t3)     at ( 1, -0.8);
            \vertex (chi)    at ( 2,  0.8) {$\chi$};
            \vertex (chibar) at ( 2, -0.8) {$\bar{\chi}$};

            \diagram*{
              (h2) -- [scalar] (v2),
              (v2) -- [scalar, edge label=$a$] (t1),
              (t1) -- [anti fermion, edge label=$\chi$] (t3),
              (t3) -- [scalar, edge label=$a$] (v2),
              (t1) -- [fermion]      (chi),
              (t3) -- [anti fermion] (chibar),
            };
          \end{feynman}
          \node at (0, -1.5) {(\textbf{b})};
        \end{tikzpicture}
    \end{minipage}
    \begin{minipage}{0.23\textwidth}
        \centering
        
        
        \begin{tikzpicture}
          \begin{feynman}
            \vertex (h3) at (-1,  0) {$h$};
            \vertex (v3) at (0,   0);
            \vertex (a3) at (1,  0.9) {$a$};
            \vertex (w3) at (1, -0.3);
            \vertex (chi3)    at (2,  0.2) {$\chi$};
            \vertex (chibar3) at (2, -0.8) {$\bar{\chi}$};
            
            \diagram*{
              (h3) -- [scalar] (v3),
              (v3) -- [scalar] (a3),
              (v3) -- [scalar, edge label'=$a^*$] (w3),
              (w3) -- [fermion]     (chi3),
              (w3) -- [anti fermion] (chibar3),
            };
          \end{feynman}

          \node at (0, -1.5) {(\textbf{c})};
        \end{tikzpicture}
    \end{minipage}
    \begin{minipage}{0.23\textwidth}
        \centering
        
        
        \begin{tikzpicture}
          \begin{feynman}
            \vertex (h4)     at (-1,  0) {$h$};
            \vertex (v4)     at ( 0,  0);
            \vertex (w4a)    at ( 1,  0.6);
            \vertex (w4b)    at ( 1, -0.6);
            \vertex (chi4a)    at (2,  0.9) {$\chi$};
            \vertex (chibar4a) at (2,  0.3) {$\bar{\chi}$};
            \vertex (chi4b)    at (2, -0.3) {$\chi$};
            \vertex (chibar4b) at (2, -0.9) {$\bar{\chi}$};

            \diagram*{
              (h4) -- [scalar] (v4),
              (v4) -- [scalar, edge label=$a^*$] (w4a),
              (v4) -- [scalar, edge label'=$a^*$] (w4b),
              (w4a) -- [fermion] (chi4a),
              (w4a) -- [anti fermion] (chibar4a),
              (w4b) -- [fermion] (chi4b),
              (w4b) -- [anti fermion] (chibar4b),
            };
          \end{feynman}
          \node at (0, -1.5) {(\textbf{d})};
        \end{tikzpicture}
    \end{minipage}

    \caption{The representative Feynman diagrams for 
    (a) $h \to aa$ , 
    (b) $h \to aa\chi\ (\text{loop}) \to \chi\bar{\chi}$ , 
    (c) $h \to aa^* \to a\chi\bar{\chi}$ , and 
    (d) $h \to a^*a^* \to \chi\bar{\chi}\chi\bar{\chi}$ .}
    \label{fig:haa}
\end{figure*}

\paragraph{The one-loop decay}
The one-loop contribution to the decay width of $h\rightarrow\chi\bar{\chi}$ is \cite{ipek_renormalizable_2014}:
\begin{equation}
    \Gamma_{\text{loop}}=\left(\frac{\lambda g_\chi^2}{32\pi^2m_a^2}\right)^2\frac{m_h}{8\pi}\,G\left(\frac{m_\chi^2}{m_a^2},\frac{q^2}{m_a^2}\right)^2\left(1-\frac{4m_\chi^2}{m_h^2}\right)^{\frac{3}{2}},
\end{equation}
where $G$ is the same loop integral as \eqref{eq:loop-g}.

\paragraph{Three and four-body decay}
For the three-body decay, we parameterize the initial state $h$ and the final-state $a, \chi$ and $\bar{\chi}$ to have momenta $p, p_1,p_2,p_3$ respectively. We then obtain the spin-averaged squared matrix element as:
\begin{equation}
    \frac{1}{4}\sum_\text{spins}\lvert\mathcal{M}\rvert^2=\frac{\lambda^2 y_\chi^2}{\lvert(p-p_1)^2-m_a^2+i\Gamma_a\rvert^2}(p_2 \cdot p_3 -m_{\chi}^2).
\end{equation}

We similarly calculate the four-body decay $h\rightarrow a^* a^*\rightarrow  \chi \bar{\chi}\chi \bar{\chi}$ by parameterizing the momenta of the initial-state $h$ and final state $\chi \bar{\chi} \chi \bar{\chi}$ to be $p, p_1, p_2, p_3, p_4$ respectively. We find:
\begin{equation}
    \frac{1}{16}\sum_\text{spins}\lvert\mathcal{M}\rvert^2 = \frac{\lambda^2 y_\chi^4}{\lvert(p_1+p_2)^2-m_a^2)(p-p_1-p_2)^2-m_a^2+i\Gamma_a\rvert^2}(p_1\cdot p_2-m_\chi^2)(p_3\cdot p_4-m_\chi^2).
\end{equation}

We use a Monte Carlo simulation to calculate the phase space factor with \textsc{Hazma} \cite{Coogan:2019qpu}: 
\begin{equation}
    d\Gamma = \frac{(2\pi)^4}{2m_h}\lvert\mathcal{M}\rvert^2 d\Phi.
\end{equation}

We then compare the decay width of the invisible decay to the upper bound of 0.105 on the invisible branching ratio of the Higgs boson from ATLAS \cite{ATLAS:2023tkt} at 95\% confidence level and the Higg boson's decay width of \qty{4.1}{\MeV} \cite{LHCHiggsCrossSectionWorkingGroup:2016ypw}.

\subsubsection{Heavy Higgs searches}
The LHC has established limits on $m_A$ and $m_H$ for the 2HDM model through the di-tau search $pp\to H/A\to \tau^+\tau^-$ \cite{Aad_2020_tau,CMS:2022goy}.\footnote{We thank Maria Olalla Olea-Romacho and Thomas Biek\"{o}tter for pointing out the importance of this search.} To adapt the limit to the 2HDM+$a$ model with the extra pseudoscalar mediator and dark fermion, we need to consider the modification on both the production of the heavy Higgs and the branching ratio of the decay of the heavy Higgs into $\tau^+\tau^-$ \cite{Arcadi:2022lpp}.

For the production of the heavy Higgs, in the 2HDM model without the additional pseudoscalar, for $\tan\beta\gg 1$, 
\begin{equation}
    \sigma_\text{2HDM}(pp\to A)=\sigma_\text{2HDM}(pp\to H)\propto \tan^2\beta,
\end{equation}
whereas in the 2HDM+$a$ model, 
\begin{equation}
    \sigma_\text{2HDM+$a$}(pp\to A)=\cos^2\theta\sigma_\text{2HDM}(pp\to A),\quad \sigma_\text{2HDM+$a$}(pp\to H)=\sigma_\text{2HDM}(pp\to H). 
\end{equation}

We now look at the branching ratio of the decay of heavy Higgs into $\tau^+\tau^-$. In our case where $\cos(\beta-\alpha)=0$, $m_A=m_H$, and $\tan\beta\gg 1$, the following channels are open in the 2HDM model without the additional pseudoscalar: 
$H\to \tau^+\tau^-$, $H\to b\bar b$, $H\to t\bar t$; $A\to \tau^+\tau^-$, $A\to b\bar b$, $A\to t\bar t$. 

In the 2HDM+$a$ model,  the open decay channels in this regime are: 
$H\to \tau^+\tau^-$, $H\to b\bar b$, $H\to t\bar t$, $H\to a Z$; $A\to \tau^+\tau^-$, $A\to b\bar b$, $A\to t\bar t$, $A\to\chi\bar\chi$, $A\to a h$. 

The partial decay widths of each of the channels are as follows \cite{Bauer:2017ota}:
\begin{align}
    \Gamma_{H\to f\bar f}&=\frac{N_c^f\xi_f^2 m_f^2}{8\pi v_{\textsc{SM}}^2}m_H\left(1-\frac{4m_f^2}{m_H^2}\right)^{3/2},\\
    \Gamma_{H\to aZ}&=\frac{1}{16\pi m_H^3v_{\textsc{SM}}^2}\left((m_H^2-m_a^2-m_Z^2)^2-4m_a^2m_Z^2\right)^{3/2}\sin^2\theta,\\
    \Gamma_{A\to f\bar f}&=\frac{N_c^f\xi_f^2 m_f^2}{8\pi v_{\textsc{SM}}^2}m_A\left(1-\frac{4m_f^2}{m_A^2}\right)^{1/2}\cos^2\theta,\\
    \Gamma_{A\to\chi\bar\chi}&=\frac{g_\chi^2}{8\pi}m_A\left(1-\frac{4m_\chi^2}{m_A^2}\right)^{1/2}\sin^2\theta,\\
    \Gamma_{A\to ah}&=\frac{(m_A^2-m_a^2)^2}{64\pi v_{\textsc{SM}}^2 m_A^3}\left((m_A^2-m_a^2-m_h^2)^2-4m_a^2m_h^2\right)^{1/2}\sin^24\theta.\label{eq:heavydecaywidths}
\end{align}

We obtain the LHC limits from Ref.~\cite{Aad_2020_tau} and recast accordingly, by rescaling both the production cross section and the branching ratio into $\tau^+ \tau^-$. 

\subsubsection{Collider searches}
In the LHC, the pseudoscalar mediator $a$ can be produced (on-shell or off-shell), and then decay into the dark fermion $\chi$. In the visible sector, this process looks like the production of a single SM particle, dubbed the ``mono-X'' search. The leading collider searches, according to ATLAS \cite{atlascollaboration20232hdm+1}, are the mono-Higgs and the mono-$Z$ search in Fig. \ref{fig:monoz-h}.

\begin{figure*}[htbp]
    \centering
    \begin{minipage}{0.45\textwidth}
        \centering
        \begin{tikzpicture}
          \begin{feynman}
            \vertex (b1) at (-1.0,  0.8) {$b$};
            \vertex (b2) at (-1.0, -0.8) {$\bar{b}$};
            \vertex (v)  at ( 0,   0) ;
            \vertex (w)  at ( 1,   0) ;
            
            \vertex (z)      at (2.0,  0.8) {$Z$};
            
            \vertex (a)      at (2.0, -0.8);
            \vertex (chi)    at (3.0, -0.3) {$\chi$};
            \vertex (chibar) at (3.0, -1.3) {$\bar{\chi}$};

            \diagram*{
              (b1) -- [fermion] (v) -- [fermion] (b2),
              (v)  -- [scalar, edge label=$h/H$] (w),
              (w)  -- [boson]   (z),
              (w)  -- [scalar, , edge label=$a$]  (a),
              (a)  -- [fermion]      (chi),
              (a)  -- [anti fermion] (chibar),
            };
          \end{feynman}
          \node at (1.0, -1.8)
            {(\textbf{a}) $b\bar{b} \to h/H \to a\,Z \to Z\,\chi\,\bar{\chi}$};
        \end{tikzpicture}
    \end{minipage}
    \begin{minipage}{0.45\textwidth}
        \centering
        \begin{tikzpicture}
          \begin{feynman}
            \vertex (b1) at (-1.0,  0.8) {$b$};
            \vertex (b2) at (-1.0, -0.8) {$\bar{b}$};
            \vertex (v)  at ( 0,   0) ;
            \vertex (w)  at ( 1,   0) ;
            
            \vertex (z)      at (2.0,  0.8) {$h$};
            
            \vertex (a)      at (2.0, -0.8);
            \vertex (chi)    at (3.0, -0.3) {$\chi$};
            \vertex (chibar) at (3.0, -1.3) {$\bar{\chi}$};

            \diagram*{
              (b1) -- [fermion] (v) -- [fermion] (b2),
              (v)  -- [scalar, edge label=$h/H$] (w),
              (w)  -- [scalar]   (z),
              (w)  -- [scalar, , edge label=$a$]  (a),
              (a)  -- [fermion]      (chi),
              (a)  -- [anti fermion] (chibar),
            };
          \end{feynman}
          \node at (1.0, -1.8)
            {(\textbf{b}) $b\bar{b} \to h/H \to a\,h \to h\,\chi\,\bar{\chi}$};
        \end{tikzpicture}
    \end{minipage}

    \caption{The representative mono-$Z$ and mono-Higgs Feynman diagrams for the $b\bar{b}$ initial state in high $\tan\beta$ regime. 
             The neutral Higgs $h$ or $H$ decays to a pseudoscalar $a$ plus a visible $Z$ or $h$, and then 
             $a$ decays invisibly to $\chi\bar{\chi}$.}
    \label{fig:monoz-h}
\end{figure*}
We explored both the mono-Higgs and  mono-$Z$ searches with \textsc{MadGraph} \cite{Alwall:2014hca} and found that for the parameter space under consideration, the mono-$Z$ search has a larger reach than the mono-Higgs search. We thus focus on the mono-$Z$ search. 

For the modeling of the mono-$Z$ search, we use \textsc{MadGraph} 5 version 2.9.21 \cite{Alwall:2014hca} to generate the hard scattering events at leading order for the $b\bar{b}$ mode of production of the DM at high $\tan\beta$, using the 2HDM Universal FeynRules Output (UFO) from Ref.~\cite{Bauer:2017ota}.
We use \textsc{NNPDF23\_lo\_as\_0130\_qed} \cite{Ball:2013hta} in \textsc{LHAPDF} 6 \cite{Buckley:2014ana} for the parton distribution function (PDF).  We then use \textsc{PYTHIA} 8.311 \cite{Bierlich:2022pfr} for parton showering and hadronization, and \textsc{DELPHES} 3.5.0 \cite{deFavereau:2013fsa} with \textsc{FastJet} 3.4.3 \cite{Cacciari:2011ma} for detector simulation. 

The cuts we employ to process the simulation data are as follows, matching the analysis of Ref.~\cite{ATLAS:2021gcn}: 
\begin{itemize}
  \item Exactly two oppositely charged electrons or muons, with:
  \begin{itemize}
    \item Leading lepton $p_{\mathrm{T}} > 30\,\mathrm{GeV}$,
    \item Subleading lepton $p_{\mathrm{T}} > 20\,\mathrm{GeV}$,
    \item Dilepton invariant mass $76 < m_{\ell\ell} < 106\,\mathrm{GeV}$.
  \end{itemize}

  \item Missing transverse momentum $E_{\mathrm{T}}^{\mathrm{miss}} > 90\,\mathrm{GeV}$, missing-$E_{\mathrm{T}}$ significance $S_{E_{\mathrm{T}}^{\mathrm{miss}}} > 9$;
  \item Leptons must be relatively close in angle: $\Delta R(\ell,\ell) < 1.8$;

  \item Events containing one or more tagged $b$-jets are rejected (to reduce top-quark backgrounds).
\end{itemize}

We then compare our result to the $3\sigma$ upper limit in Ref.~\cite{ATLAS:2021gcn}. 

\section{Results}
\label{sec:results}
Having laid out how to calculate the constraints on the secluded hypercharge model and the 2HDM+$a$ model, we will now scan the parameter space to obtain the updated exclusion plots. In particular, we use the following updated constraints:
\begin{itemize}
    \item Direct detection: the latest result from the LZ experiment in 2024 \cite{LZ:2024zvo};
    \item CMB bound: the latest Planck result in 2018 \cite{planck_collaboration_planck_2021};
    \item Accelerator and beam dump search for dark photon: the summary plot in \cite{Batell:2022dpx};
    \item Gamma-ray line search: a recent analysis using data from the Fermi Gamma-Ray Space Telescope result in 2023 \cite{foster_search_2023}; 
    \item Collider measurements: the measurements from LHC run 2 of the invisible Higgs decay \cite{higgsdecay2022}, the $B_s^0\to\mu^+\mu^-$ decay \cite{Aaij_2022}, the mono-Higgs search \cite{mono-higgs-bb-2021}, and the mono-Z search \cite{atlascollaboration20232hdm+1,ATLAS:2021gcn} for the 2HDM+$a$ model. 
\end{itemize}
\subsection{The secluded hypercharge model}

\begin{figure}
    \centering
    \includegraphics[width=\linewidth]{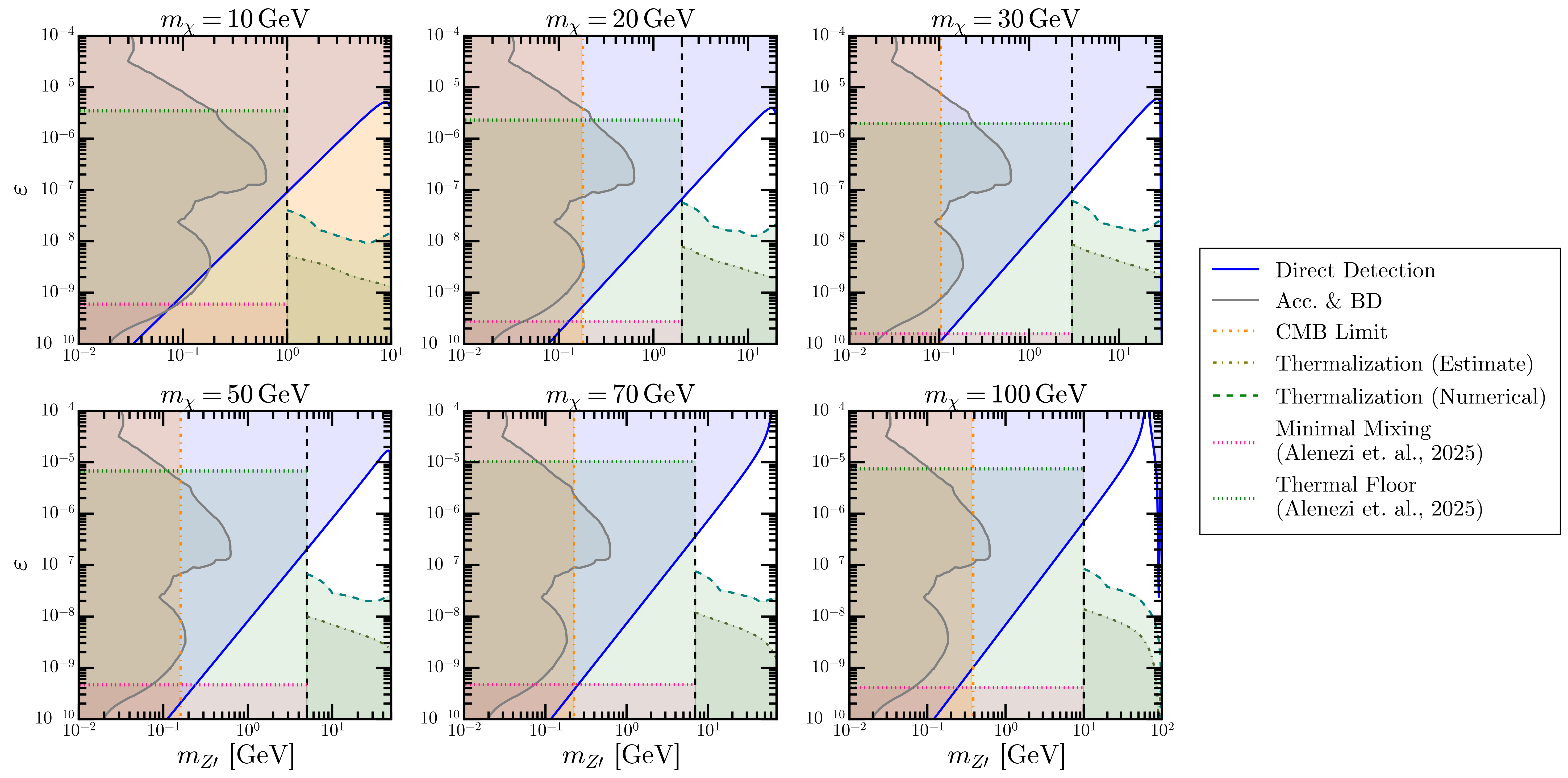}
    \caption{Constraints for the secluded hypercharge model at different $m_\chi$. Note that for each $m_\chi$, we plotted the corresponding $m_{Z'}$ from $\qty{1e-2}{\GeV}$ up to $m_\chi$. The black dotted lines indicate where $m_{Z'}=0.1m_\chi$. At $m_{Z'}\sim0.1m_\chi$, our result is similar to the thermalization floor in Ref.~\cite{Evans:2017kti}, which also did not include in medium effect (insignificant at $m_{Z'}\gtrsim T$). We also show the limit from Refs. \cite{Alenezi:2025kwl} for comparison. In general, there is still parameter space open at higher $m_{Z'}$.}
    \label{fig:hypercharge-result}
\end{figure}

After scanning the parameter space of the secluded hypercharge model, we obtain the exclusion plot for the secluded hypercharge model in Fig. \ref{fig:hypercharge-result}. 

We see that large parts of the phase space are excluded, but there is still some available phase space at higher $m_{Z'}$. The accelerator and beam dump search is always sub-dominant to the direct detection search in our parameter space (although relatively modest changes to the scenario, such as making $\chi$ a pseudo-Dirac fermion with a modest mass splitting between its components, may alter this conclusion). Assuming the standard cosmological history and the thermal relic annihilation cross section, the CMB bound excludes all parameter space where $m_\chi\lesssim \qty{10}{\GeV}$ while posing a modest constraint for low $m_{Z'}$ at larger $m_\chi$ due to the Sommerfeld enhancement at low $m_{Z'}/m_\chi$. In general, our constraints disfavor a large hierarchy between $m_{Z'}$ and $ m_\chi$. 

The thermalization bound, together with direct detection, constrains the viable range of $\varepsilon$ for higher $m_{Z'}$, beyond the current reach of accelerators. For lighter $m_{Z'} < 0.1 m_\chi$, the results of Ref.~\cite{Alenezi:2025kwl} imply that the DM must always be in the ``leak-in'' or ``reannihilation'' regimes, rather than undergoing standard thermal freezeout. This implies that $\alpha_D$, and consequently the present-day DM annihilation cross-section, is {\it suppressed} relative to the standard thermal relic value. The degree of suppression found in Ref.~\cite{Alenezi:2025kwl} is typically up to a factor of few in $\alpha_D$ (corresponding to up to $\sim 1$ order of magnitude in the cross section); this region may thus have difficulty matching higher estimates of the cross section needed to explain the GCE, although it could potentially alleviate the tension with bounds from dwarf galaxies.

For larger $m_{Z'} > 0.1 m_\chi$, we have provided both an estimate of the thermalization floor based on the $Z'$ decay lifetime, and the results of solving the full Boltzmann equations. The refined thermalization threshold for $\varepsilon$ from the latter calculation, as a function of $m_\chi$ and $m_{Z'}$, is slightly higher than our estimate based on the $Z'$ decay rate, but in qualitatively good agreement, and confirms that for $m_{Z'}$ within an order of magnitude of $m_\chi$, there can be open parameter space for several orders of magnitude in $\varepsilon$. As in the case with $m_{Z'} < 0.1 m_\chi$, the thermalization floor is also not a hard exclusion; points below this floor may still accommodate the GCE, but will correspond to a lower present-day annihilation cross section.

Our results show that the CMB constraint is powerful for $m_\chi \lesssim 10$ GeV, but at higher masses rapidly becomes irrelevant (unless the $Z'$ mass is small enough to allow for large Sommerfeld enhancement, but for this model this region is excluded by direct detection). 

At high $m_{Z^\prime}$, there is still a window of parameter space that evades direct detection limits while still having a large enough mixing to ensure full equilibration between DM and SM in the early universe. We offer a discussion of the approximate dependence of different constraints in Table \ref{tab:secluded}. 

 Allowing an annihilation cross section modestly lower than the canonical thermal relic cross section--which might still produce the correct thermal relic density, due to a modified cosmological history or the presence of additional annihilation channels--weakens the CMB constraint and marginally relaxes the direct detection limit. Nevertheless, the allowed parameter region remains essentially unchanged. 

 Note that when $m_{Z'}\ll m_\chi$, the annihilation might in any case have difficulty matching the GCE spectrum (and certainly once $m_{Z'} \lesssim 300$ MeV, our assumption that quarks are the appropriate final-state degrees of freedom will break down, due to the QCD phase transition \cite{Guenther:2020jwe}). However, from Fig. \ref{fig:hypercharge-result}, small values of $m_{Z'}$ are already strongly constrained by direct detection (and in particular the whole region with $m_{Z'} \lesssim 300$ MeV is excluded; we show it only for completeness). While the GCE spectrum has significant uncertainties, Ref.~\cite{hooper_systematic_2020} found that the region with $m_{Z'}$ not too much smaller than $m_\chi$ provided the best fit to the gamma-ray data; this region is also the least constrained by terrestrial experiments.


\begin{table}[htbp]
\centering
\caption{The approximate dependence of different constraints for the secluded hypercharge model, assuming $m_{Z'}\ll m_\chi$.}
\label{tab:secluded}
\begin{tabular}{lcccc}
\toprule
Process & $\varepsilon$ & $m_{Z'}$ & $m_\chi$ & $g_D$ \\
\midrule
$\langle\sigma v\rangle$ 
  & $\times$ 
  & $\times$
  & $1/m_\chi^{2}$ 
  & $g_D^4$ \\
CMB
  & $\times$ 
  & $\times$
  & $1/m_\chi^3$ 
  & $g_D^4$ \\
Direct Detection
  & $\varepsilon^2$ 
  & $1/m_{Z'}^4$
  & $\times$ 
  & $g_D^2$ \\
Standard Cosmology
  & $\varepsilon^2$ 
  & $m_{Z'}$
  & $\times$ 
  & $\times$ \\
\bottomrule
\end{tabular}
\end{table}

\subsection{The 2HDM+$a$ model}

\begin{figure}
    \centering
    \includegraphics[width=0.65\linewidth]{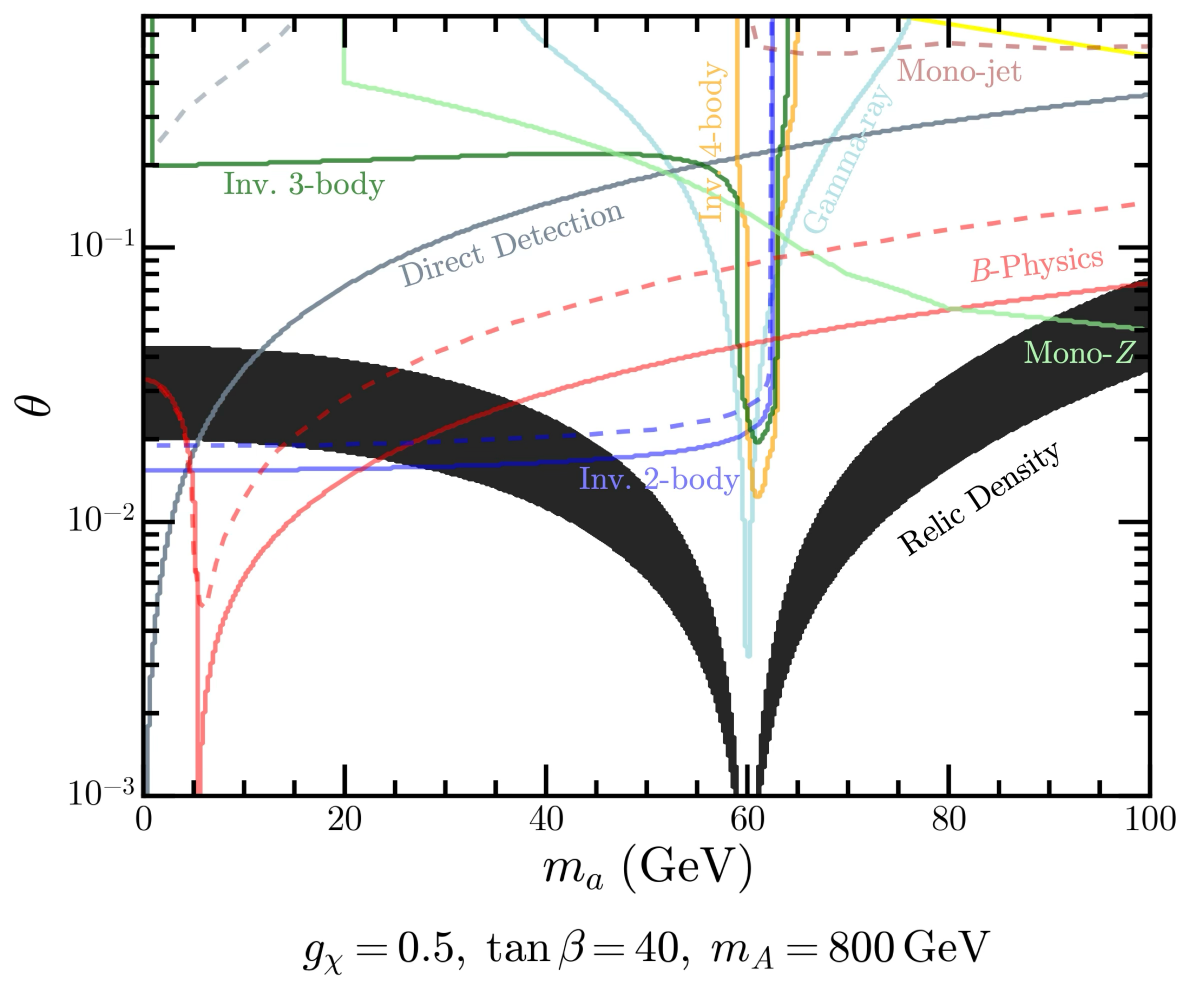}
    \caption{Comparison between the updated constraints (except the heavy Higgs decay bound) and the constraints from \cite{ipek_renormalizable_2014} in 2014. The dashed lines indicate old constraints while the solid lines indicate updated constraints. The black regions indicate parameters that reproduces the thermal relic density. The yellow line on the top-right of the plot is the mono-Higgs search constraint, which is much weaker than the mono-$Z$ search constraint in our parameter space. The parameter we hold constant for this plot are: $g_\chi=0.5$, $m_\chi=\qty{30}{\GeV}$, and $\tan\beta=40$. Note that the heavy Higgs decay bounds completely exclude this region; we include this plot to show the advancement of other constraints.}
    \label{fig:2hdm-compare}
\end{figure}

For the 2HDM+$a$ model, we first compare our updated result to the exclusion plot from Ref.~\cite{ipek_renormalizable_2014}, for the benchmark parameter point they chose in that work of $g_\chi=0.5$, $m_\chi=\qty{30}{\GeV}$, and $\tan\beta=40$.  Fig.~\ref{fig:2hdm-compare} shows the direct analogue of that exclusion plot, with all our new and updated exclusions {\it except} those arising from decay of the heavy $A$ and $H$ fields.

Compared to the old constraints from Ref.~\cite{ipek_renormalizable_2014}, the direct detection bound improves significantly, but it is still generally subleading compared to the $B$-physics bound which is improved with data from the LHC Run 2. The $B$-physics bound excludes regions with high $\theta$ and low $m_a$, although heavier $m_a$ regions will likely have improved coverage with the data sets collected at the high-luminosity LHC (HL-LHC).  There is a marginal improvement for the invisible Higgs decay including 3-body and 4-body decay, which excludes regions of moderate-to-high $\theta$ at $m_a<\frac{1}{2}m_h$. Collider searches exclude regions of very high $\theta$ and high $m_a$. The Mono-$Z$ collider search from LHC Run 2 gives a more stringent bound compared to the mono-jet collider bound in \cite{ipek_renormalizable_2014}. However, with only these bounds, there is still a substantial amount of open parameter space.  

\begin{figure}
    \centering
    \includegraphics[width=0.65\linewidth]{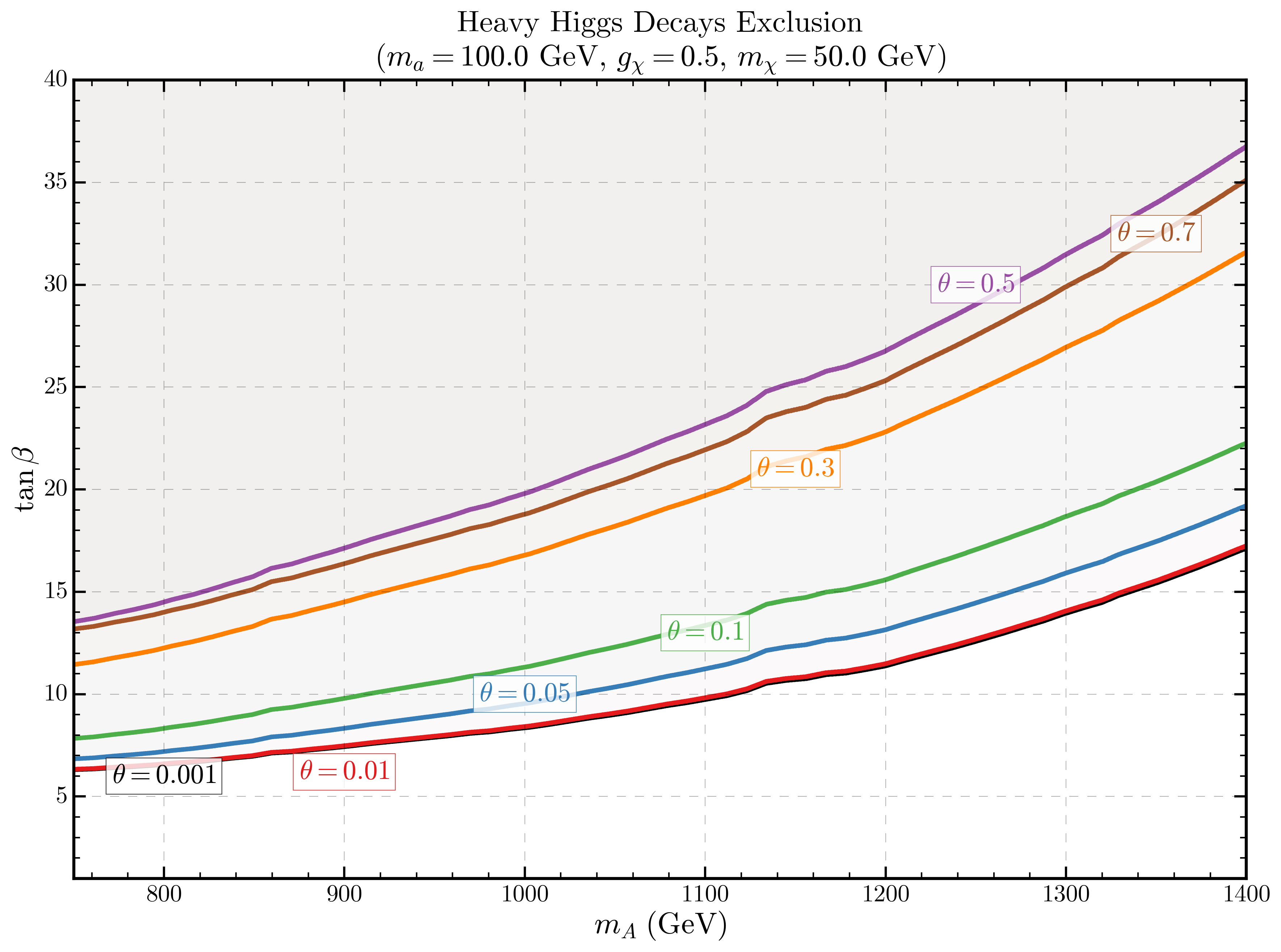}
    \caption{Constraints on the parameter space of the 2HDM+$a$ model based on recasting LHC limits \cite{Aad_2020_tau} on the decay of the heavy $A$, $H$ states to $\tau^+\tau^-$ at $95\%$ confidence level, for the sample point $m_a=100$ GeV, $g_\chi=0.5$, $m_\chi=50$ GeV, for different values of $\theta$. Values of $\tan \beta$ above the colored lines are excluded. We assume $m_H=m_A$.}
    \label{fig:2hdm-heavyhdecay}
\end{figure}

Considering new constraints: the gamma-ray line search is not particularly constraining at the large $\tan \beta$ value used in Fig. \ref{fig:2hdm-compare}, although it becomes more important at lower $\tan\beta$, effectively because the signal is enhanced by the larger coupling to the top quark loop at lower $\tan\beta$. However, the search for decay of the $A$ and $H$ to $\tau^+\tau^-$ has the opposite behavior, and is {\it very} constraining at large $\tan \beta$,  fully excluding the benchmark $\tan\beta=40$ region shown in Fig. \ref{fig:2hdm-compare}. More generally, we show in Fig.~\ref{fig:2hdm-heavyhdecay} that these bounds can probe significant parts of the region with $\tan \beta \gtrsim 6$ and $m_A \lesssim 800-1200$ GeV, with the lower limit on $\tan \beta$ depending on the mixing angle $\theta$ and $m_A$ (we retain the assumption that $m_A=m_H$ throughout).\footnote{Ref.~\cite{Arcadi:2022lpp} shows a constraint from heavy Higgs decays that cuts off rapidly and becomes non-constraining on $\tan \beta$ for $m_A=m_H \gtrsim 1200$ GeV. However, our understanding is that this cutoff reflects a breakdown in the theoretical treatment in that work, due to the loss of perturbativity at this mass scale; the original ATLAS analysis \cite{Aad_2020_tau} shows no such upturn, and so we do not believe these limits can be evaded by pushing $m_A=m_H$ to modestly higher values. We obtain comparable resutls to Ref.~\cite{Arcadi:2022lpp} at masses well below this cutoff.} These bounds are essentially independent of the lighter masses $m_a, m_\chi$, provided these are small compared to $m_A=m_H$, which can be inferred from Eq.~\ref{eq:heavydecaywidths}. The limits also depend only weakly on $g_\chi$, which controls the branching ratio for the $A\rightarrow \bar{\chi} \chi$ decay, as this is only one of the competing branching ratios (and even if it comes to dominate the $A$ decay, there will still be a signal from the $H$ decay).
 
\begin{figure}
    \centering
    \includegraphics[width=0.8\linewidth]{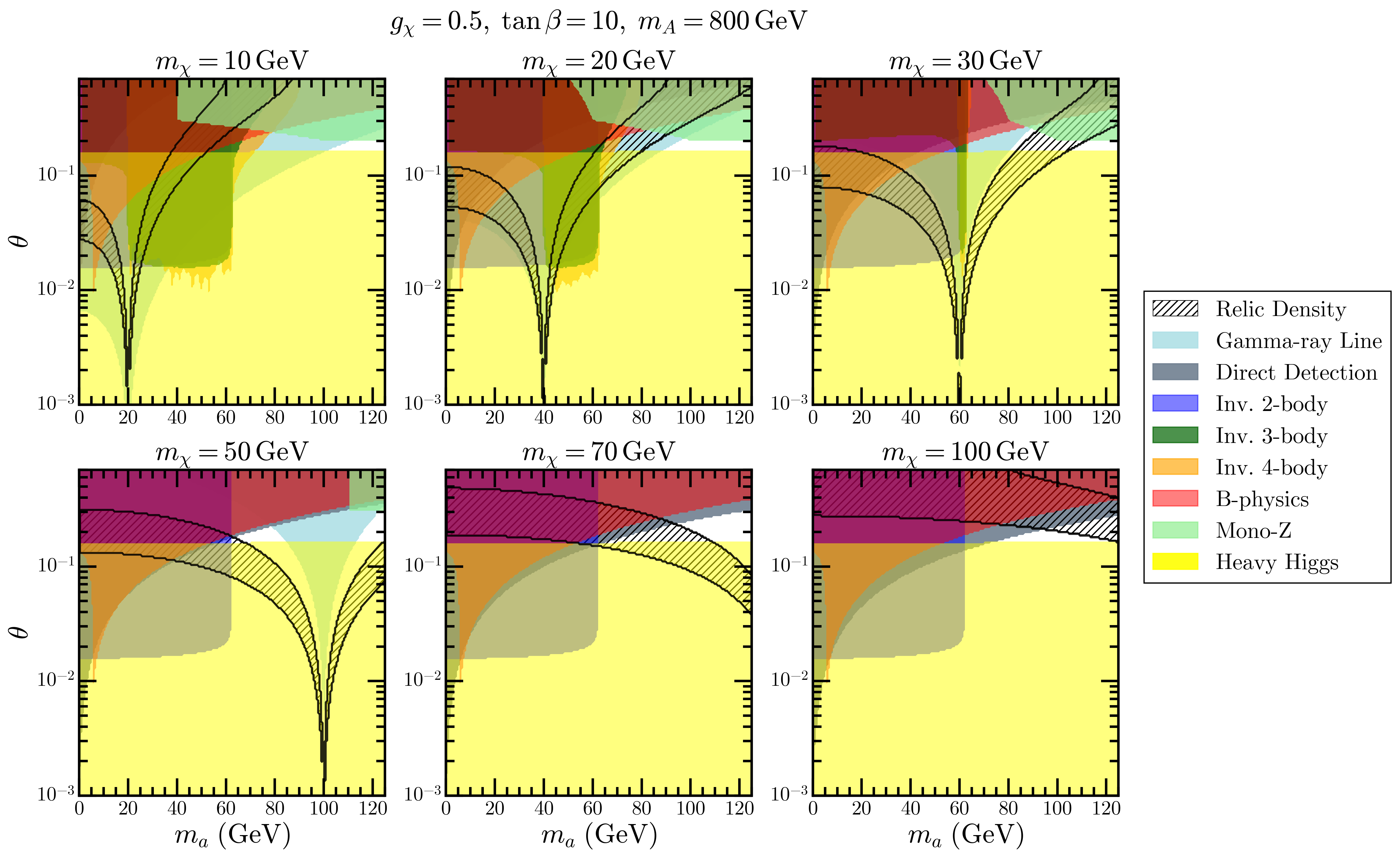}
    
    \includegraphics[width=0.8\linewidth]{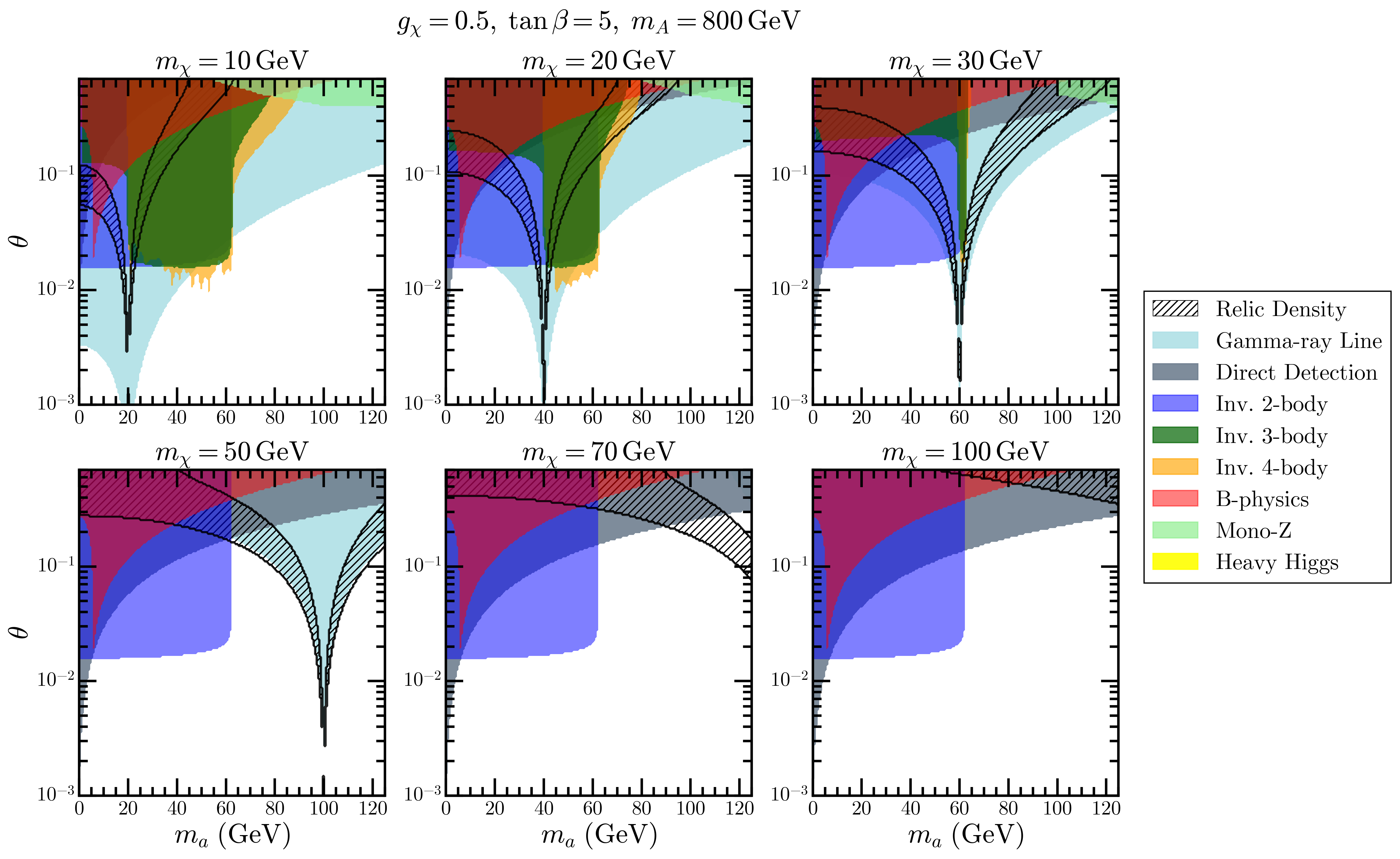}
    \caption{2HDM+$a$ model at $g_\chi=0.5$. The mono-$Z$ search constraints are only shown for $m_a\geq20$ GeV as complications might arise at very low mediator mass; additionally, the mono-$Z$ search constraints are not leading for $m_a\leq20$ GeV. The 3-body and 4-body invisible decay have numerical instabilities at $m_a\leq1/2m_h$. In general, the available parameter space is: $\tan\beta\lesssim 10$, moderate-to-high $m_a\sim\mathcal{O}(2m_\chi)$, $m_\chi \gtrsim 30$ GeV, and $\theta\sim0.1$.}
    \label{fig:2hdm-result1}
\end{figure}

\begin{figure}
    \centering
    \includegraphics[width=0.8\linewidth]{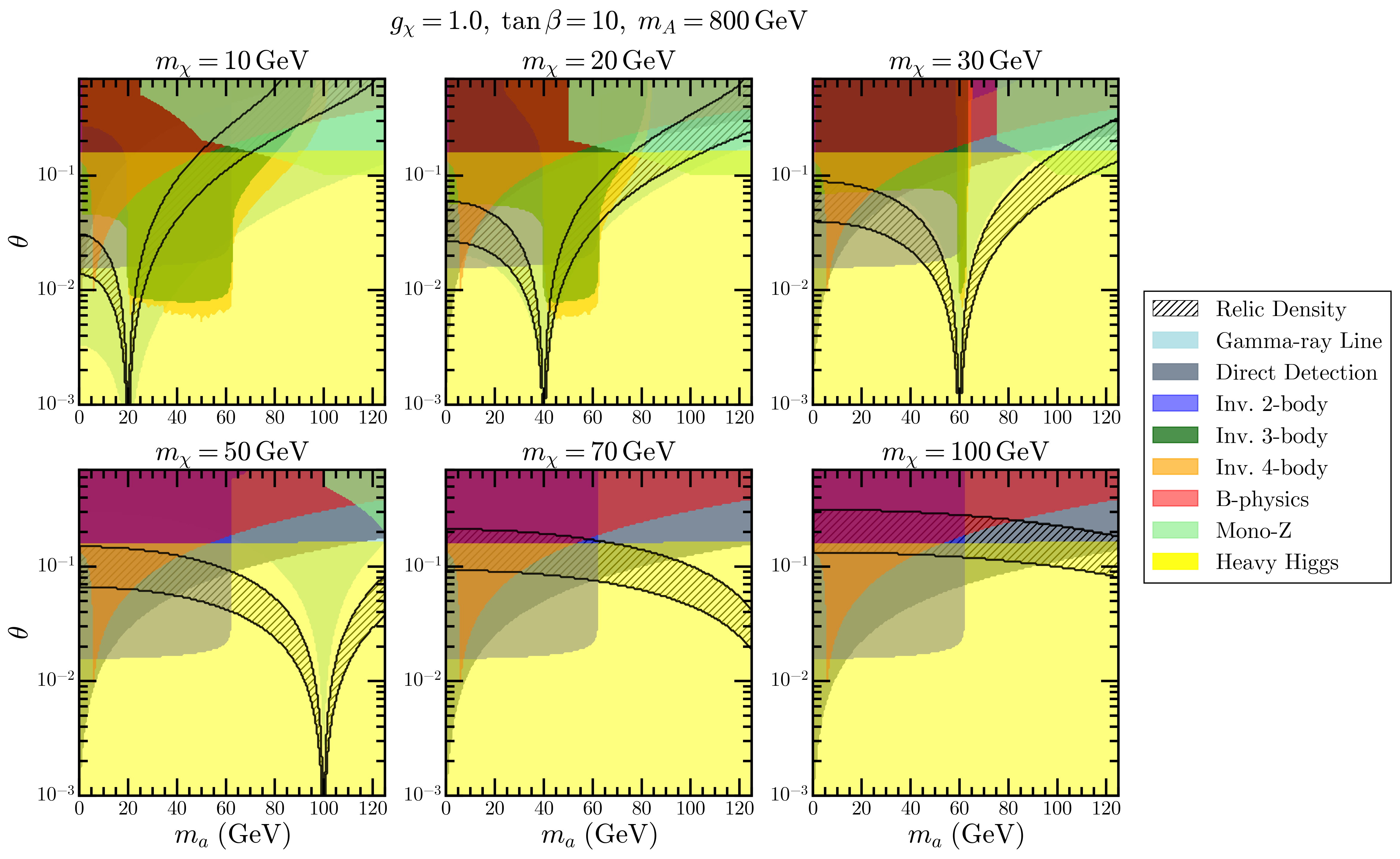}
    
    \includegraphics[width=0.8\linewidth]{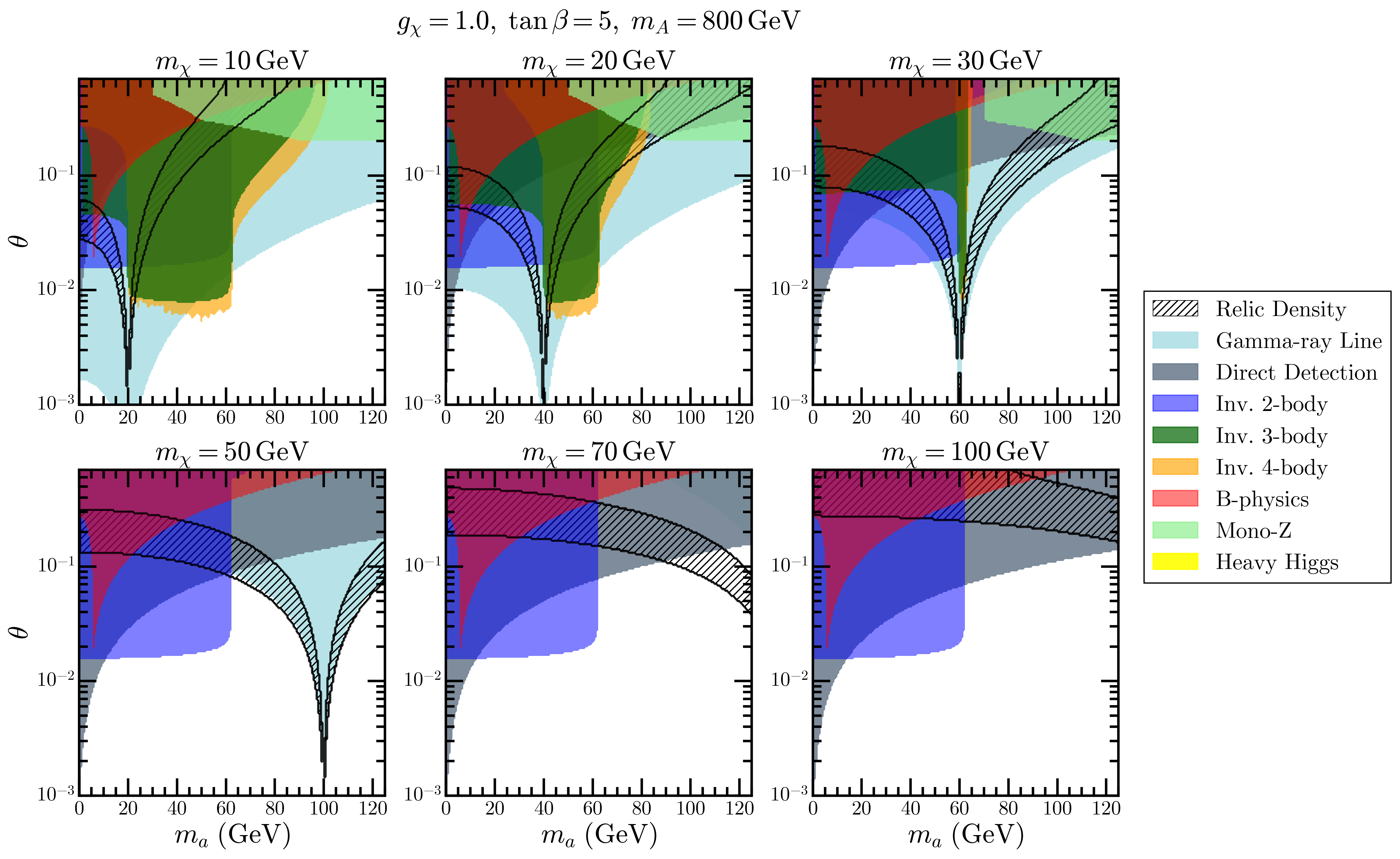}
    \caption{Constraints for the 2HDM+$a$ model at $g_\chi=1$, with the same plotting considerations and similar available parameter space as Fig. \ref{fig:2hdm-result1}, except that the previous narrow window at $\tan \beta=10$ is now closed.}
    \label{fig:2hdm-result2}
\end{figure}

For this reason, in the rest of this section we focus on the parameter space with modest $\tan \beta = 5-10$, noting that somewhat higher $\tan \beta$ values will be viable for higher values of the mixing angle $\theta$. We scan through the parameter space and obtain the results for different values of $g_\chi$, $\tan\beta$, and $m_\chi$ in Fig. \ref{fig:2hdm-result1} and \ref{fig:2hdm-result2}. 

We can see that at relatively lower $\tan\beta$, the gamma-ray line constraint is particularly strong (especially at lower $m_\chi$). The combination of the gamma-ray line limit and direct detection bounds are already enough to largely eliminate the thermal relic parameter space we have tested at $\tan\beta=5$, except in the mass range around $m_\chi \approx70$ GeV, where there is some available parameter space around the resonance region where $2m_\chi \sim \mathcal{O}(m_a)$. The invisible decay constraints are also quite stringent for $m_a <\frac{1}{2}m_h$. Going to larger $\tan \beta$ alleviates the gamma-ray line limits, allowing for viable thermal relic parameter space for $m_\chi \gtrsim 30$ GeV. However, at this value of $m_A$ and $\tan \beta$, we see the limits shown in Fig.~\ref{fig:2hdm-heavyhdecay} place quite stringent lower bounds on $\theta$, with only a sliver of parameter space remaining open and consistent with the relic density. Pushing $\tan \beta$ much higher than 10 would eliminate the parameter space due to requiring a too-high $\theta$ to be consistent with limits from direct detection; conversely, pushing $\tan \beta$ lower than 5 would strengthen the gamma-ray line bounds and eliminate parameter space at low DM masses.
Thus the combined constraints suggest a preferred parameter space with $5 \lesssim \tan \beta \lesssim 10$, $\theta \lesssim 0.3$, $m_\chi \gtrsim 30$ GeV (at sufficiently high masses, fitting the GCE may also become challenging), and $m_a\sim\mathcal{O}(2m_\chi)$ (not necessarily very close to resonance).  Increasing $g_\chi$ to 1.0 would close the remaining open parameter space at $\tan\beta=10$, forcing somewhat lower $\tan \beta$. Within this window, higher $\tan \beta$ allows somewhat lower $m_\chi$ (avoiding gamma-ray line limits) but places a more stringent lower limit on $\theta$ (to avoid constraints from the heavy Higgs decays). While there is still a non-negligible allowed region, it is now clearly more constrained than initially proposed in Ref.~\cite{ipek_renormalizable_2014}.   Allowing higher $m_A=m_H$, up to the perturbativity limit around 1.2 TeV, would alleviate the constraints from heavy Higgs decays and thus allow solutions with somewhat larger $\tan \beta$; we show the results for this case in Appendix~\ref{app:12tev}, but focus in the main text on the $m_A=m_H=800$ GeV case where there are no concerns about perturbativity/unitarity.

\begin{figure}
    \centering
    \includegraphics[width=0.8\linewidth]{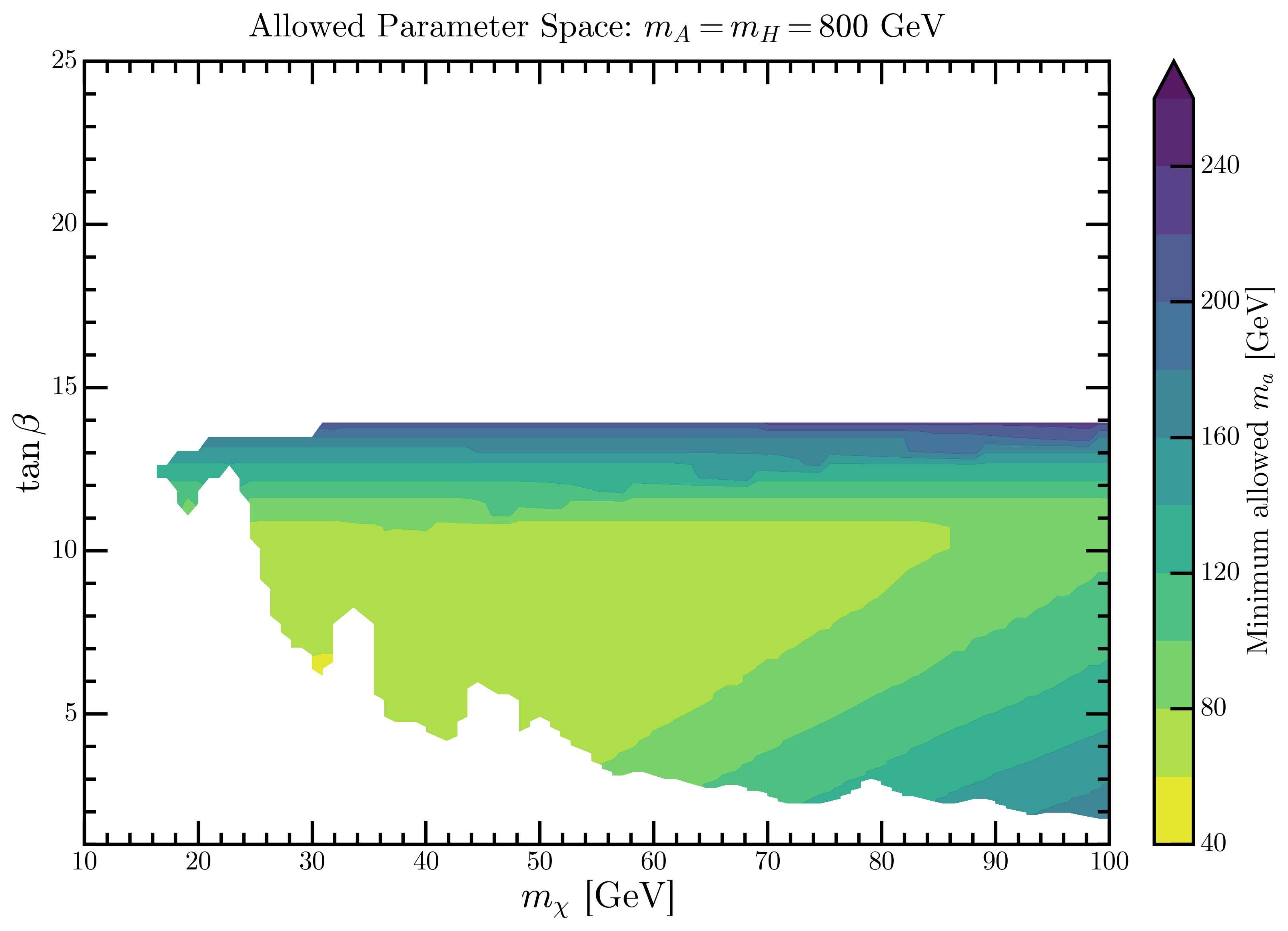}
    \caption{The values of $m_\chi$ and $\tan\beta$ that allow for simultaneous evasion of all current constraints while obtaining the correct relic density. We scan across $0.1\leq g_\chi\leq1$, $20\text{ GeV}\leq m_a\leq 250\text{ GeV}$, $10^{-3}\leq\theta\leq 1$ at $m_A=m_H=800$ GeV. In general, there is very limited parameter space at $m_\chi\lesssim 20$ GeV.}
    \label{fig:scan}
\end{figure}

To explore the broader parameter space, we have numerically scanned possible values of ($m_\chi$,$\tan\beta$), marginalizing over the other parameters $(g_\chi,\theta,m_a)$. For each value of $(m_\chi,\tan\beta)$, we identify the minimum $m_a$ (if any) for which there is parameter space that satisfies the relic density requirement and does not violate any constraints. We show the results in Fig.~\ref{fig:scan}. As inferred from the previous figures, we see that there is a fairly hard cutoff at $\tan \beta \approx 13-14$, and for lower $\tan \beta$ there is a tradeoff that lower $\tan \beta$ forces $m_\chi$ to higher values, but in general there is a broad swath of parameter space open for DM masses $\gtrsim 30$ GeV (or $\gtrsim 20$ GeV for $m_A=1200$ GeV).

To help build intuition, we also offer a discussion of the approximate dependence of different constraints at high $\tan\beta$ in Table \ref{tab:high-tanb}. This table omits couplings that are suppressed by relative factors of $\text{cot}\beta$, so will be less accurate as $\tan \beta$ approaches 1.

\begin{table}[htbp]
\centering 
\caption{The approximate dependence of different constraints of the 2HDM+$a$ model at small $\theta$, high $\tan\beta$, and $m_\chi\ll m_a \ll m_A=m_{H^\pm}$. Note that the B physics constraint also is dependent on $m_{H^\pm}$.}
\label{tab:high-tanb}
\begin{tabular}{lcccccc}
\toprule
Process & $\theta$ & $\tan\beta$ & $m_a$ & $m_\chi$ & $g_\chi$ & $m_A$ \\
\midrule
$\langle\sigma v\rangle$ 
  & $\theta^2$ 
  & $\tan^4\beta$ 
  & $1/m_a^{4}$ 
  & $m_\chi^2$ 
  & $g_\chi^2$ 
  & $\times$ \\
Direct Detection
  & $\theta^4$ 
  & $\times$ 
  & $\times$ 
  & $m_\chi^2$ 
  & $g_\chi^4$ 
  & $m_A^4$ \\
Gamma-ray Line 
  & $\theta^2$ 
  & $\tan^2\beta$ 
  & $1/m_a^{8}$ 
  & $m_\chi^4$ 
  & $g_\chi^2$ 
  & $\times$ \\
$B_s^0\to\mu^+\mu^-$    
  & $\theta^4$ 
  & $\tan^4\beta$ 
  & $1/m_a^{4}$ 
  & $\times$ 
  & $\times$ 
  & $\times$ \\
$h\to aa$ onshell 
  & $\theta^4$ 
  & $\times$ 
  & $\times\,(\text{if }m_a\ll \frac{1}{2}m_h)$
  & $\times$ 
  & $\times$ 
  & $m_A^4$ \\
Invisible One-Loop Decay   
  & $\theta^4$ 
  & $\times$ 
  & $1/m_a^{4}$ 
  & $\times$ 
  & $g_\chi^4$
  & $m_A^4$ \\
  Invisible 3-Body Decay 
  & $\theta^4$
  & $\times$
  & $1/m_a^2$
  & $\times$
  & $g_\chi^2$
  & $m_A^4$ \\
  Invisible 4-Body Decay 
  & $\theta^4$
  & $\times$
  & $1/m_a^4$
  & $\times$
  & $g_\chi^4$
  & $m_A^4$ \\
Mono-$Z$
  & $\theta^2$ 
  & $\tan^2\beta$ 
  & $1/m_a^2$ 
  &  $\times$
  & $g_\chi^2$ 
  & $\times$ \\
\bottomrule
\end{tabular}
\end{table}

 Close to resonance (i.e. $m_a \approx 2 m_\chi$), a smaller mixing $\theta$ is required to obtain the thermal relic cross section and so this region is generically less constrained. However, sufficiently close to the resonance the $s$-wave cross section may be enhanced at low velocities relative to freezeout, potentially overproducing the GCE. Roughly speaking, near the Breit-Wigner resonance the denominator of the (squared) propagator has the form $(2 m_\chi + m_\chi v^2)^2 - m_a^2 \approx (2 m_\chi)^2 - m_a^2 + 4 m_\chi^2 v^2$ (where $v$ is the velocity of one of the DM particles in the center-of-momentum frame), so the velocity-dependent term is important for $v^2 \gtrsim |1 - (m_a/(2 m_\chi))^2|$. The characteristic velocity at freeze-out corresponds to $v^2 \sim 0.1$, so we can expect there to be non-trivial velocity evolution between freeze-out and the present day for $m_a/(2 m_\chi) \sim 0.95-1.05$. Since there is substantial unconstrained parameter space available without relying on such close proximity to the resonance, we do not explore this region further in this work.

\section{Future Projections}
\label{sec:forecast}

Finally, we will project the prospects for improvements on these constraints--or for a discovery--as the experimental sensitivity continues to advance in the coming years. 

\begin{itemize}
    \item We will consider the neutrino floor as the direct detection forecast \cite{Billard_2022}. This can be achieved in the next-generation experiments such as DARWIN \cite{DARWIN:2016hyl} or XLZD \cite{XLZD:2024nsu}. 

\item For the beam dump and accelerator search for the secluded hypercharge model, we include the latest projection for SHiP experiment \cite{SHiP:2021nfo}, which has the highest sensitivity to the parameter space we are searching for \cite{Alenezi:2025kwl}. 

\item For the LHC mono-$Z$ search, we roughly estimate that the LHC increases the luminosity by one order of magnitude with the HL-LHC \cite{CERNHLHCPDR2015}; we then scale the current collider bounds to match this estimate. 

\item For the $B$-physics search, the HL-LHC is expected to improve the uncertainty of the branching ratio $\operatorname{Br}\left(B_s^0 \to \mu^+ \mu^- \right)$ to around 10\% \cite{ATL-PHYS-PUB-2025-016}. The resulting upper bound is not improved compared to the current constraint. 

\item For the search for the $\tau^+\tau^-$ decays of the heavy Higgses, Ref~\cite{ATL-PHYS-PUB-2018-050} forecasts the sensitivity for the HL-LHC.

\item For the invisible Higgs decay branching ratio, the HL-LHC will give a constraint of $2.5\%$ at 95\% confidence \cite{Cepeda:2019klc}. 

\item For the gamma-ray line search, we compare the Fermi telescope bound to the sensitivity forecast of the upcoming Cherenkov Telescope Array (CTA) \cite{CTAO:2024wvb} and the proposed Advanced Particle-astrophysics Telescope (APT) \cite{APT:2021lhj,Alnussirat:2021tlo}. We find that at the DM candidate mass range of $\sim\qty{50}{\GeV}$ range that we are interested for the GCE, the bound of the CTA is not improved compared to our current Fermi telescope limit, while the proposed APT improves the constraint by more than an order of magnitude \cite{Alnussirat:2021tlo, Xu:2023zyz}. Because the APT is still at a relatively early design stage, rather than perform a detailed forecast, we simply examine the effect of improving the cross section sensitivity in gamma rays by an order of magnitude.

\end{itemize}

\begin{figure}
    \centering
    \includegraphics[width=0.65\linewidth]{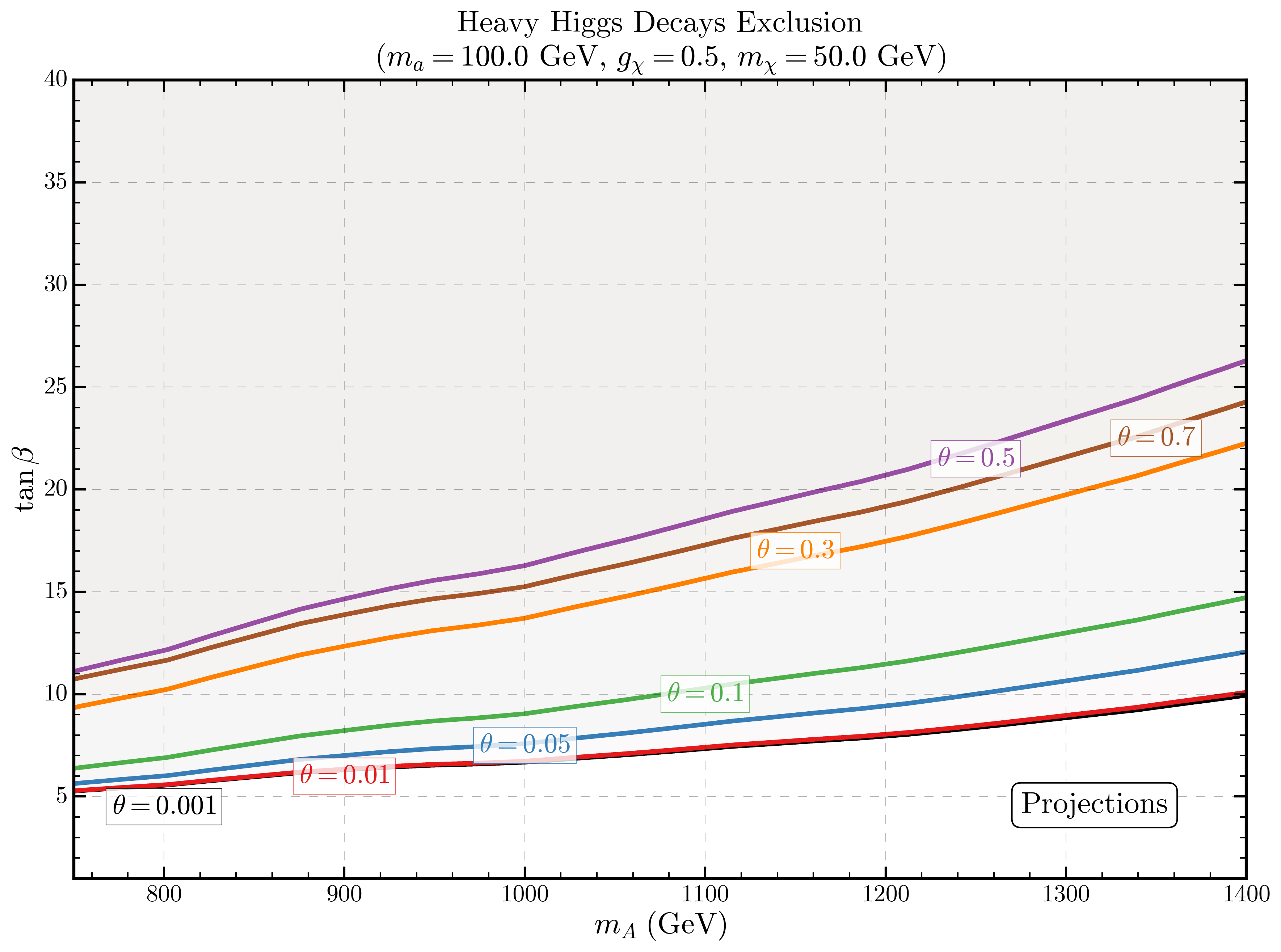}
    \caption{Forecast constraints on the parameter space of the 2HDM+$a$ model based on recasting forecast HL-LHC limits on the decay of the heavy $A$, $H$ states to $\tau^+\tau^-$ at $95\%$ confidence level, for the sample point $m_a=100$ GeV, $g_\chi=0.5$, $m_\chi=50$ GeV, for different values of $\theta$. Values of $\tan \beta$ above the colored lines will be excluded by a null result. We assume $m_H=m_A$.}
    \label{fig:2hdm-heavyhdecay-forecast}
\end{figure}

\begin{figure}
    \centering
    \includegraphics[width=\linewidth]{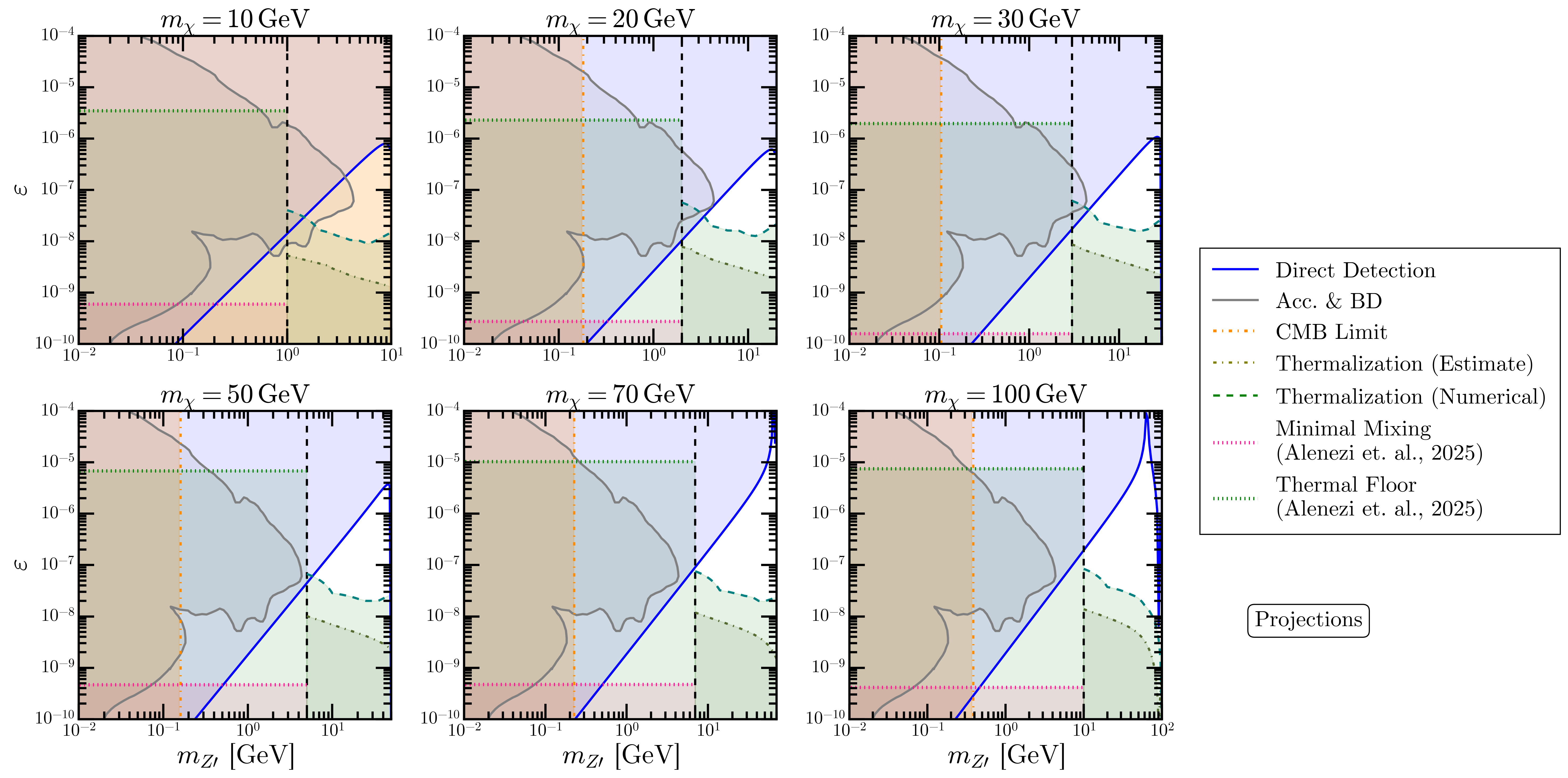}
    \caption{Forecast of constraints for the secluded hypercharge model at different $m_\chi$. In general, there would still be some parameter space left open at higher $m_{Z'}$.}
    \label{fig:hypercharge-forecast}
\end{figure}

\begin{figure}
    \centering
    \includegraphics[width=0.8\linewidth]{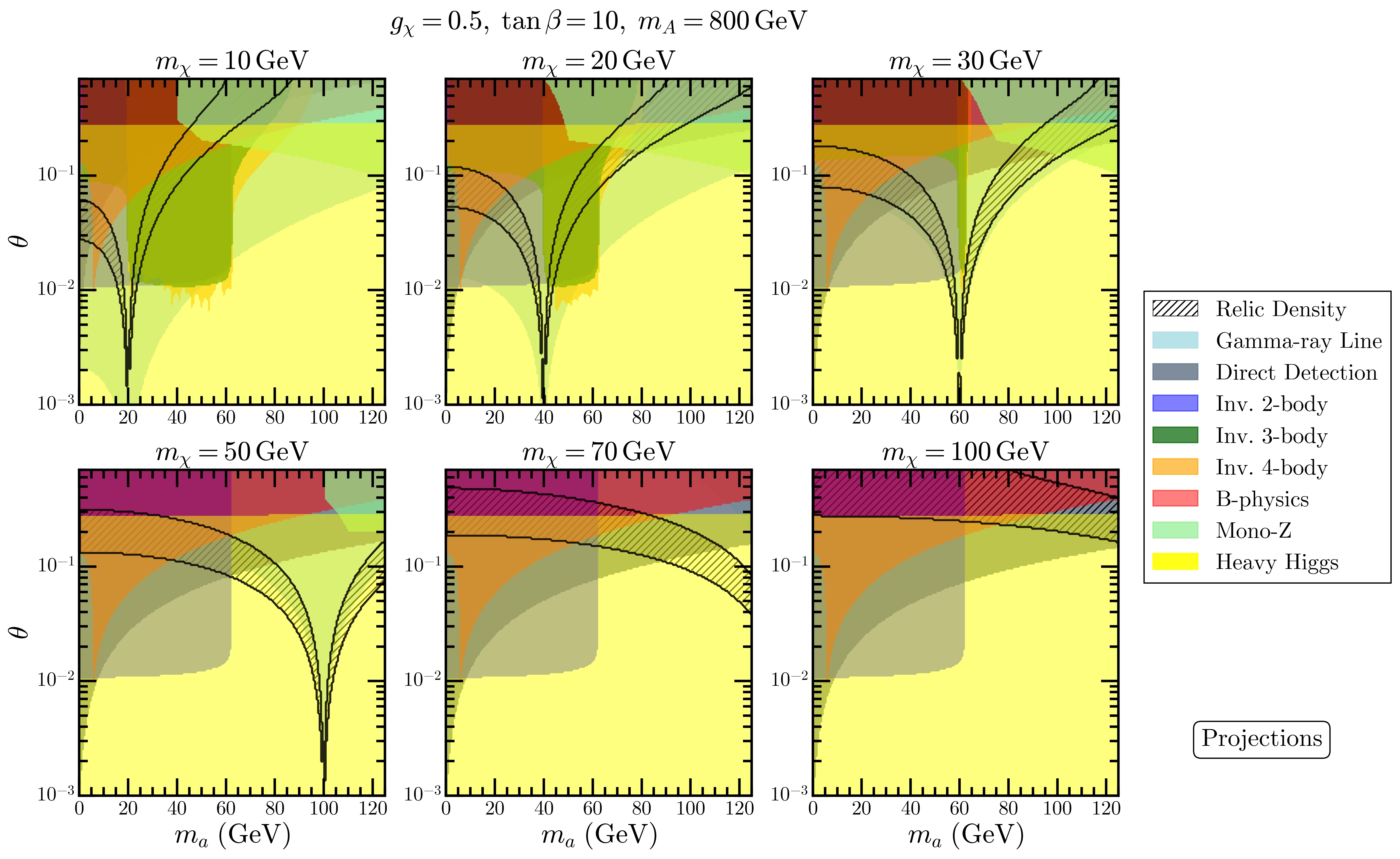}

    \includegraphics[width=0.8\linewidth]{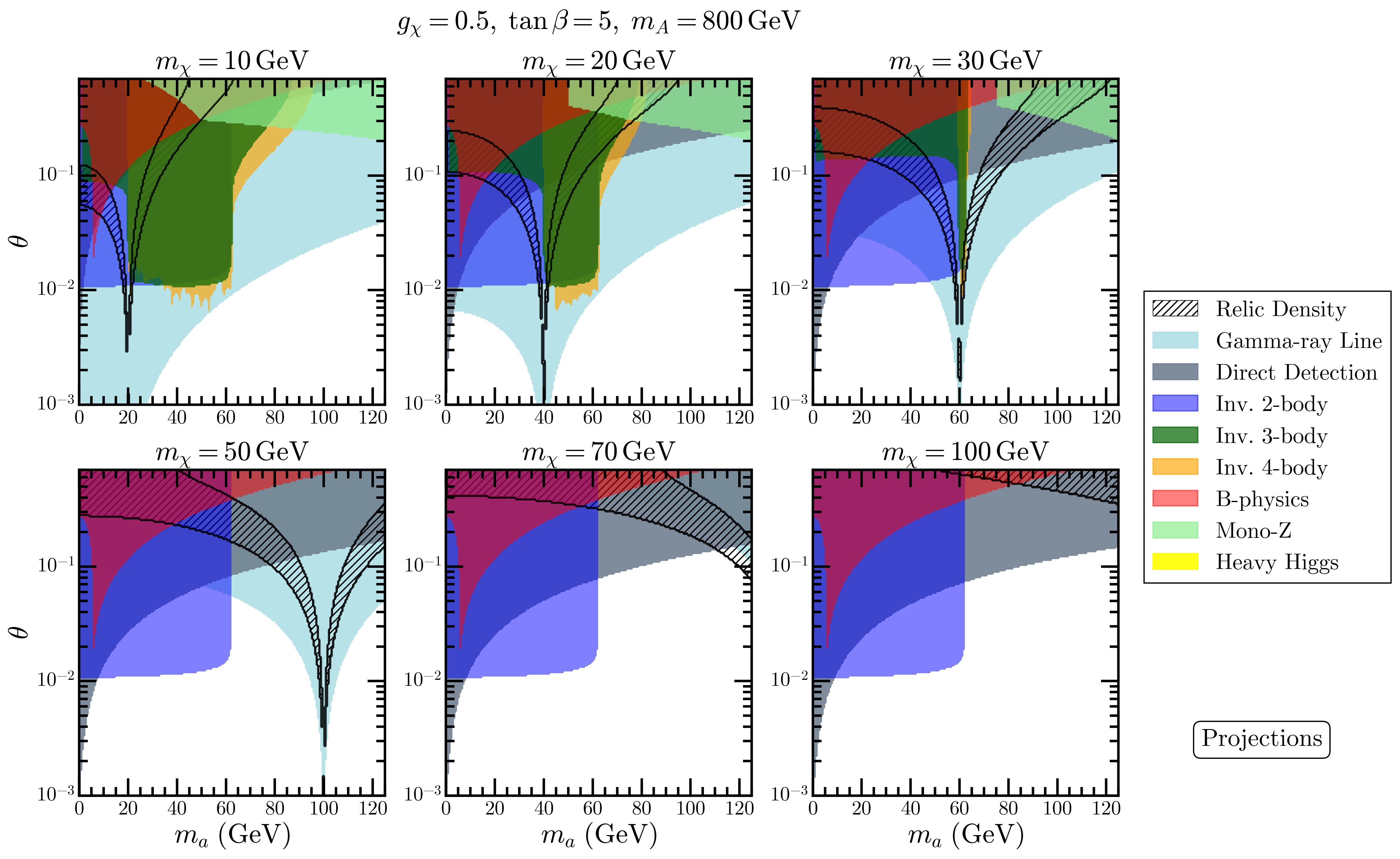}
    \caption{Forecast of constraints for the 2HDM+$a$ model at $g_\chi=0.5$. The viable parameter space would be significantly decreased by null results at the HL-LHC for decay signals from the heavy $H$ and $A$ fields, and by null results from direct detection and gamma-ray line searches in particular. A narrow window would survive for $\tan \beta \sim 5-10$ at $m_\chi \sim 70$ MeV, $m_a\sim\mathcal{O}(2m_\chi)$. 
}
    \label{fig:2hdm-forecast1}
\end{figure}

\begin{figure}
    \centering
    \includegraphics[width=0.8\linewidth]{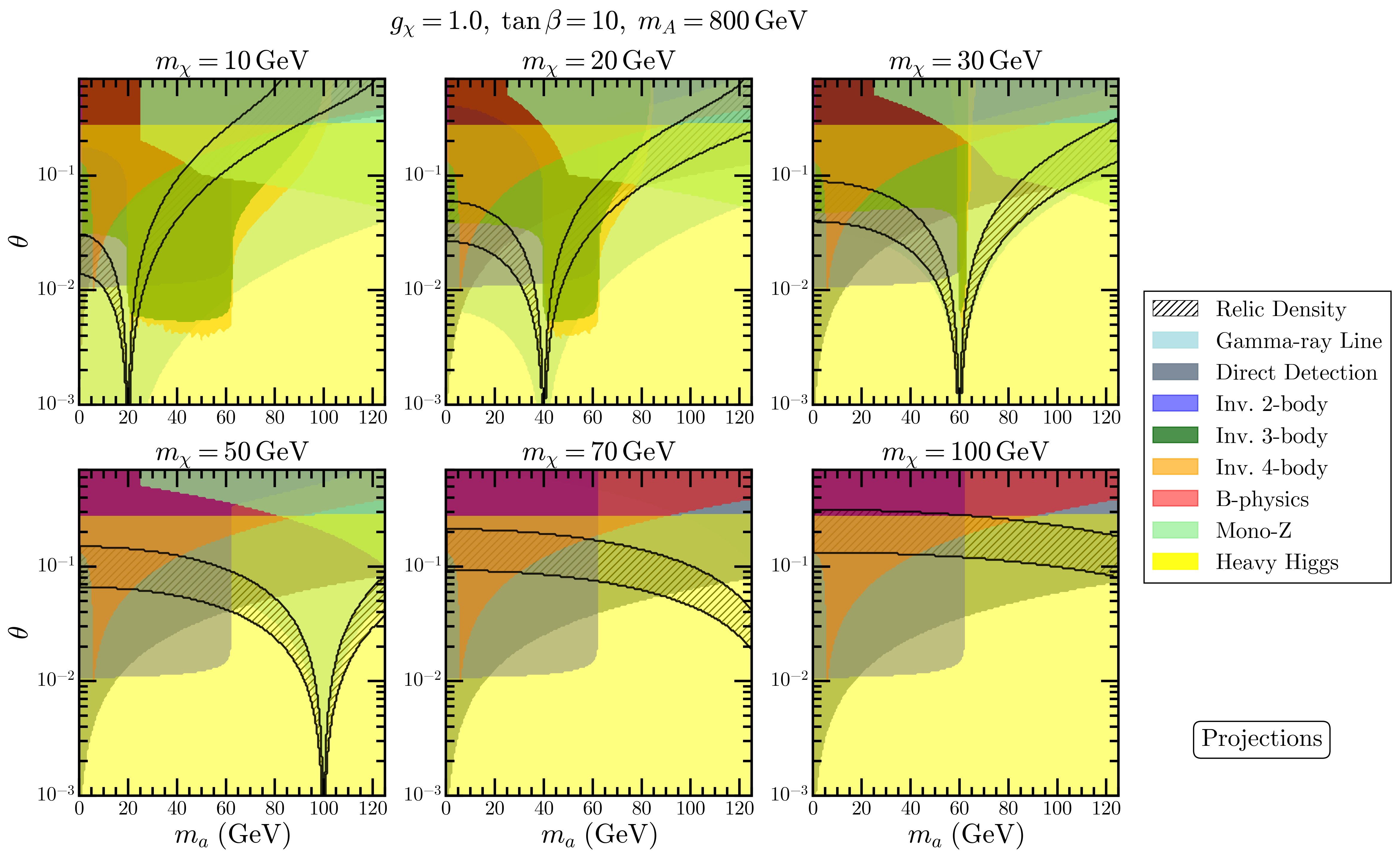}

    \includegraphics[width=0.8\linewidth]{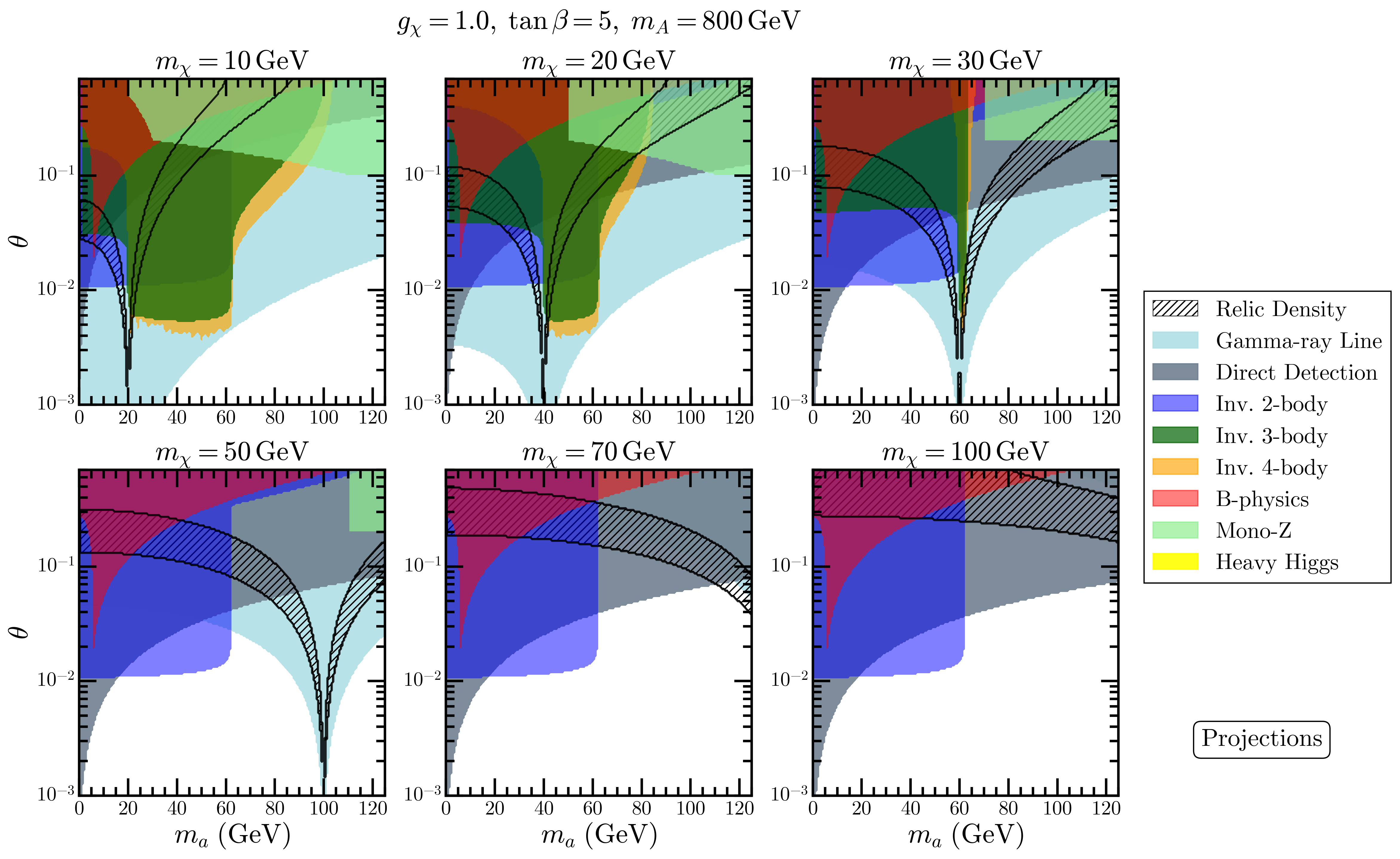}
    \caption{Forecast of constraints for the 2HDM+$a$ model at $g_\chi=1$, with similar available parameter space as Fig. \ref{fig:2hdm-forecast1}.}
    \label{fig:2hdm-forecast2}
\end{figure}

\begin{figure}
    \centering
    \includegraphics[width=0.8\linewidth]{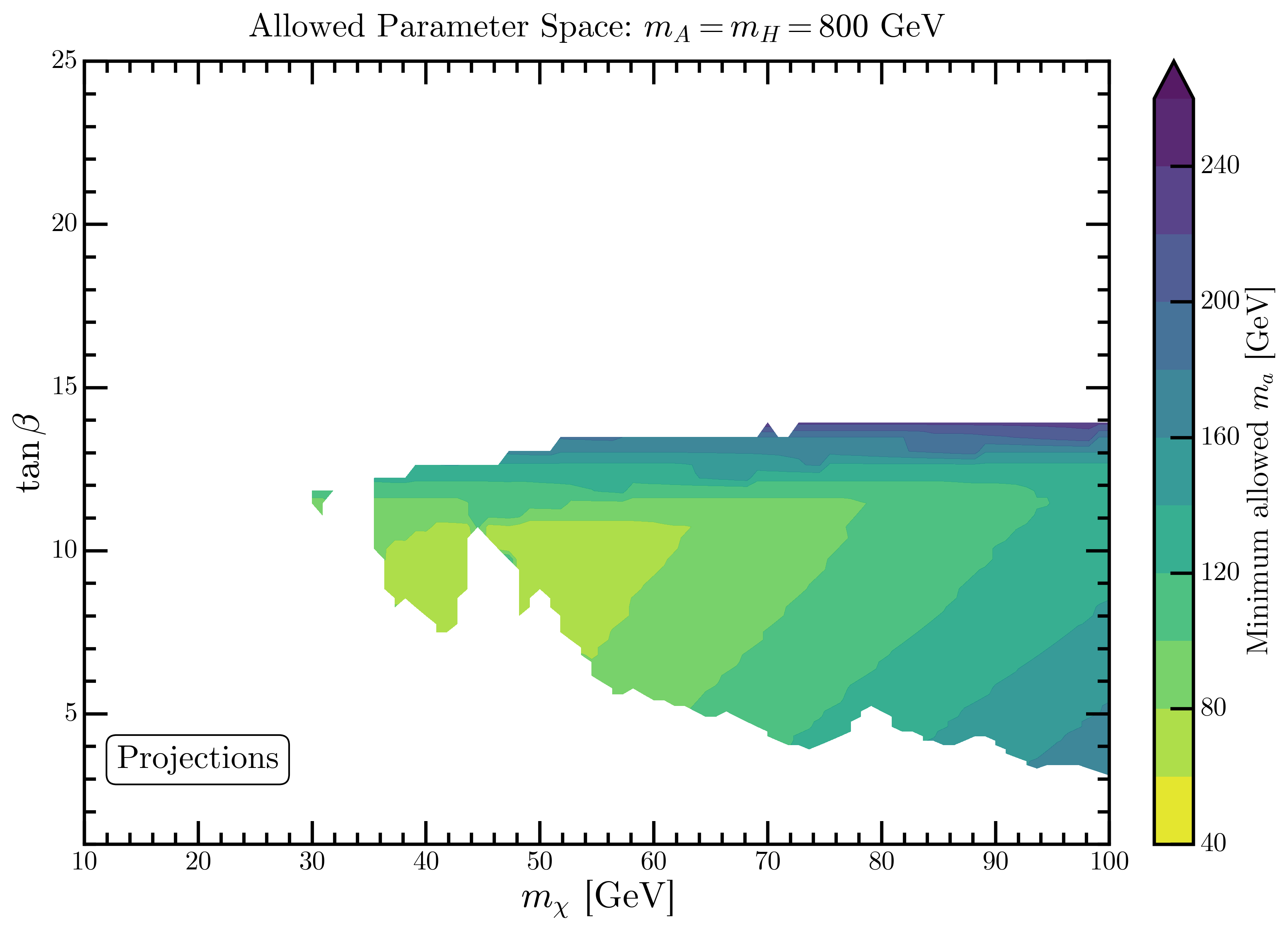}
    \caption{The values of $m_\chi$ and $\tan\beta$ that allow for simultaneous evasion of all forecast constraints while obtaining the correct relic density. We scan across $0.1\leq g_\chi\leq1$, $20\text{ GeV}\leq m_a\leq 250\text{ GeV}$, $10^{-3}\leq\theta\leq 1$ at $m_A=m_H=800$ GeV. With the forecast constraints, lower $m_\chi$ values are more severely limited; in particular, there is no open parameter space for $m_\chi\lesssim35$ GeV.}
    \label{fig:scan_forecast}
\end{figure}

We therefore forecast the constraints of the secluded hypercharge model and the 2HDM+$a$ model in Fig.~\ref{fig:2hdm-heavyhdecay-forecast}, \ref{fig:hypercharge-forecast}, \ref{fig:2hdm-forecast1}, \ref{fig:2hdm-forecast2}, and \ref{fig:scan_forecast}.

For the secluded hypercharge model, we can see that the SHiP search would still remain subdominant to the direct detection search. There would be a marginal improvement from the direct detection limit at the neutrino floor; however, there would still be some parameter space left open at higher $m_{Z^\prime}$. 

For the 2HDM+$a$ model, we can see that at $\tan\beta\gtrsim 10$, the open parameter space would  be entirely closed by the improved heavy Higgs decay constraints for our default choice of $m_A=m_H=800$ GeV (for $m_A=m_H=1200$ GeV, approaching the limit from perturbativity, there would be a sliver of parameter space left open).
At $\tan \beta=5$, this level of strengthening of direct-detection and gamma-ray constraints would leave only a small region of open parameter space at high masses $m_\chi \approx 70$ GeV (note, however, that a factor-of-ten improvement in sensitivity to gamma-rays could independently possibly exclude or confirm the DM hypothesis for the GCE based on dwarf galaxy observations). Going to modestly higher $\tan\beta$ should allow for a somewhat wider viable DM mass range.

We show the forecast for the $\tan \beta-m_\chi$ parameter space surviving all constraints in Fig.~\ref{fig:scan_forecast}. The lower bound on the viable mass range would be driven up to $\sim 40-50$ GeV (or $\sim 30$ GeV for $m_A=m_H=1200$ GeV), with a modest improvement to the upper limit on $\tan \beta$. As previously, the favored parameter space to avoid exclusion (assuming null results) would correspond to $\theta \lesssim 0.2$ and $2 m_\chi \sim \mathcal{O}(m_a)$.

\section{Conclusion}
After updating the constraints for the secluded hypercharge model and the 2HDM+$a$ model, we see that there is still open parameter space that is broadly consistent with a thermal relic DM origin for the GCE. However, we note that the advent of experiments have eliminated a large part of this closing parameter space. Future experiments have the potential to further test new regions of parameter space, but will not fully exclude these scenarios in the absence of a detection (with the possible exception of improved gamma-ray sensitivity allowing model-independent tests of the GCE through observations of dwarf galaxies).

For the secluded hypercharge model, indirect bounds from the cosmic microwave background are constraining at the low end of the mass range ($m_\chi \lesssim 10$ GeV). At the opposite DM mass regime, while the CMB limits do place bounds on the $Z^\prime$ mass due to the presence of Sommerfeld enhancement, for thermal relics the constraints from direct detection (and to a lesser degree, accelerators) are always  stronger (similar results are found in \cite{Alonso-Gonzalez:2025xqg}). In general, these limits require the $Z^\prime$ mediator to be heavier than $1-2$ GeV, in order to simultaneously maintain the assumption of a standard cosmological history. The limit on the mixing of the $Z^\prime$ with the SM hypercharge depends strongly on its mass, being constrained above by direct detection and below by the requirement of full equilibrium prior to freeze-out, but broadly speaking there is parameter space available when the DM mass is 10s of GeV, the $Z^\prime$ mass is in the range from a few GeV to the DM mass (with heavier $m_{Z'}$, closer to $m_\chi$, being both less constrained and likely to give a better fit to the GCE spectrum), and the mixing is in the range $\epsilon \sim 10^{-8}-10^{-6}$.

For the 2HDM+$a$ model, it is harder to provide a simple summary as the parameter space is quite high-dimensional, but we can again make some general statements. Decays of the heavy Higgs fields $H$ and $A$ are strongly constrained by the LHC, and constrain $\tan\beta \lesssim 30$ in general, with the constraint being stronger for lower mixing angles $\theta$ (and lower $m_A=m_H$). This complements the constraints involving the low-scale physics, in particular those from direct detection, which are generally more constraining at high $\theta$. In general, we find $\theta \lesssim 0.3$ is preferred, which means the heavy Higgs decay limits imply $\tan \beta \lesssim 13-14$ for $m_A=800$ GeV ($\lesssim 22-23$ for $m_A=1200$ GeV), and we also require $2 m_\chi$ to be of the same order as $m_a$ to obtain the correct relic density. The gamma-ray line limits are especially constraining at low $\tan \beta$ and low DM mass, whereas limits from invisible decays of the Higgs become very strong for $m_a < m_h/2$. Putting these bounds together, we find the island of viable parameter space for $m_A=m_H=800$ GeV satisfies $\tan \beta \lesssim 12-13$, $m_\chi \gtrsim$ 30 GeV, with higher $\tan \beta$ allowing a wider range of masses, and higher $m_A=m_H$ opening up more space at high $\tan \beta$. Future improvements in direct detection, gamma-ray line searches, and heavy Higgs decays would simultaneously push down the limit on $\theta$, modestly reduce the allowed range of $\tan \beta$, and constrain the DM mass to $m_\chi \gtrsim 40-50$ GeV, but are unlikely to fully exclude the model's parameter space.


A conclusion that we can draw from both of the benchmark models is that the union of cosmological and terrestrial experiments are necessary for the study of DM. While the fundamental origin of the GCE is not yet known, the next generation of experiments will provide important insight towards understanding the particle content and observable nature of DM in the universe. At the same time, we see in both of these examples that while terrestrial experiments have significantly narrowed the viable parameter space for a DM explanation of the GCE (and will continue to do so), there remains parameter space that will be very difficult to exclude with any set of null results at terrestrial experiments, emphasizing the importance of more model-independent tests of the GCE in dwarf galaxies and other astrophysical systems.

\section*{Acknowledgments}
We would like to thank Gordan Krnjaic and Jessie Shelton for useful feedback and discussions, and in particular we are grateful to Maria Olalla Olea-Romacho and Thomas Biek\"{o}tter  for pointing out the relevance of the constraints on heavy Higgs decays for the 2HDM+$a$ model. This work was 
supported by the U.S. Department of Energy (DOE) Office of High Energy Physics under
Grant Contract No. DE-SC0012567. Y.H. was supported by the MIT Undergraduate Research Opportunities Program. T.R.S. and C.C. were supported in part by the Simons Foundation (Grant No. 929255). T.R.S. was supported during the course of this work by a Guggenheim Fellowship; the Edward, Frances, and Shirley B. Daniels Fellowship of the Harvard Radcliffe Institute; and the Bershadsky Distinguished Fellowship of the Harvard Physics Department.

\appendix

\section{Calculation of the GCE spectra}
\label{sec:spectra}
With the parameters of the DM models, we can generate the corresponding GCE spectra. We obtain the photon fluxes $dN_a/dE_\gamma$ at production per annihilation event from \cite{cirelli_pppc_2011,Ciafaloni:2010ti}, where $E_\gamma$ is the energy of the photon. To convert the photon flux at production per annihilation event of Dirac fermions to the GCE signal $E_\gamma^2dN_\gamma/dE$, we use \cite{Fermi-LAT:2017opo,elor_multi-step_2015}
\begin{equation}
E_\gamma^2\frac{dN_\gamma}{dE_\gamma}= E_\gamma^2\frac{\langle\sigma v\rangle}{16\pi m_\chi^{2}}J(\Delta\Omega)\frac{dN_a}{dE_\gamma}.
\end{equation}

$J(\Delta\Omega)$ is the $J$-factor of the region of interest (ROI) $\Delta\Omega$ in the Galactic Center, which accounts for the distribution and the density of the DM halo in the Galactic Center. 

The $J$-factor is calculated with \cite{BERGSTROM1998137}
\begin{equation}
J(\Delta\Omega)=\int_\text{ROI}d\Omega\int_0^\infty ds\rho^{2}(r(s,l,b)),
\end{equation}
where
\begin{equation}
r(s,l,b)=\sqrt{r_{\odot}^{2}+s^{2}-2r_{\odot}s\cos l\cos b},
\end{equation}
$s$ is the distance along the line of sight (LOS), $r_{\odot}\simeq8.5\text{ kpc}$ is the distance between the sun and the Galactic Center, $l$ and $b$ are the galactic coordinates, and $\rho$ is the density of the DM. 

In the generalized Navarro-Frenk-White (gNFW) profile, 
the density of the DM can be written as \cite{Navarro:1996gj}
\begin{equation}
    \rho(r)=\rho_{s}\left(\frac{r}{r_{s}}\right)^{-\gamma}\left(1+\frac{r}{r_{s}}\right)^{\gamma-3},
\end{equation}
where $\gamma$ is the inner slope, $r_s$ is the scale radius, and $\rho_s$ is the scale density.  We take $r_{s}\simeq20\text{ kpc}$, and the scale density $\rho_s$ fixed by the local density $\rho(r_\odot)\approx\qty{0.4}{\GeV\cm^{-3}}$. 

For the secluded hypercharge model, the dominant annihilation channel is $\chi\bar{\chi}\to Z' Z'$ followed by $Z'\to f\bar{f}$. We then calculate the kinematics of the cascade decay \cite{elor_multi-step_2015}: given the decay spectrum of $dN_a/dE_{\gamma,Z'}$ where $E_{\gamma,Z'}$ is the energy of photons in the rest frame of $Z'$, we can convert the spectrum to the energy of photons in the rest frame of $\chi$, $E_\gamma$, by \cite{elor_multi-step_2015}:
\begin{equation}
    \frac{dN_a}{dE_\gamma}=\frac{2m_{Z'}}{m_\chi}\int^{E_\text{max}}_{E_\text{min}}\frac{dE_{\gamma,Z'}}{E_{\gamma,Z'}\sqrt{1-(2m_{Z'}/m_\chi)^2}}\frac{dN_a}{dE_{\gamma,Z'}},
\end{equation}
where the integration limits
\begin{equation}
\begin{aligned}
    &E_\text{max}=\min\left[\frac{m_{Z'}}{2},\frac{E_\gamma m_\chi}{2m_{Z'}}\left(1+\sqrt{1-(2m_{Z'}/m_\chi)^2}\right)\right],\\
    &E_\text{min}=\frac{E_\gamma m_\chi}{2m_{Z'}}\left(1-\sqrt{1-(2m_{Z'}/m_\chi)^2}\right).
\end{aligned}
\end{equation}

We then find the branching ratio from \eqref{eq:decay_z} and the annihilation spectra from \textsc{CosmiXs} \cite{Arina:2023eic,DiMauro:2024kml} and verify the spectra against \textsc{PPPC4DMID} \cite{cirelli_pppc_2011,Ciafaloni:2010ti}.

For the 2HDM+$a$ model, the available phase space favors $s$-channel annihilation of $\chi\chi\to b\bar{b}$. We then directly use the annihilation spectra from \textsc{CosmiXs} \cite{Arina:2023eic,DiMauro:2024kml}.

We obtain the possible GCE spectra across different background models from Ref.~\cite{Dinsmore:2021nip}, with an ROI of $|l|<20^\circ,  2^\circ<|b|<20^\circ$. We use a gNFW profile with $\gamma=1.25$. The $J$-factor is then calculated to be 
\begin{equation}
    J(|l|<20^\circ,  2^\circ<|b|<20^\circ)\simeq\qty{1.02e23}{\GeV^2\cm^{-5}}.
\end{equation}

For the secluded hypercharge model, we select $m_\chi=50\text{ GeV}$, $m_{Z'}=15\text{ GeV}$, and $\braket{\sigma v}=\qty{4.4e-26}{\cm^3\s^{-1}}$. For the 2HDM+$a$ model, we select $m_\chi=50\text{ GeV}$ and $\braket{\sigma v}=\qty{4.4e-26}{\cm^3\s^{-1}}$. 

\begin{figure}
    \centering
    \includegraphics[width=0.8\linewidth]{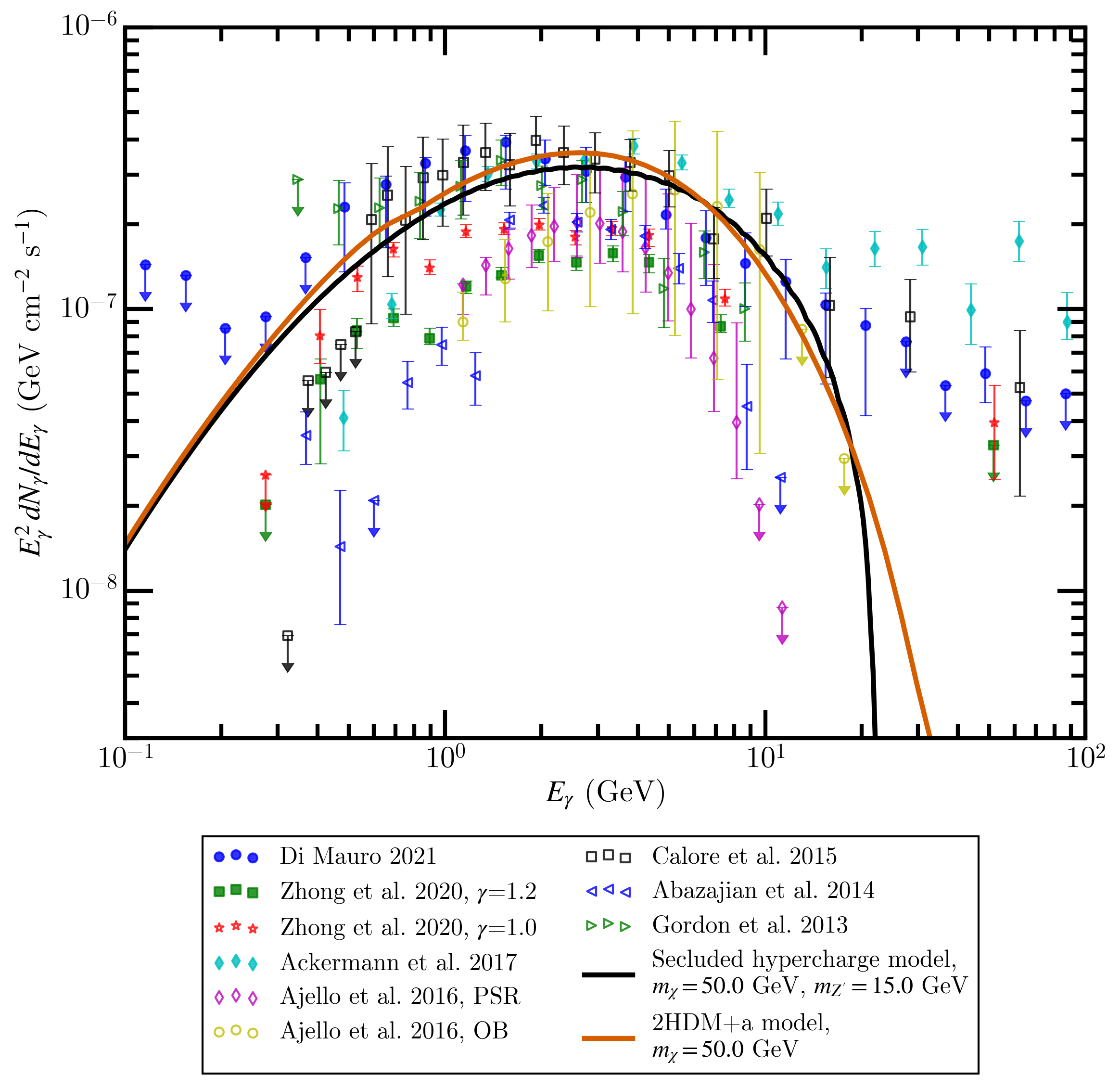}
    \caption{Example spectra of the allowed parameters in our model. The possible GCE spectra with different background models are obtained from \cite{DiMauro:2021raz,Calore:2014xka,Zhong:2019ycb,Gordon:2013vta,Fermi-LAT:2015sau,Fermi-LAT:2017opo,Abazajian:2014fta}, adapted from \cite{Dinsmore:2021nip}. For the secluded hypercharge model, we select $m_\chi=50\text{ GeV}$, $m_{Z'}=15\text{ GeV}$; for the 2HDM+$a$ model, we select $m_\chi=50\text{ GeV}$. The annihilation cross section is at $\braket{\sigma v}_\text{th}=\qty{4.4e-26}{\cm^3\per\s}$. The $J$-factor $J=\qty{1.02e23}{\GeV^2\per\cm^5}$ using a gNFW profile of $\gamma=1.25$ and a ROI of $|l|<20^\circ,  2^\circ<|b|<20^\circ$. }
    \label{fig:spectrum}
\end{figure}

We calculate the example spectra in Fig. \ref{fig:spectrum}. The spectra analyses are obtained from Ref.~\cite{DiMauro:2021raz,Calore:2014xka,Zhong:2019ycb,Gordon:2013vta,Fermi-LAT:2015sau,Fermi-LAT:2017opo,Abazajian:2014fta}. We can see that the generated spectra of both the secluded hypercharge model and the 2HDM+$a$ model are within the envelope of the possible GCE spectra. Due to the large statistical uncertainties of the GCE and the DM profile, we are only able to offer a qualitative comparison and not use the fitting as a quantitative constraint. 

\section{Supplemental results for $m_A=m_H=1.2$ TeV}
\label{app:12tev}

In this appendix we show the analogues of Figs.~\ref{fig:2hdm-result1}-\ref{fig:scan} (constraints),  Figs.~\ref{fig:2hdm-forecast1}-\ref{fig:scan_forecast} (forecasts) for $m_A=m_H=1.2$ TeV, approaching the limit from perturbativity/unitarity. We observe that the main effect of increasing $m_A=m_H$ is to open more parameter space at modestly higher $\tan \beta$, via relaxation of the bounds from heavy Higgs decays. 

\begin{figure}
    \centering
    \includegraphics[width=0.8\linewidth]{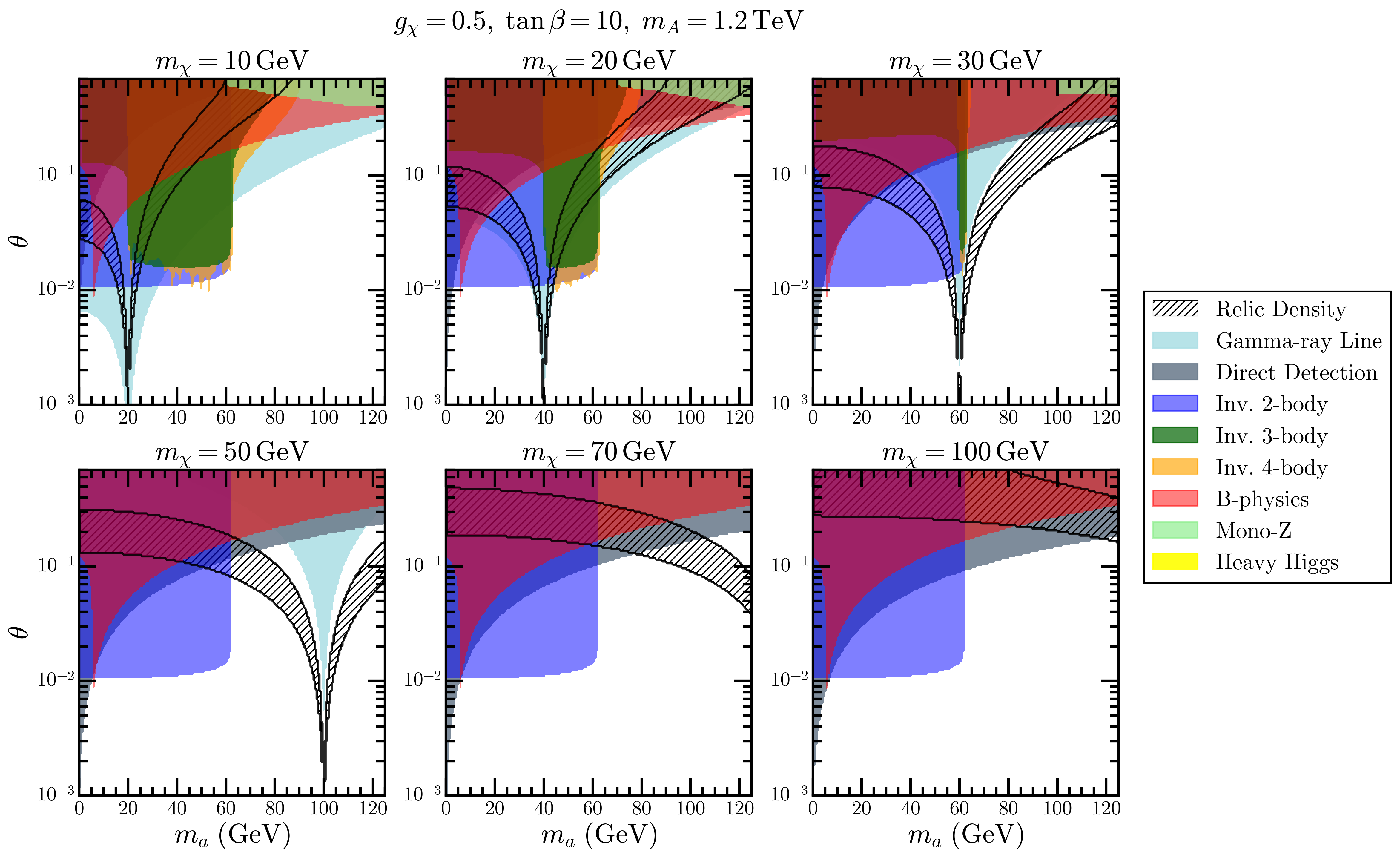}
    
    \includegraphics[width=0.8\linewidth]{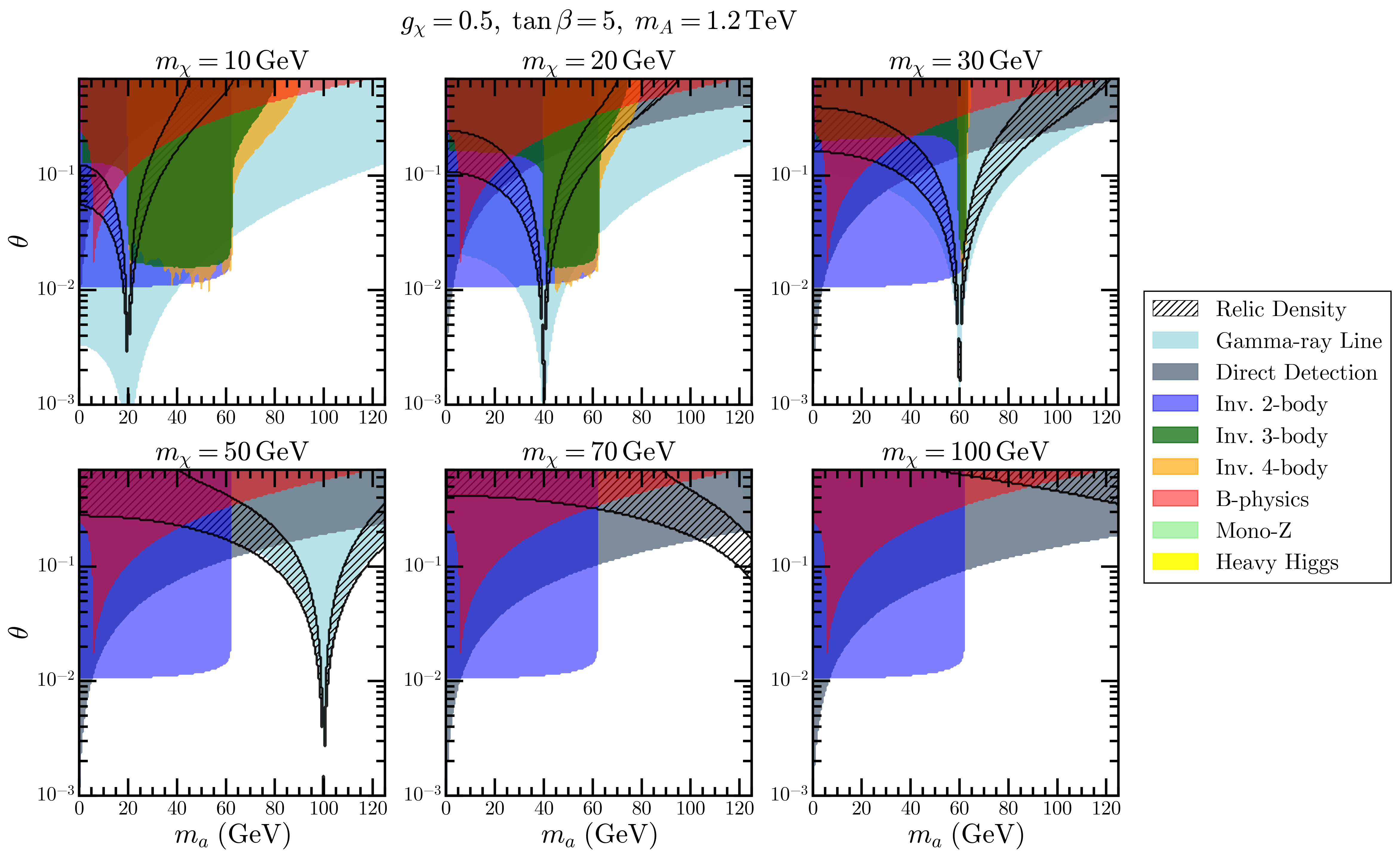}
    \caption{Analogue of Fig.~\ref{fig:2hdm-result1}, but with $m_A=m_H=1.2$ TeV.}
    \label{fig:2hdm-result1-12tev}
\end{figure}

\begin{figure}
    \centering
    \includegraphics[width=0.8\linewidth]{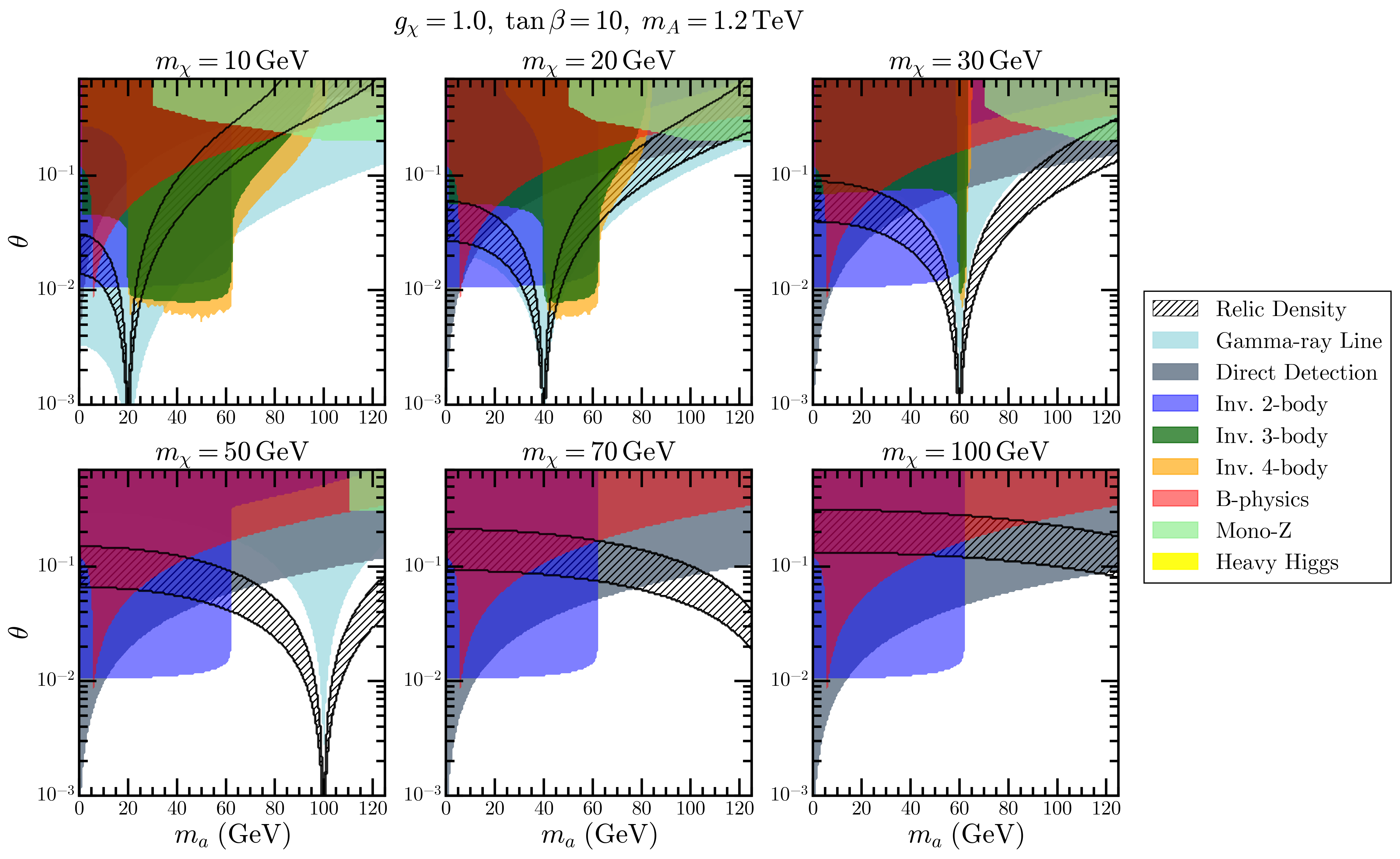}
    
    \includegraphics[width=0.8\linewidth]{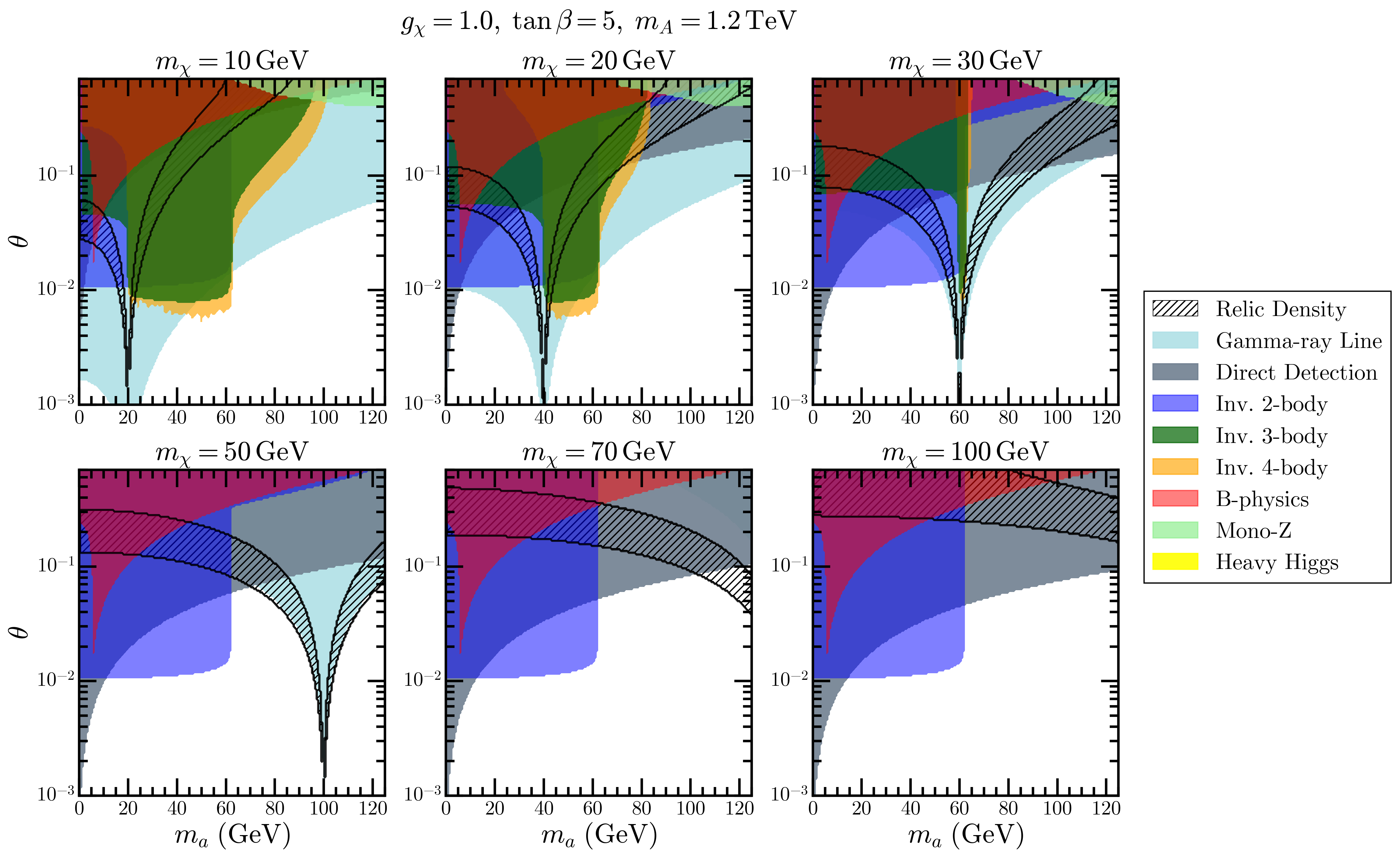}
    \caption{Analogue of Fig.~\ref{fig:2hdm-result2}, but with $m_A=m_H=1.2$ TeV.}
    \label{fig:2hdm-result2-12tev}
\end{figure}

\begin{figure}
    \centering
    \includegraphics[width=0.8\linewidth]{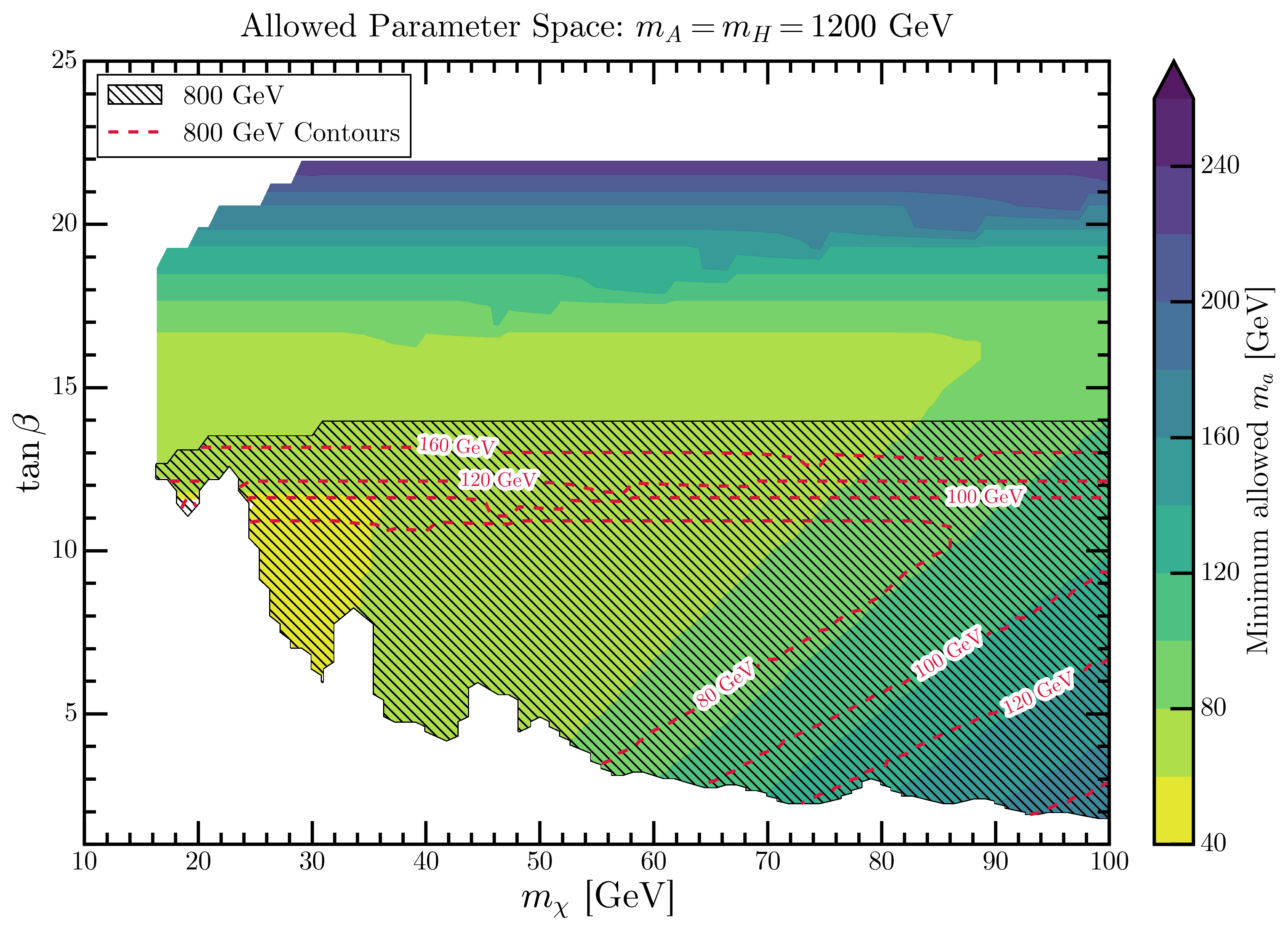}
    \caption{Analogue of Fig.~\ref{fig:scan}, but with $m_A=m_H=1.2$ TeV. We include the $m_A=m_H=800$ GeV contours for comparison.}
    \label{fig:scan-12tev}
\end{figure}

\begin{figure}
    \centering
    \includegraphics[width=0.8\linewidth]{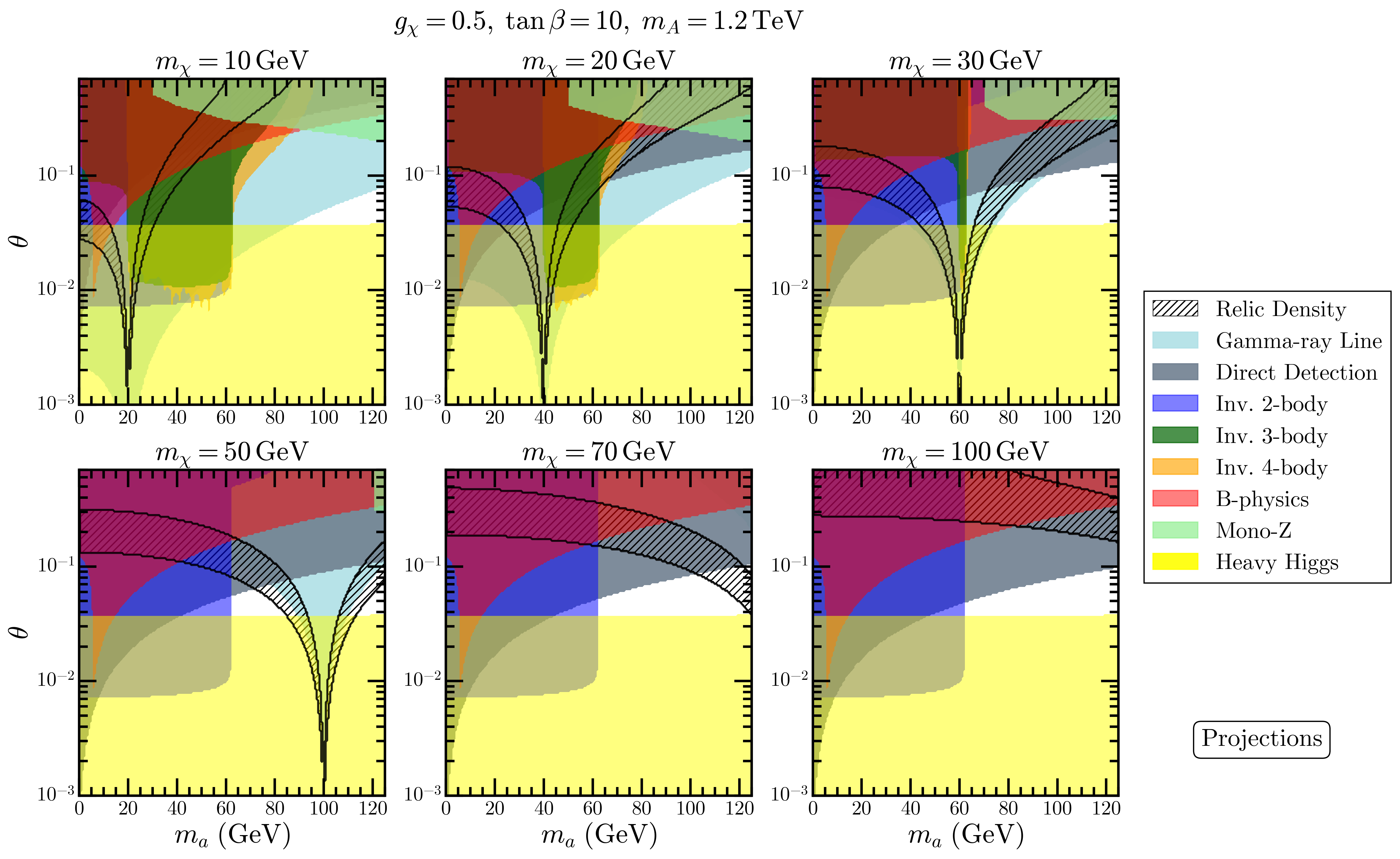}

    \includegraphics[width=0.8\linewidth]{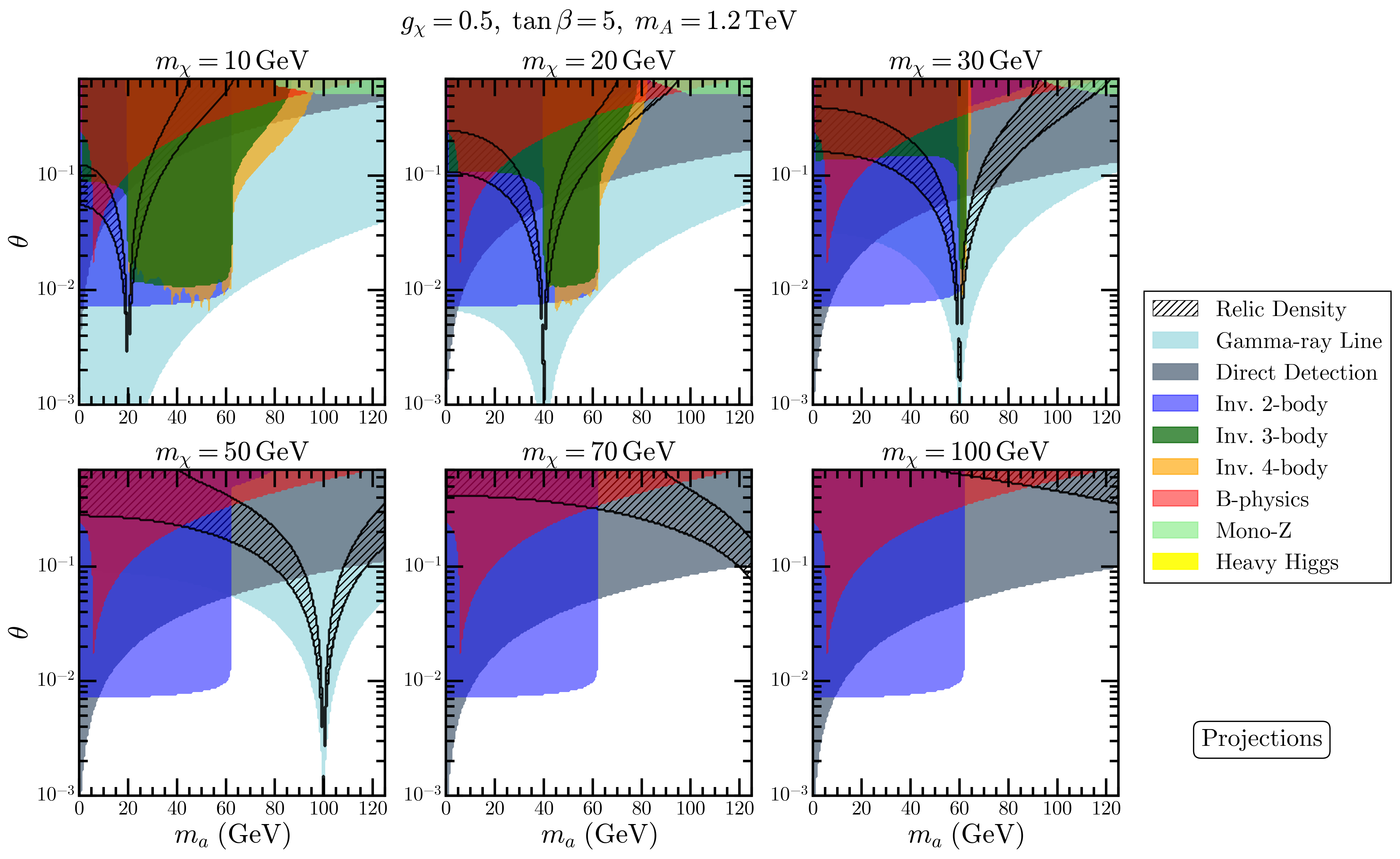}
    \caption{Analogue of Fig.~\ref{fig:2hdm-forecast1}, but with $m_A=m_H=1.2$ TeV.
}
    \label{fig:2hdm-forecast1-12tev}
\end{figure}

\begin{figure}
    \centering
    \includegraphics[width=0.8\linewidth]{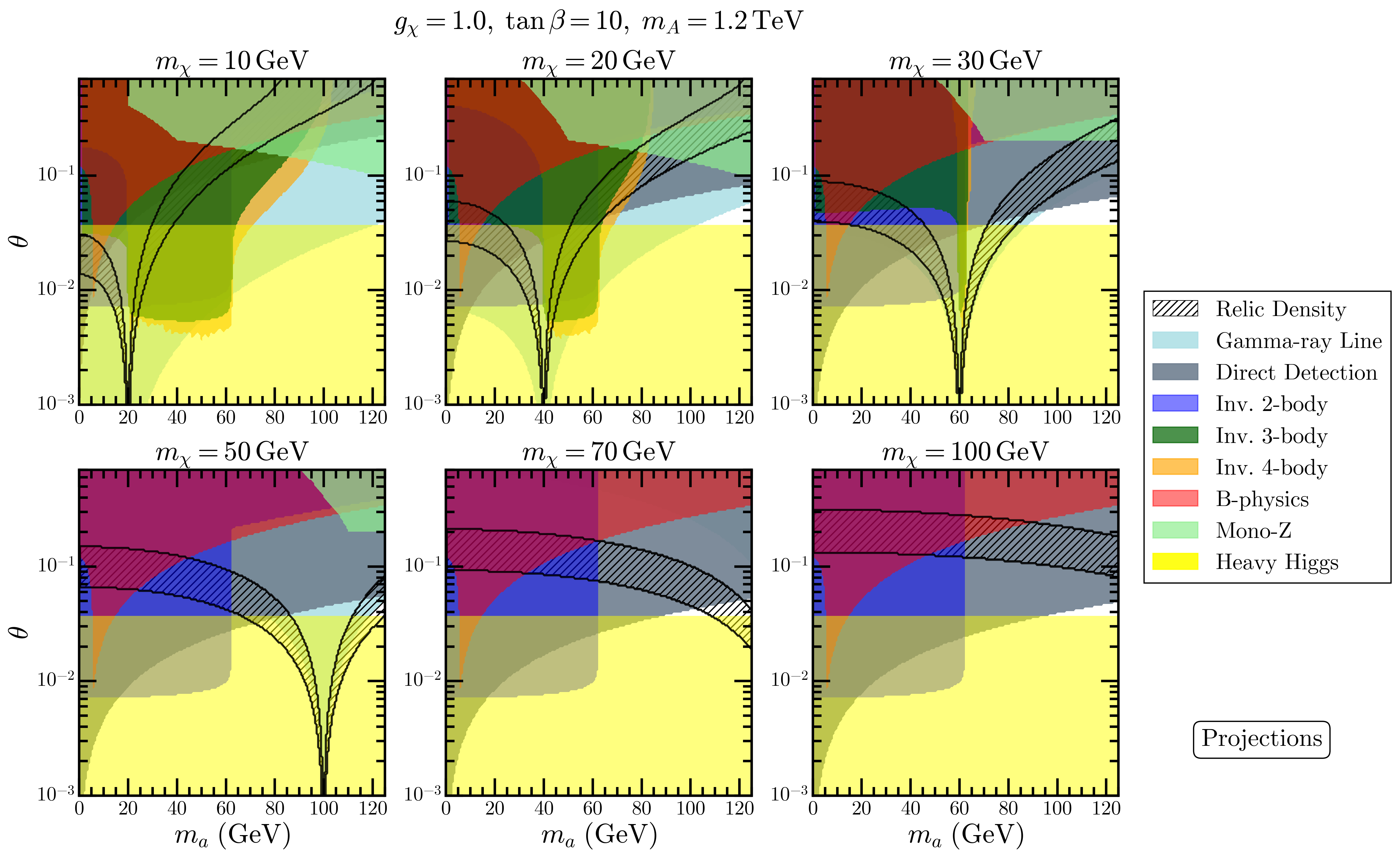}

    \includegraphics[width=0.8\linewidth]{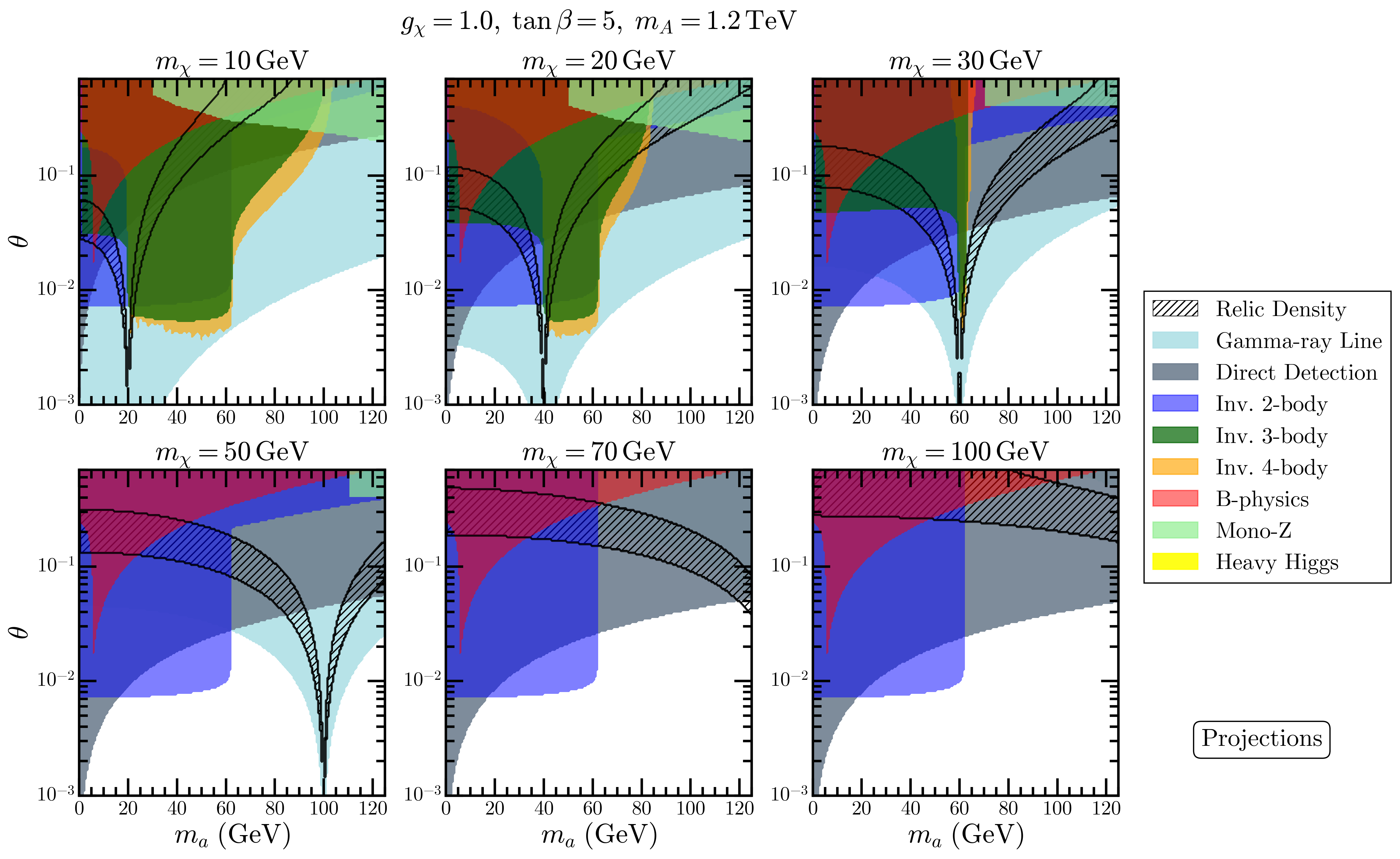}
    \caption{Analogue of Fig.~\ref{fig:2hdm-forecast2}, but with $m_A=m_H=1.2$ TeV.}
    \label{fig:2hdm-forecast2-12tev}
\end{figure}

\begin{figure}
    \centering
    \includegraphics[width=0.8\linewidth]{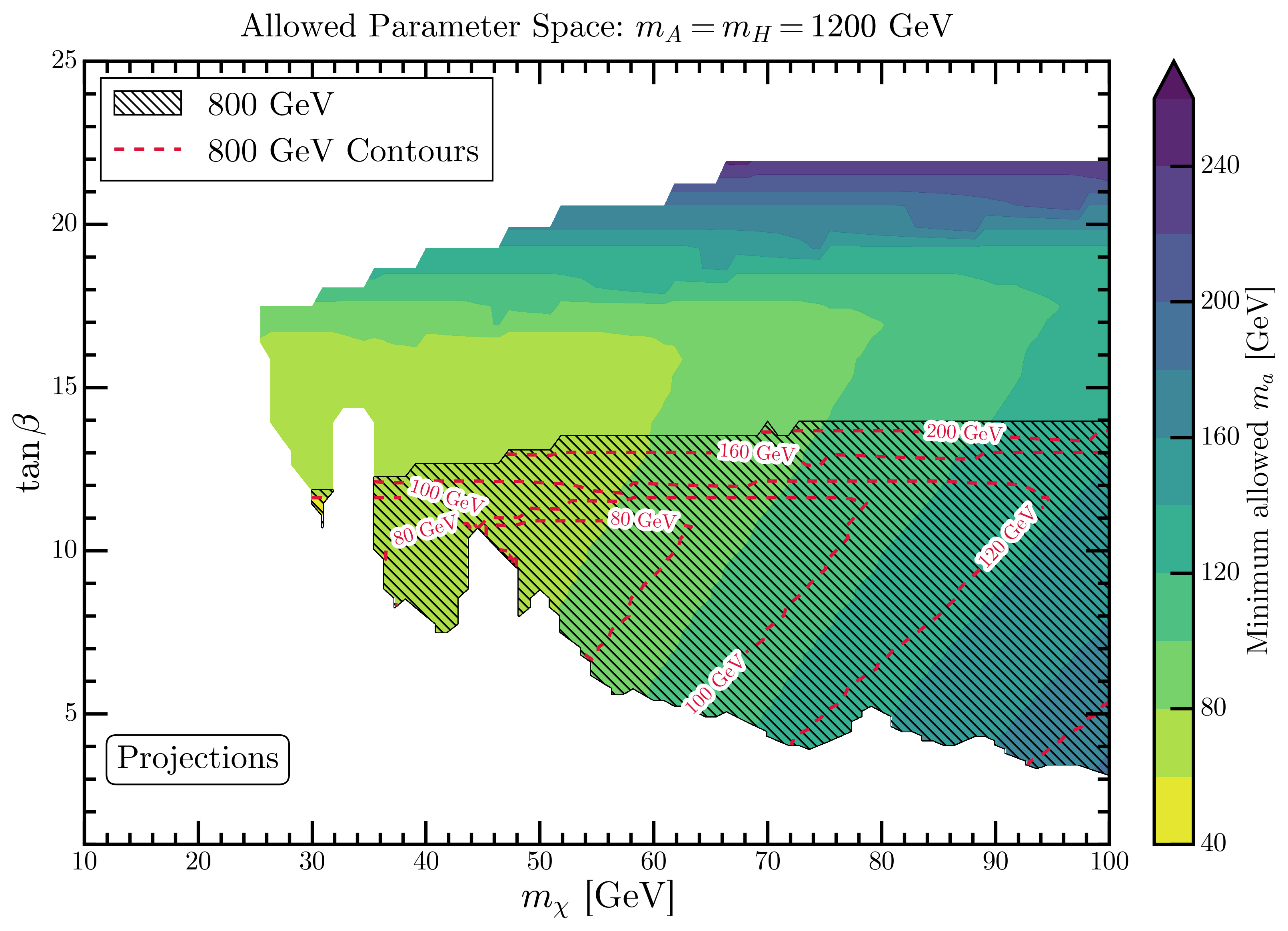}
    \caption{Analogue of Fig.~\ref{fig:scan_forecast}, but with $m_A=m_H=1.2$ TeV. We include the $m_A=m_H=800$ GeV contours for comparison.}
    \label{fig:scan_forecast-12tev}
\end{figure}

\clearpage

\bibliographystyle{JHEP}
\bibliography{ref}

\end{document}